\documentclass[]{jfm}

\usepackage{graphicx}
\usepackage{newtxtext}
\usepackage{newtxmath}
\usepackage{natbib}
\usepackage{bm}
\usepackage{hyperref}
\usepackage{ragged2e}
\hypersetup{
    colorlinks = true,
    urlcolor   = blue,
    citecolor  = blue,
}

\newcommand{\RomanNumeralCaps}[1]
\linenumbers

\title{Convective scalar transport from spherical drops in complex shearing flows}

\author{Sabarish V. Narayanan\aff{1}  
 \and Ganesh Subramanian\aff{2} \corresp{\email{sganesh@jncasr.ac.in}}}

\affiliation{\aff{1} Robert Frederick Smith School of Chemical and Biomolecular Engineering, Cornell University, Ithaca, NY, 14853, USA.
\aff{2}Engineering Mechanics Unit, Jawaharlal Nehru Centre for Advanced Scientific Research, Jakkur, Bengaluru 560064, India.}

\begin{document}
\maketitle

\begin{abstract}
We calculate the scalar\,(heat or mass) transport rate, as characterized by the Nusselt number\,($Nu$), from a neutrally buoyant spherical drop in an ambient linear flow, in the absence of inertia and in the strong convection limit. This corresponds to the regime $Re \ll 1, Pe \gg 1$, where $Re$ and $Pe$ are the Reynolds and Péclet numbers, and denote the ratios of the diffusive and convective time scales associated with momentum and scalar transport, respectively. The focus of the analysis is on the exterior problem, with the drop-phase transport resistance assumed negligible, and the scalar field therefore being a constant on the drop surface. While $Nu \propto Pe^{\frac{1}{2}}$ for $Pe \gg 1$, owing to the transport occurring across a thin $O(a Pe^{-\frac{1}{2}})$ boundary layer\,($a$ being the drop radius), the proportionality factor in this relation depends sensitively on ambient flow geometry via the surface-streamline topology. Unlike a rigid sphere, a variety of surface-streamline topologies can drive transport across the large-$Pe$ boundary layer at widely differing rates. In contrast to earlier studies which almost exclusively focus on axisymmetric ambient flows, we
calculate $Nu$ for a pair of non-axisymmetric linear flow families: (i) 3D extensional flows with aligned vorticity and (ii) Axisymmetric extensional flows with inclined vorticity; taken together, the families span the entire gamut of surface-streamline topologies in the space of incompressible linear flows. The boundary layer analysis is carried out in a surface-streamline-aligned non-orthogonal coordinate system, and nontrivial features of the calculated $Nu/Pe^{\frac{1}{2}}$-surfaces are correlated to corresponding changes in the surface-streamline topology. Independent numerical simulations of the interior transport problem reveal, for the first time, the emergence of an $O(aPe^{-\frac{1}{2}})$ boundary layer beneath the drop surface, driven by chaotic streamlines, thereby pointing to the possibility of $Nu \propto Pe^{\frac{1}{2}}$ for the conjugate problem, for sufficiently large $Pe$.
\end{abstract}


\section{Introduction}\label{1}
Transport of heat and mass in disperse multiphase systems is ubiquitous, being relevant to non-equilibrium processes that occur both in nature and industry. Examples of natural processes involving scalar\,(heat and/or mass) transport across an interface include the uptake of nutrients and biochemicals by planktonic microorganisms in the complex marine environment\,\citep{Stocker12,StockerARFM}, and condensation-induced growth of sub-Kolmogorov droplets in a turbulent cloud\,\citep{Pruppacher71,Stampfer71}. Improving the efficiency of industrial equipment such as bubble column reactors\,\citep{Joshi01} and spray driers\,\citep{Patel2009}, and that of combustion in internal combustion engines\,\citep{Law82}, also requires a fundamental understanding of transport from/to suspended drops and particles. While the examples cited above involve multiple complicating factors including unsteadiness due to a time dependent flow field\,(turbulence, vortex shedding), shape oscillations for drops, hydrodynamic interaction between drops in dense sprays, importance of transient dynamics, etc, considerable insight can nevertheless be obtained by the use of single-particle or single-drop models that assume a fixed shape and quasi-steady transport\,\citep{Ganesh24}. Calculating the transport rate for such model problems is, in any case, the first step towards incorporating the aforesaid complexities.

In this work, we examine scalar transport from a single neutrally buoyant spherical drop of viscosity $\hat{\mu}$, suspended in the shearing flow of an ambient fluid of viscosity $\mu$; the ratio $\hat{\mu}/\mu$ is denoted by $\lambda$. Both fluids are Newtonian, with the drop assumed to be sufficiently small for inertial effects to be negligible. In contrast, the Péclet number\,($Pe$) is assumed large, so that convection is dominant, and the transport of scalar occurs across a thin boundary layer on the drop surface; here, $Pe =  Ua/D \gg 1$, with $U$ being an appropriate ambient velocity scale, $a$ the drop radius and $D$ the scalar diffusivity. The assumptions above imply a large momentum-to-scalar diffusivity ratio, which is true for mass transport in liquids. Mass diffusivities in liquids are typically three orders of magnitudes smaller than momentum diffusivities, so one can have $Pe \sim O(10^3)$ even while $Re \lesssim O(1)$ in these systems; for highly viscous liquids and macromolecular or colloidal solutes, $Pe$ can often be much larger. Note that the small size of the drop, in relation to a characteristic length scale of the ambient flow, implies that the latter can be approximated as a linear flow in its vicinity\,\citep{Leal07}. A linear flow is characterized by a time scale alone, that may be taken as the inverse of the shear rate\,($\dot{\gamma}$), implying that the velocity scale $U \sim \dot{\gamma}a$, and therefore, $Pe = \dot{\gamma}a^2/D$.
The objective here is to calculate the non-dimensional scalar transport rate, defined as the ratio of the actual transport rate
 to the diffusive rate of transport\,($Pe = 0$), for specific classes of ambient 3D linear flows, in the convection-dominant limit mentioned above\,($Pe \gg 1$), and when the drop-phase transport resistance is negligibly small\,(the exterior problem). The non-dimensional transport rate above is termed the Nusselt number\,($Nu$) for heat transport, and the Sherwood number\,($Sh$) for mass transport, although we will use the former terminology for both scenarios; the normalization implies $Nu = 1$ for $Pe = 0$. 

Broadly speaking, $Nu-Pe$ relationships for the exterior problem involving spherical drops and particles are of two types. The first type arises for ambient linear flows that lead to an open-streamline topology around the particle or drop, and is characterized by $Nu$ increasing as an algebraic power of $Pe$ for $Pe \gg 1$. The enhanced large-$Pe$ transport in this case is mediated by a boundary layer on the particle or drop surface, with the $Pe$-exponent directly related to the boundary layer thickness. The latter may be obtained from a balance of convection and diffusion time scales. Assuming $\delta$ to be the boundary layer thickness, the convective time scale associated with the near-surface simple shear flow on the particle surface is $O[a/(U\delta/a)]$, and that associated with diffusion is $O(\delta^2/D)$; equating the two yields $\delta \sim a Pe^{-\frac{1}{3}}$, with $Nu \propto Pe^{\frac{1}{3}}$ for $Pe \gg 1$. The convective time scale for a drop is $O(a/U)$ which yields $\delta \sim a Pe^{-\frac{1}{2}}$, and $Nu \propto Pe^{\frac{1}{2}}$ in the same limit\citep{Leal07}. The second type of $Nu-Pe$ relationship pertains to ambient linear flows that lead to a closed-streamline topology around the particle or drop. In this case, $Nu$ increases from unity with increasing $Pe$, to begin with, but eventually saturates in a $Pe$-independent plateau for $Pe \gg 1$. The order unity plateau value is a function of the closed-streamline geometry, which in turn depends on the ambient linear flow-type parameters; for a drop, the closed-streamline geometry also depends on $\lambda$. The one-parameter family of planar linear flows\citep{Bentley_86,Leal07}, which includes simple shear flow as one of its members, is an important example in this regard. A spherical particle or drop, when immersed in a subset of this family, is enveloped by a region of closed streamlines; for a particle, the subset includes all members of the family except planar extension, while for a drop, the size of this subset depends on $\lambda$. The transport at large $Pe$ now requires the scalar to diffuse across the closed-streamline region, and this diffusion-limited transport leads to $Nu \sim O(1)$ for $Pe \rightarrow \infty$. There are exceptions to this generic two-fold classification - examples of ambient linear flows leading to $Nu \propto Pe^{\frac{1}{3}}$\,(rather than $Pe^{\frac{1}{2}}$) for drops will be discussed in section \ref{Nu:inclinedvort} of this paper\,(also see \citet{Sabarish22}).

\subsection{Literature survey} \label{sec:litsurvey}
\begin{figure}
    \centering
    \includegraphics[scale = 0.75]{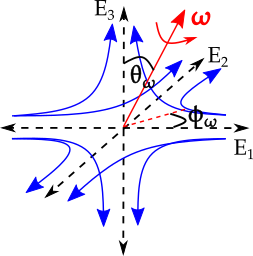}
    \caption[\textwidth]{\justifying{The geometry of an incompressible linear flow characterized by four flow-type parameters: $[\epsilon,\alpha, \theta_\omega,\phi_\omega]$.} }   \label{fig:3dlinflow}
\end{figure}

A detailed survey of the literature on scalar transport from suspended particles and drops has been presented in earlier articles\,\citep{Deepak18a,Sabarish24,Ganesh24}, and in what follows, we focus mainly on those that examine convectively enhanced transport from drops in ambient linear flows. To begin with, note that the coefficients of proportionality in the $Nu-Pe^{\frac{1}{3}}$ or $Nu-Pe^{\frac{1}{2}}$ relationships for the open-streamline scenario, and the infinite-$Pe$ $Nu$-plateau in the closed streamline scenario, are functions of the ambient linear flow-type parameters. An incompressible linear flow is defined by $\bm{u} = \bm{\Gamma} \cdot \bm{x}$, with (transpose of)\,the velocity gradient tensor $\bm{\Gamma} = \bm{E} + \bm{\Omega}$, $\bm{E}$ and $\bm{\Omega}$ being the ambient rate-of-strain and vorticity tensors. The linear flow geometry is characterized by the following flow-type parameters: $\epsilon=E_3/E_2$ measuring the non-axisymmetry of ${\bm E}$, with $E_i$ being the principal extensions\,($\sum_{i=1}^3 E_i = 0$); $\hat{\alpha}=|\bm{\omega}|/E_2$ measuring the relative magnitudes of vorticity and extension; and the polar\,($\theta_\omega$) and azimuthal\,($\phi_\omega$) angles that define the orientation of $\bm{\omega}$ in ${\boldsymbol E}$-aligned coordinates; see Figure \ref{fig:3dlinflow}. In terms of these parameters, $\bm{\Gamma}$ is given by:
\begin{align}
\bm{\Gamma} = \begin{bmatrix}
-(1+\epsilon) & -\frac{\hat{\alpha} \cos \theta_{\omega}}{2} & -\frac{\hat{\alpha} \sin \theta_{\omega} \sin \phi_{\omega}}{2}\\
\frac{\hat{\alpha} \cos \theta_{\omega}}{2} & 1 & -\frac{\hat{\alpha} \sin \theta_{\omega} \cos \phi_{\omega}}{2}\\
\frac{\hat{\alpha} \sin \theta_{\omega} \sin \phi_{\omega}}{2} & \frac{\hat{\alpha} \sin \theta_{\omega} \cos \phi_{\omega}}{2} & \epsilon 
\end{bmatrix}.  \label{Gamma:exp}
\end{align}
Accounting for axes relabeling and invariance to an overall sign change, all linear flow topologies for purposes of the $Nu$-calculation, may be shown to be covered for $\epsilon\!\in\![-2,0]$, $\hat{\alpha}\!\geq\!0$, $\theta_\omega,\phi_\omega \!\in\![0,\pi/2]$; the invariance to a sign change arises owing to $Nu$ being unchanged by a reversal of the ambient flow\,\citep{Brenner67}. In the description below, we will also use the parameter $\alpha$ as an alternative to $\hat{\alpha}$ in (\ref{Gamma:exp}); the two being related as $\alpha = \frac{1 - \hat{\alpha}/2}{1 + \hat{\alpha}/2}$, so that $\hat{\alpha}\in[0,\infty)$ maps to $\alpha \in [-1,1]$. It is worth pointing out that the number of parameters in (\ref{Gamma:exp}) is two more than the number typically use to organize incompressible linear flow topologies - in the context of 3D incompressible turbulence, sub-Kolmogorov linear-flow topologies are organized on the $QR$-plane, with $Q$ and $R$ being the quadratic and cubic invariants of $\bm{\Gamma}$\,\citep{Chong90}. The larger number of parameters is because (\ref{Gamma:exp}) also recognizes the structure of the linear flow as, for instance, characterized by the relative orientations of the (real)\,eigenvectors.

For transport from spherical particles pertaining to the open-streamline scenario, ambient linear flows in the limit $Pe \rightarrow \infty$ can be conveniently demarcated into those with and without vorticity\,\citep{Batchelor79,Ganesh24}. For the first group, which are the ambient extensional flows, the near-surface streamline topology is organized by six stagnation\,(fixed) points - pairs of stable and unstable nodes, and saddles, at diametrically opposite locations - that divide the unit sphere into identical octants, allowing for a boundary layer analysis in an orthogonal surface-stressline-aligned coordinate system, that is the analog of the Von Mises coordinates originally used in boundary layer theory\,\citep{Batchelor79}. One finds $Nu = \mathcal{F}(\epsilon) Pe^{\frac{1}{3}}$, with $\mathcal{F}(\epsilon)$ increasing monotonically from to $1.01$ to $1.22$, as the ambient flow changes from a planar\,($\epsilon=-1$) to an axisymmetric extension\,($\epsilon=-2$); see Fig.11 in \citet{Deepak18a}. For the vortical ambient linear flows, the streamlines within the boundary layer on the rotating sphere have a tightly spiralling character, which allows for an analysis in spherical polar coordinates, with transport being driven by the component of the extension along the ambient vorticity direction; one finds $Nu = 0.968 Pe_\omega^{\frac{1}{3}}$, where $Pe_\omega = E_\omega a^2/D$ is based on the component of the extension rate projected along $\bm \omega$, defined by $E_\omega = \frac{{\bm \omega} \cdot {\bm E} \cdot {\bm \omega}}{|\bm \omega|^2}$. This simple two-fold classification of linear flows is not possible for drops, since the vortical and extensional components of the near-surface flow are comparable in magnitude, leading to surface-streamline topologies considerably more complicated than the aforementioned near-surface streamline\,(stressline) topologies for spherical particles. Despite this, existing transport rate calculations for spherical drops are largely restricted to highly symmetrical scenarios. For instance, $Nu$-calculations in the regime of interest\,($Re \ll 1, Pe \gg 1$) exist for:
\begin{enumerate}
    \item An ambient uniform flow, for which $Nu = 0.461 \left(Pe/(1+\lambda)\right)^{\frac{1}{2}}$, with $Pe = Ua/D$; see Chap.9 in \citet{Leal07}.
    
    \item An axisymmetric extensional flow, for which $Nu = 0.977 \left(Pe/(1+\lambda)\right)^{\frac{1}{2}}$, with $Pe = |\dot{\gamma}|a^2/D $\,\citep{Gupalo72}. Note that the sign of $\dot{\gamma}$ determines whether the extensional flow has a uniaxial or biaxial character, with $Nu$ being invariant to this change in sign, as already mentioned above.
    
    \item A combination of the two flows above, with the uniform flow aligned with the symmetry axis of the extensional component, which leads to\,\citep{Gupalo75}:
    \begin{align}
    Nu =& \frac{1}{(8\pi)^{\frac{1}{2}}} \left[\left(1+\left|\frac{U}{6\dot{\gamma}a}\right|\right)^{\frac{3}{2}} \left(\left|\frac{6\dot{\gamma}a}{U}\right|-\frac{1}{3}\right)^{\frac{1}{2}} + \left(1-\left|\frac{U}{6\dot{\gamma}a}\right|\right)^{\frac{3}{2}} \left(\left|\frac{6\dot{\gamma}a}{U}\right|+\frac{1}{3}\right)^{\frac{1}{2}} \right] (\frac{Pe}{1+\lambda})^{\frac{1}{2}}, \,\,\text{for}\,\, \left|\frac{6\dot{\gamma}a}{U}\right|>1, \label{Nu:extdom} \\
    Nu =& 0.461 (\frac{Pe}{1+\lambda})^{\frac{1}{2}}, \,\,\text{for}\,\, \left|\frac{6\dot{\gamma}a}{U}\right| \leq1, \label{Nu:unifdom}
    \end{align}
    where the threshold $\dot{\gamma}a/U = \pm 1/6$ marks a change in the streamline topology from a two-roll interior pattern corresponding to uniform flow being dominant, to a four-roll pattern expected for a dominant extensional component; the dominant flow component determines the nature of the inlet and wake flows, as well as the wake locations, in the large-$Pe$ exterior boundary layer analysis. Note that $Nu$ remains unchanged on a reversal of the extensional component. Further, and rather surprisingly, for $|\dot{\gamma}a/U| < 1/6$, $Nu$ turns out to be exactly the same as that for uniform flow\,(compare (\ref{Nu:unifdom}) with item (i)), despite the boundary layer thickness profile being a function of $\dot{\gamma}a/U$. Interestingly, this independence is also seen in a closely related transport scenario, that from a translating spherical squirmer\,\citep{MagarPedley_2003}, where the rate of transport is independent of the dipolar component of the disturbance field\,(that plays the role of the extensional component above) below a finite threshold\,\citep{Ganesh24}.
    
    \item A planar extensional flow, for which $Nu = 0.737 \left(Pe/(1+\lambda)\right)^{\frac{1}{2}}$, with $Pe = |\dot{\gamma}|a^2/D$; see \citet{Polyanin84}. 
\end{enumerate}

The surface-streamline topologies for (i)-(iv) mimic the near-surface stream\,(stress)lines mentioned for spherical particles above. In that, the surface-streamline pattern either has an eight-fold symmetric character, as for planar extension\,(item (iv)), allowing for the use of the analog of the Von-Mises coordinates employed by \citet{Batchelor79}; or, is axisymmetric and without swirl, allowing for the straightforward use of spherical polar coordinates, as true for the remaining ambient flows\,(items (i), (ii) and (iii)). 

\cite{Deepak18a} were the first to extend the large-$Pe$ $Nu$-calculation to drops in more general ambient linear flows, based on the insight that surface streamlines on a spherical drop are described by an equation of the same form as the one describing the trajectories of an axisymmetric particle, both in an ambient linear flow\,\citep{jeffery_1922,Bretherton_1962}. The equation is given by $\dot{\bm n} = \bm{\Omega} \cdot \bm{n} + {\mathcal G}[\bm{E} \cdot \bm{n} - (\bm{E}:\bm{n}\bm{n})\bm{n}]$, where the unit normal $\bm n$ replaces the particle orientation vector, and the overdot denotes the time rate of change. Further, the geometry-dependent Bretherton constant in the original axisymmetric particle problem, is now replaced by ${\mathcal G}$ that is a function of $\lambda$. The above insight allowed the boundary layer analysis, and the subsequent $Nu$-calculation, to be carried out in a novel non-orthogonal surface-streamline-aligned coordinate system. For a spheroid rotating along closed Jeffery orbits in (a subset of)\,ambient planar linear flows, the aforementioned coordinate system reduces to $(C,\tau)$ coordinates, with $C$ and $\tau$ being the orbit constant and phase, respectively. \citet{Hinch71} and \citet{Leal72} solved the Smoluchowski equation in these coordinates, in order to analytically characterize the orientation distribution of a Brownian spheroid in the convection-dominant limit\,(large rotational Peclet numbers), with the above coordinates allowing them to account separately for the orientation dependence along and across Jeffery orbits. Note that $\mathcal G = \frac{\kappa^2-1}{\kappa^2+1}$ for spheroids, $\kappa$ being the spheroid aspect ratio, with the subset of planar linear flows that lead to rotation along closed orbits corresponding to $\alpha < 1/\kappa^2\,(< \kappa^2)$ for prolate\,(oblate) spheroids. 

\citet{Deepak18a} used the above coordinate system to calculate $Nu$ for a pair of one-parameter linear flow families - 3D extensional flows parameterized by $\epsilon$\,(with $\alpha = 1$), and planar linear flows parameterized by $\alpha$\,(with $\epsilon = \theta_\omega =0)$; note that $\alpha = 1$\,(or $\hat{\alpha} = 0$), with $\epsilon = \theta_\omega = 0$, corresponds to planar extension, a limiting member of both families examined. Their calculation for ambient extensional flows generalized the effort of \citet{Polyanin84}, in accounting for all flows intermediate between axisymmetric and planar extension. The planar linear flow calculation was for the interval $\alpha\in[\frac{\lambda}{2+\lambda},1]$ which corresponds to flows that lead to an open surface-streamline topology. While surface streamlines are still organized by six fixed points, in contrast to extensional flows, the pattern in ambient planar linear flows only exhibits a four-fold symmetry, owing to the diametrical segments connecting the pairs of stable and unstable nodes not being orthogonal to each other; the segments coincide at the marginal value, $\alpha = \frac{\lambda}{2+\lambda}$, when all surface streamlines having a degenerate meridional character. For smaller $\alpha$\,(or, alternatively, with $\alpha$ fixed, for $\lambda > 2\alpha/(1-\alpha) = \lambda_c$), the surface streamlines transform to closed Jeffery orbits with an effective aspect ratio that is now a function of $\alpha$ and $\lambda$, being given by $\kappa_e = \left[\frac{\lambda_c+\alpha\lambda}{\alpha(\lambda-\lambda_c)}\right]^{1/2}$\citep{Torza71,Powell83,Deepak18b}. As a result, the $Nu-Pe$ relationship transitions to the second type mentioned earlier, with $Nu$ asymptoting to an $(\alpha,\lambda)$-dependent plateau in the limit $Pe \rightarrow \infty$. The singular effect of weak inertia on diffusion-limited transport at large $Pe$, for $\alpha\in(-1,\frac{\lambda}{2+\lambda})$ was examined in \citet{Deepak18b}. Using $C$ and $\tau$ as tangential coordinates, and the parameter labelling the exterior closed stream-surfaces as the radial coordinate, the authors showed that the spiralling streamlines induced by inertia eliminated the diffusion limitation, and $Nu \propto (Re Pe)^{\frac{1}{2}}$ in the limit $Re \ll 1, RePe \gg 1$. 

Although less symmetric than the ambient linear flows examined in earlier efforts mentioned above\,\citep{Leal07,Gupalo72,Gupalo75,Polyanin84}, the ambient flows examined by \citet{Deepak18a} are nevertheless restrictive when considered in the context of the full 4D parameter space mentioned above. This is evident from the interior streamline topology for extensional flows and planar linear flows - almost all interior streamlines in all such flows are regular closed curves\,\citep{Singeetham24}, in contrast to the chaotic nature of the interior streamlines in a generic linear flow\,\citep{Stone91,Sabarish21}. Herein, we move beyond this restriction by extending the $Nu$-calculation in \citet{Deepak18a} to the following two-parameter linear flow families:\\
(i) 3D extensional flows with vorticity aligned along one of the principal axes, that are parameterized by $(\epsilon,\alpha)$, and\\ (ii) axisymmetric extensional flows with the vorticity vector inclined to the axis of symmetry, that are parameterized by $(\alpha,\theta_\omega)$.\\
These families include the one-parameter families examined in \citet{Deepak18a} as limiting cases, but are shown to have chaotically wandering streamlines, thereby being more representative of a generic linear flow. The boundary layer analysis for these flows is carried out in the aforementioned surface-streamline-aligned coordinate system. Expressions for these coordinates\,(again denoted by $C$ and $\tau$), for families (i) and (ii) above, are obtained by invoking the notion of an auxiliary linear flow with a velocity gradient tensor that is also a function of $\lambda$\,\citep{Deepak18a,Ganesh24}; streamlines on the drop surface in the actual ambient flow are projections, onto the unit sphere, of the streamlines of the auxiliary linear flow. Following earlier efforts on the classification of sub-Kolmogorov linear-flow topologies in turbulence\,\citep{Chong90}, a pair of $\lambda$-dependent scalar invariants of the auxiliary velocity gradient tensor may be used to organize the drop surface-streamline topologies. The $\lambda$-dependence of the said invariants is such that a changing viscosity ratio only leads to an overall shift in the flow-type parameter intervals corresponding to different topologies, without the introduction of new ones. This in turn suggests that the exterior problem transport-rate calculations, for the above linear flow families, are best characterized in terms of $Nu/Pe^{\frac{1}{2}}$ surfaces, plotted as a function of the pair of flow-type parameters appropriately rescaled by $\lambda$.

In addition to the boundary layer analysis for the exterior problem, we perform Langevin tracer simulations for the interior problem. The transport resistance in the ambient phase is now assumed to be negligibly small, so that the constant-scalar boundary condition at infinity is directly imposed on the drop surface. The simulations show that, in sharp contrast to the diffusion-limited transport scenario that arises for interior closed-streamline topologies\,\citep{Singeetham24}, maximally chaotic streamline topologies\,(as characterized by Poincare sections) lead to the emergence of an $O(Pe^{-\frac{1}{2}})$ interior boundary layer at the largest $Pe$ values, with $Nu \propto Pe^{\frac{1}{2}}$ in this limit. For these cases, the transport rate calculation for the exterior problem may be extended to that for the conjugate problem, with comparable transport resistances in the drop and ambient phases, and where the large-$Pe$ transport is mediated by both interior and exterior boundary layers\,\citep{Deepak18a}. 

\subsection{Organisation of the paper}
We begin in section \ref{sec:setup} by writing down the governing convection-diffusion equation and defining the Nusselt number. Next, in Section \ref{sec:flow_org}, based on the scalar invariants of an auxiliary velocity gradient tensor, we organize the different surface-streamline topologies associated with families (i) and (ii) above. This is followed by a large-$Pe$ boundary-layer analysis in $C-\tau$ coordinates in section \ref{sec:Nusselt_calc}, which simplifies the convective terms, enabling a solution for the scalar field by means of a similarity transformation, and thence, an expression for $Nu$. For both families, $Nu \propto Pe^{\frac{1}{2}}$ almost everywhere in the relevant parameter space. The boundary layer analysis yields closed form expressions for the proportionality factor that is a function of $(\alpha,\epsilon,\lambda)$ for the first family, and of $(\theta_\omega,\alpha, \lambda)$ for the second. Interestingly, the analysis for the second family highlights a special class of linear flows, the {\it eccentric} elliptic flows\,\citep{Sabarish22}, for which surface streamlines are closed and resemble eccentric versions of the Jeffery orbits above, but near-surface streamlines are not. As a result, drops in these flows mimic spherical particles, with $Nu \propto Pe^{\frac{1}{3}}$ for $Pe \rightarrow \infty$. Finally, in Section \ref{sec:conc}, we discuss the broader implications of the work and future directions. Appendix \ref{auxiliaryflow:map} discusses the relation between the solution for the surface streamlines in spherical coordinates, and that obtained using the auxiliary linear flow interpretation. In Appendix \ref{App:interior_topo}, we examine the interior problem for the aforementioned linear flow families in some detail. This involves first characterizing the interior streamlines for members of these families, using Poincare sections, and then using Langevin simulations to determine the $Nu-Pe_i$ relationships\,($Pe_i$ being the internal Péclet number). The Poincare sections reveal chaotically wandering streamlines, which lead to $Nu$ growing algebraically with $Pe_i$. Although the exponent is in general smaller than $1/2$ and ambient flow-type dependent, it equals $1/2$ for nearly space-filling Poincare sections, in which case one also sees the emergence of an interior boundary layer. Appendices \ref{SmallNu_asy} and \ref{LargeNu_asy} obtain analytical approximations for $Nu$ in the limits of small and large $\hat{\alpha}$, respectively, which serve to validate the numerical evaluation of the $Nu$-integral, obtained from the boundary layer analysis, for general values of the flow-type parameters.

\section{Problem definition and governing equations} \label{sec:setup}
The scalar field $T(\bm{x},t)$ is governed by a convection-diffusion equation, with the convecting velocity field independently determined by momentum conservation, as is typical of forced convection problems. Further, assuming constancy of fluid properties, and using $a$ and $\dot{\gamma}a$ as the length and velocity scales, the governing equation takes the following dimensionless form:
\begin{align}
    Pe (\bm{u} \cdot \nabla \Theta) = \nabla^2 \Theta, \label{HEQ1}
\end{align}
at steady state, where $\Theta = (T - T_\infty)/(T_0 - T_\infty)$ and $Pe = \dot{\gamma}a^2/D$. The boundary conditions at the drop surface, and at infinity, take the form:
\begin{align}
    &\Theta = 1 \text{ at } r = 1, \label{HBC1} \\
    &\Theta = 0 \text{ as } r \rightarrow \infty, \label{HBC2} 
\end{align}
for the exterior problem. The velocity field  in (\ref{HEQ1}) is the Stokes flow around a spherical drop in an ambient linear flow, and is given by\,\citep{Leal07}
\begin{align}
    &\bm{u} = \bm{\Omega}\cdot \bm{r} +  \left[ 1 - \frac{\lambda}{(1+\lambda)r^5} \right] \bm{E} \cdot \bm{r} + \left[ \frac{5 \lambda}{2(1+\lambda)r^7} - \frac{5\lambda +2}{2(1+\lambda)r^5} \right](\bm{E}:\bm{rr})\bm{r}, \label{Sol1}
\end{align}
where $\bm{E} = (\bm{\Gamma} + \bm{\Gamma}^\dag)/2$ and $\bm{\Omega} = (\bm{\Gamma} - \bm{\Gamma}^\dag)/2$ are the rate of strain and (transpose of)\,vorticity tensors, with $\bm{\Gamma}$ having been defined in (\ref{Gamma:exp}). Having determined $\Theta$, the Nusselt number is calculated using:
\begin{align}
    Nu = -\frac{1}{4 \pi}\displaystyle\int \left( \frac{\partial \Theta}{\partial r} \right)_{r=1}d\Omega, \label{Nudef}
\end{align}
where $d\Omega$ is a differential areal element on the unit sphere\,(drop surface). We only examine the large-$Pe$ transport for linear flows leading to an open surface-streamline topology. The transport in almost all these cases occurs across an $O(Pe^{-\frac{1}{2}})$ boundary layer, and as a result, $\frac{\partial \Theta}{\partial r}|_{r=1}$ in (\ref{Nudef}) is $O(Pe^{\frac{1}{2}})$, leading to $Nu \propto Pe^{\frac{1}{2}}$. As mentioned earlier, the focus is on the proportionality factor in this relation which is a function of the flow-type parameters\,($\epsilon$, $\alpha$ or $\hat{\alpha}$, $\theta_\omega$ and $\phi_\omega$) and $\lambda$.

The velocity field on the drop surface that, at leading order, convects the scalar field within the boundary layer, is obtained by setting $r = 1$ in \eqref{Sol1}. Recognizing that the surface velocity field is the rate of change of the unit normal $\bm{n}$, one then obtains:
\begin{align}
    \dot{\bm{n}} = \bm{\Omega} \cdot \bm{n} + \mathcal G(\lambda) [\bm{E}\cdot \bm{n}-(\bm{E}:\bm{nn})\bm{n})], \label{surffield}
\end{align}
with $\mathcal G(\lambda) = 1/(1+\lambda)$. The surface streamlines are obtained by solving (\ref{surffield}), which leads to closed-form expressions for the $C$ and $\tau$ coordinates. Now, for the linear extensional flows and planar linear flows examined in \cite{Deepak18a}, substitution of the specific forms of $\bm{E}$ and $\bm{\Omega}$ leads to equations that are readily solved analytically. For a general linear flow, a closed-form solution of the coupled nonlinear system (\ref{surffield}) is still possible, since \eqref{surffield} is the projection, onto the unit sphere, of the streamlines of the auxiliary linear flow defined by:
\begin{align}
    \dot{\bm{r}} = \hat{\bm{\Gamma}} \cdot \bm{r}, \label{auxfield}
\end{align}
where (transpose of)\,the auxiliary velocity gradient tensor is given by $\hat{\bm{\Gamma}} = \bm{\Omega} + \mathcal{G}(\lambda) \bm{E}$. For a bubble\,($\lambda = 0$), $\hat{\bm{\Gamma}} = \bm{\Gamma}$, and the auxiliary flow is identical to the ambient one. Being a linear system with constant coefficients, $\bm{r}(t)$ is readily obtained in closed form, and the solution of (\ref{surffield}) is then obtained as $\bm{n} = \bm{r}/|\bm{r}|$. 
For the linear flow families examined here, (\ref{surffield}) leads to coupled nonlinear equations in spherical coordinates, and the auxiliary-flow-based approach is therefore better suited to obtaining expressions for $C$ and $\tau$.\\ 

With the above background, we introduce the two families of linear flows to be analyzed:
\begin{itemize}
    \item The first is the two-parameter family of non-axisymmetric extensional flows with vorticity aligned along one of the principal axes. Depending on the choice of axis, the aligned orientation can correspond to any of $\theta_\omega = 0$, $[\theta_\omega,\phi_\omega] \equiv [\pi/2,0]$ or $[\theta_\omega,\phi_\omega] \equiv [\pi/2,\pi/2]$, with the $\epsilon$-interval, needed to include all members of the family, dependent on this choice; see Figure \ref{fig:3dlinflow}. Without loss of generality, one may choose $\theta_\omega = 0$, so (\ref{Gamma:exp}) takes the form:
    \begin{align}
    \bm{\Gamma} = \begin{bmatrix}
        -(1+\epsilon) &-\frac{\hat{\alpha}}{2} &0 \\
        \frac{\hat{\alpha}}{2} &1 &0\\
        0 &0 &\epsilon
    \end{bmatrix}, \label{alignedvort_Gamma}
\end{align}
with $\epsilon \in [-2,0]$. Regardless of the choice of principal axis, it is sufficient to examine $\hat{\alpha} \in [0, \infty)$, owing to $Nu$ being invariant to $\hat{\alpha} \leftrightarrow -\hat{\alpha}$. As $\epsilon$ increases from $-2$ to $-1$, the linear flow transitions from an axisymmetric biaxial extension\,($E_1 = E_2 = 1$, $E_3 = -2$) with the vorticity vector aligned along the axis of symmetry, to a planar extension\,($E_1 = 0$, $E_2 = 1$, $E_3 = -1$) with vorticity aligned with the compressional axis. As $\epsilon$ increases further to $-1/2$, the flow transitions to a uniaxial extension\,($E_1 = -1$, $E_2 = 1$, $E_3 = 0$) with vorticity orthogonal to the symmetry axis. Finally, for $\epsilon \rightarrow 0^-$, one approaches the one-parameter family of planar linear flows examined by \cite{Deepak18a}, with vorticity perpendicular to the plane of extension; $\hat{\alpha} = 0\,(\alpha = 1)$, $\hat{\alpha} = 2\,(\alpha = 0)$ and $\hat{\alpha} = \infty\,(\alpha = -1)$ correspond to planar extension, simple shear flow and solid-body rotation, respectively, in this limit.  For $\epsilon$ values outside the interval $[-2,0]$, one obtains linear flows that (to within a scaling factor)\,are time-reversed versions of those in the said interval. Specifically, the flows for $\epsilon \in [0,\infty)$ are time-reversed versions of those for $\epsilon \in [0,-1)$, while those for $\epsilon \in (-\infty,-2]$ are time-reversed versions of the ones in $(-1,-2]$.

The above family reduces to the one-parameter family of linear extensional flows for $\hat{\alpha} = 0$, in which case it suffices to consider the subset $\epsilon \in [-2,-1]$. One now traverses the sequence of non-axisymmetric 3D extensional flows that interpolate between biaxial\,($\epsilon = -2$) and planar\,($\epsilon = -1$) extensions. A rotated or time-reversed version of this sequence arises for $\epsilon \in [-1,-1/2]$, and again for $\epsilon \in [-1/2,0]$. The flows in these latter intervals are not repetitions for $\hat{\alpha} \neq 0$ owing to the differing vorticity orientation. Note also that, for their extensional flow $Nu$-calculation, \citet{Deepak18a} used $\epsilon \in [0,1]$ which comprises time-reversed versions of the flows in $[-2,-1]$.

As will be seen below in Section \ref{sec:flow_org}, for a drop immersed in any of the linear flows with a non-zero $\epsilon$, the surface streamlines change from a non-spiralling to a spiralling topology across a threshold $\hat{\alpha}$. For $\epsilon \rightarrow 0^-$, the threshold is given by $\hat{\alpha} = \frac{2}{1+\lambda}$, and the spiralling streamlines approach closed Jeffery orbits. Thus, for $\epsilon = 0$, one transitions from a non-spiralling to a closed Jeffery-orbit topology across $\hat{\alpha} = \frac{2}{1+\lambda}$; as already described in section \ref{1}, the transport above the said threshold is diffusion limited in the limit $Pe \rightarrow \infty$\,\citep{Deepak18b}.\\

\item Second is the two-parameter family of axisymmetric extensions with the vorticity vector inclined to the symmetry axis of the extensional component. The streamline pattern, to within a rotation of axes, is the same for any $\phi_\omega$, and we choose $\phi_\omega = 0$, so (\ref{Gamma:exp}) takes the form: 
    \begin{align}
    \bm{\Gamma} = \begin{bmatrix}
        1 &-\frac{\hat{\alpha}\cos \theta_\omega}{2} & 0 \\
        \frac{\hat{\alpha}\cos\theta_\omega}{2} &1 & -\frac{\hat{\alpha}\sin\theta_\omega}{2}\\
         0 & \frac{\hat{\alpha}\sin\theta_\omega}{2} & - 2
    \end{bmatrix}. \label{inclinedvort_Gamma}
\end{align}
Accounting for invariance of the surface-streamline topology, and $Nu$, to a reversed vorticity orientation\,(this corresponded to the transformation $\hat{\alpha} \leftrightarrow -\hat{\alpha}$ in the aligned-vorticity case), all members of this family are covered for $\theta_\omega \in [0, \pi/2]$ and $\hat{\alpha} \in [0, \infty)$. The limiting cases of $\hat{\alpha} = 0$ and $\hat{\alpha} = \infty$ correspond to axisymmetric extension and solid-body rotation, respectively, with $Nu$ for the former flow having been calculated by \citet{Gupalo72}, as mentioned in section \ref{sec:litsurvey}. The complementary limits of $\theta_\omega = 0$ and $\frac{\pi}{2}$, with $\hat{\alpha}$ arbitrary, are already covered by the aligned-vorticity family above. Note that $\theta_\omega = 0$ leads to an axisymmetric spiralling streamline configuration on the drop surface, and may therefore also be analyzed using spherical polar coordinates with the symmetry axis chosen as the polar axis; for $\hat{\alpha}$ large, this is similar to the boundary layer analysis of \citet{Batchelor79} for a spherical particle in vortical linear flows. The conclusion is that the transport rate only depends on the rate of extension, and $Nu$ for $\theta_\omega = 0$ is therefore independent of $\hat{\alpha}$ in the limit $Pe \rightarrow \infty$. 

As mentioned in section \ref{1}, surface streamlines are (trivially)\,closed circles on a rotating spherical particle, and lead to diffusion-limited transport in planar linear flows with vorticity. A more nontrivial analog of this scenario will be seen in section \ref{Nu:inclinedvort} where, for the inclined-vorticity family, the locus $\theta_\omega = \frac{1}{2}\cos^{-1}\left[-\frac{16 + ((1+\lambda) \hat{\alpha})^2}{3 ((1+\lambda)\hat{\alpha})^2}\right];\, \hat{\alpha} \in [\sqrt{12},\infty)$ defines ambient linear flows that lead to a closed surface-streamline, but spiralling near-surface streamline, topology. The surface streamlines here are not circles, but instead constitute a generalization of the Jeffery orbits encountered for $\epsilon = 0$ in the aligned-vorticity family above\,\citep{Sabarish21,Sabarish22}. Although not diffusion-limited, the closed surface-streamline topology will nevertheless be shown to result in an asymptotically smaller transport rate for large $Pe$. 
\end{itemize}

\section{Organization of surface-streamline topologies} \label{sec:flow_org}

Herein, we delineate all possible streamline topologies on the drop surface for the linear flow families defined above, and then proceed to develop expressions for the flow-aligned coordinates\,($C$ and $\tau$) to be used for the boundary layer analysis. Surface-streamline topologies are classified based on the scalar invariants of $\hat{\bm{\Gamma}}$, defined by $P = Tr(\hat{\bm{\Gamma}}) = \hat{\bm{\Gamma}}\!:\!{\bm I}$, $Q = (1/2)[Tr(\hat{\bm{\Gamma}})^2 - Tr(\hat{\bm{\Gamma}}^2)]$, and $R = \det(\hat{\bm{\Gamma}})$, and that are the coefficients of the characteristic equation satisfied by the eigenvalues: $\xi^3 + P \xi^2 + Q \xi + R = 0$. As mentioned in section \ref{1}, the above invariants, evaluated using the instantaneous turbulent velocity gradient tensor, have earlier been used to examine the Lagrangian structure of sub-Kolomogorov turbulence\citep{Chong90,Girimaji_2010}. Since the Kolmogorov scale defines the smallest eddies, the flow on smaller scales can be approximated as a temporally fluctuating linear flow. For incompressible linear flows, $P = 0$, and the key element in the above classification is the cubic discriminant, $\Delta = 4Q^3 + 27R^2$, that changes sign\,(from negative to positive) as the streamlines transition from a non-spiralling to a spiralling topology. This coincides with a change from $\hat{\bm \Gamma}$ having three real eigenvalues, to having one real eigenvalue and a complex-conjugate pair. In what follows, plots of $\Delta$, as a function of the auxiliary linear flow parameters, are therefore used to organize the drop-surface-streamline topologies since the latter mirror the topology of the auxiliary flow\,(defined by $\hat{\bm \Gamma}$). The $\Delta$-plots are accompanied by plots of the cubic invariant $R$. Zero crossings of the latter help identify (auxiliary)\,planar linear flows. As discussed in the introduction, such flows are often associated with closed surface-streamline topologies and thence, reduced transport rates. Note that $\Delta \propto Q$ for planar linear flows, and it is therefore the sign of $Q$ that differentiates between open\,($Q<0$) and closed\,($Q>0$) streamline topologies.

\subsection{Spherical drop in an ambient non-axisymmetric extension with aligned vorticity} \label{sec:alignedfamily}

\subsubsection{Surface streamline topology} \label{topology_alignedvort}
Using $\hat{\bm \Gamma} = \bm{\Omega} + \mathcal G(\lambda) \bm{E}$, with $\bm \Gamma$ given by (\ref{alignedvort_Gamma}), one obtains:
\begin{align}
    &\Delta' = -\frac{[(2+\epsilon)^2 - \hat{\alpha}'^2][4(1 + \epsilon - 2\epsilon^2) - \hat{\alpha}'^2]^2}{16}, \label{disc1} \\
    &R' = \frac{\epsilon(4 + 4 \epsilon - \hat{\alpha}'^2)}{4}, \label{R1}
\end{align}
where $\Delta' = \Delta(1+\lambda)^6$, $\hat{\alpha}' = \hat{\alpha}(1+\lambda)$ and $R' = R(1+\lambda)^3$; the dependence of $\Delta'$ and $R'$ on $\hat{\alpha}'^2$ being consistent with the $\hat{\alpha} \leftrightarrow -\hat{\alpha}$ invariance of the streamline topology. The ability to write down (\ref{disc1}) and (\ref{R1}) solely in terms of $\lambda$-rescaled invariants, and a similarly rescaled vorticity parameter, implies that the viscosity ratio only acts to shift the sequence of surface-streamline topologies along the $\hat{\alpha}$-axis. As will be seen below, increasing $\lambda$ for a fixed $\hat{\alpha}$ amplifies the effect of ambient flow vorticity, leading in general to a larger region in parameter space that corresponds to spiralling surface-streamline topologies. In the limit $\lambda \rightarrow \infty$, any $\hat{\alpha} \neq 0$ leads to circular surface streamlines, consistent with those on a rotating spherical particle in an ambient vortical 
flow\,\citep{Batchelor79}. Importantly, (\ref{disc1}) and (\ref{R1}) imply that organizing the streamline topologies only requires one to examine $\Delta'$ as a function of $\hat{\alpha}'$ and $\epsilon$, rather than $\Delta$ as a function of $\hat{\alpha}$, $\epsilon$ and $\lambda$.
 
We begin by noting that $\Delta' = 0$, the locus that separates spiralling and non-spiralling surface-streamline topologies in general, has solutions given by:
\begin{align}
    &\hat{\alpha}'_{th1} = (2 + \epsilon), \\
    &\hat{\alpha}'_{th2} = 2(1 + \epsilon -2\epsilon^2)^{\frac{1}{2}}.
\end{align}
Here, $\hat{\alpha}'_{th1}$ is a simple zero-crossing for $\epsilon > -2$ and is always real valued. In contrast, $\hat{\alpha}'_{th2}$ is real valued only for $-1/2 \leq \epsilon \leq 0$, and corresponds to a point of tangency at the $\hat{\alpha}'$-axis, so that $\Delta' = d\Delta'/d\hat{\alpha}' = 0$ at $\hat{\alpha}' =  \hat{\alpha}'_{th2}$. 
For $\epsilon = 0$, $\hat{\alpha}'_{th1} = \hat{\alpha}'_{th2} = 2$, leading to a triply degenerate point\,($\Delta' = d\Delta'/d\hat{\alpha}' = d^2\Delta'/d^2\hat{\alpha}' = 0$) on the $\hat{\alpha}'$-axis that separates non-spiralling and closed-streamline topologies for planar linear flows - this corresponds to the threshold $\alpha = \lambda/(2+\lambda)$, or $\hat{\alpha} = 2/(1+\lambda)$, mentioned in section \ref{1}, and helps highlight the role of $\lambda$ in merely shifting the interval of open-streamline topologies, without qualitative alteration. Along similar lines, $R' = 0$ has a solution given by:
\begin{align}
    \hat{\alpha}'_{th3} = 2 (1 + \epsilon)^{\frac{1}{2}},
\end{align}
which is real-valued for $-1 \leq \epsilon < 0$. Thus, for any $\epsilon$ in this range, one has two intervals of 3D auxiliary linear flows, separated by a planar linear flow at $\hat{\alpha}'=\hat{\alpha}'_{th3}$. For $\epsilon = 0$ alone, $R'$ is zero for all $\hat{\alpha}'$, so that both the auxiliary and actual ambient flows are always planar linear flows.\,Based on the above, it is convenient to demarcate the $\Delta'-\hat{\alpha}'$ plots into the following cases:
(i) $-2 \leq \epsilon < -1/2$, 
(ii) $-1/2 \leq \epsilon < 0$, 
and (iii) $\epsilon = 0$.
 
\begin{figure}
    \centering
    \includegraphics[scale = 0.35]{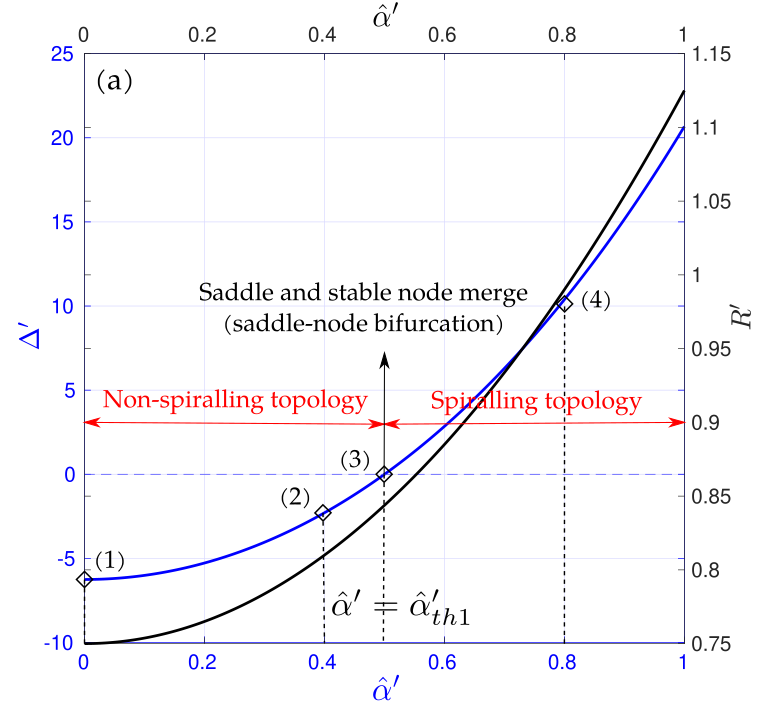}
    \includegraphics[scale = 0.34]{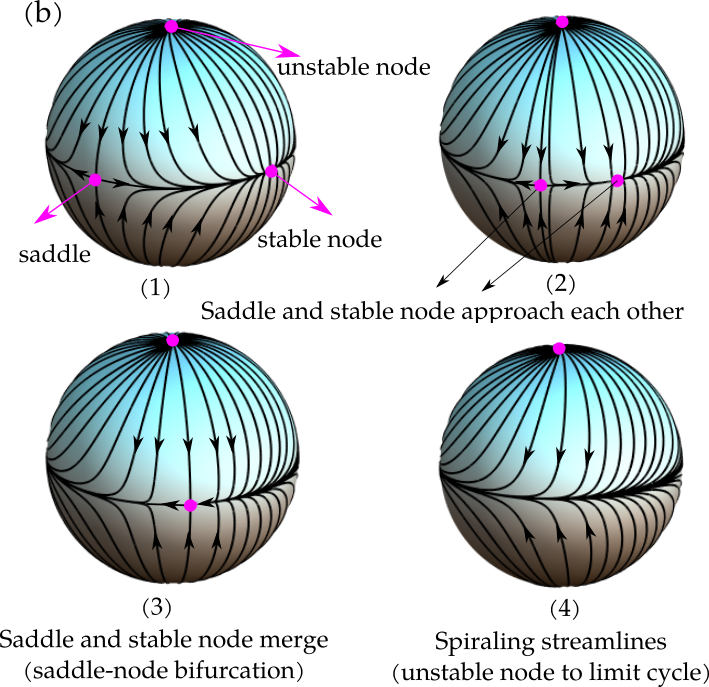}\\
    \includegraphics[scale = 0.335]{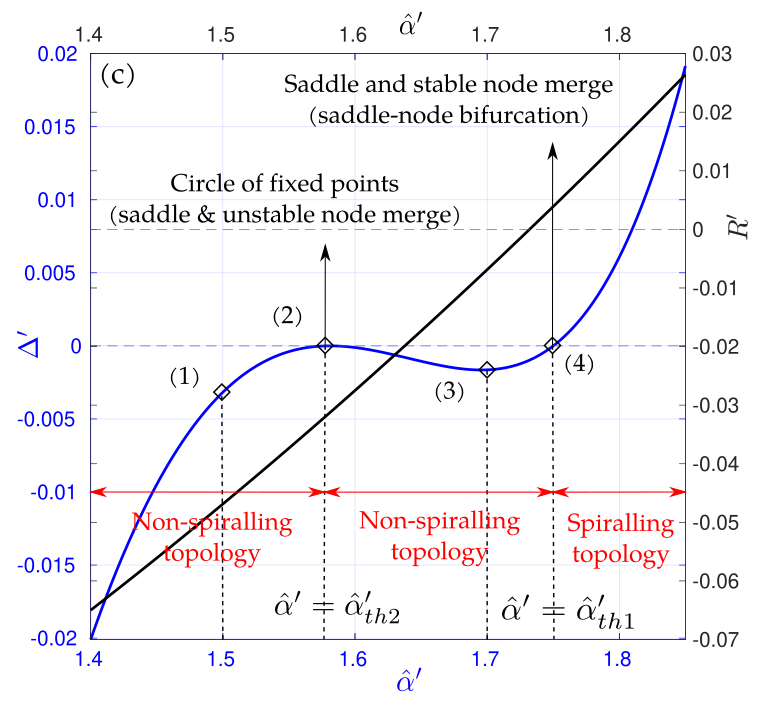}
    \includegraphics[scale = 0.34]{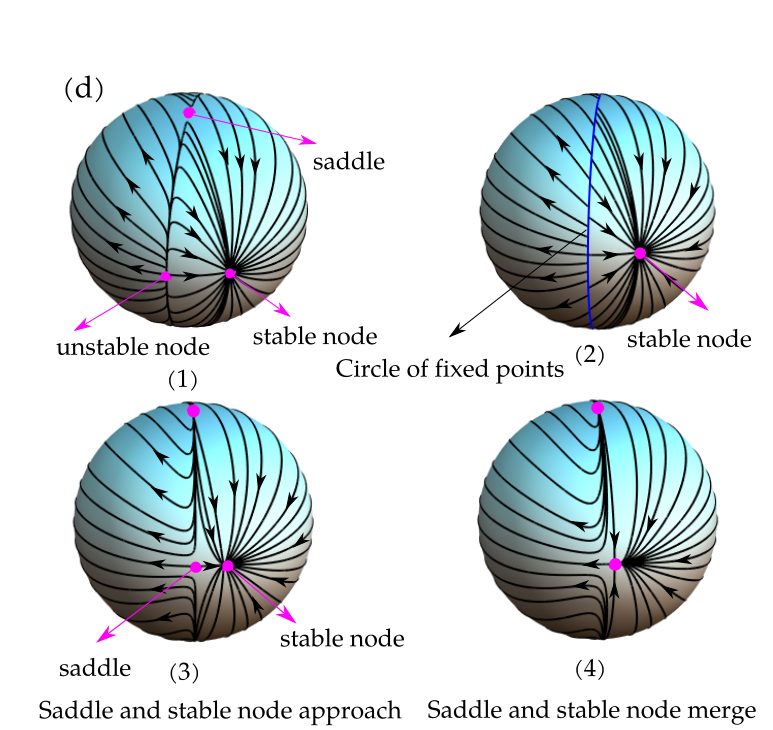}\\
    \includegraphics[scale = 0.335]{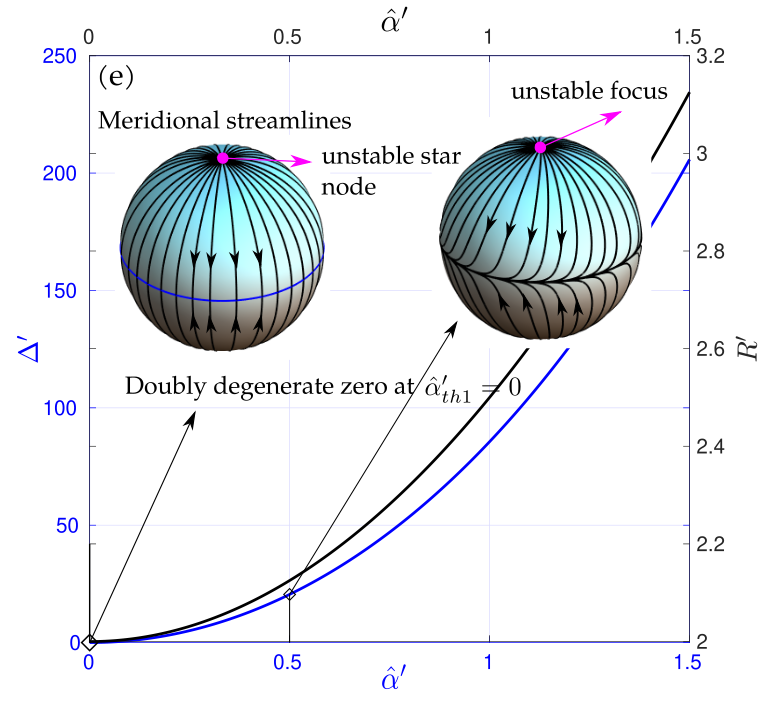}
    \includegraphics[scale = 0.335]{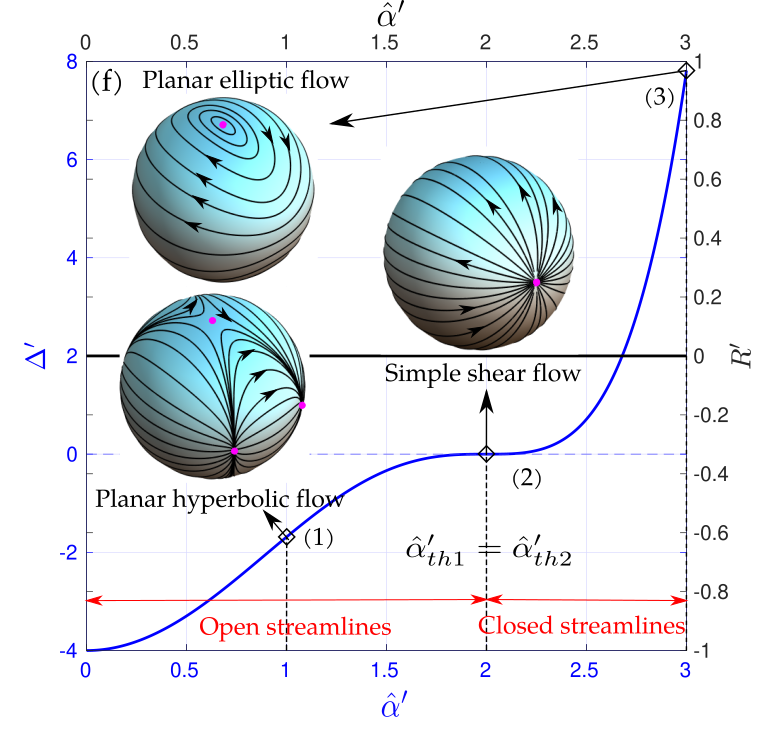}
    \caption[\textwidth]{\justifying{Organization of surface-streamline topologies for the aligned-vorticity family. The figures show plots of $\Delta'$ and $R'$, as a function of $\hat{\alpha}'$, for different $\epsilon$: (a) $\epsilon = -3/2$ (case (i):\,$-2 \leq \epsilon < -1/2$); (b) shows surface-streamline topologies corresponding to the marked points\,(open diamond symbols) in (a). (c) $\epsilon = -1/4$ (case (ii):\,$-1/2 \leq \epsilon < 0$); (d) shows surface-streamline topologies corresponding to the marked points in (c). (e) and (f) correspond to $\epsilon = -2$ and $\epsilon = 0$\,(case (iii)), respectively. Surface-streamline topologies are embedded within these two plots; those in (f) show the transition from open to closed surface streamlines for planar linear flows with increasing vorticity.}}
    \label{fig:discandstream}
\end{figure}

Figs.\ref{fig:discandstream}a and b, with $\epsilon\in(-2,-1)$, depict the $\Delta'-\hat{\alpha}'$, $R'-\hat{\alpha}'$ plots and surface-streamline topologies for case (i). In Fig.\ref{fig:discandstream}a, $\Delta'$ has a single zero crossing at $\hat{\alpha}' = \hat{\alpha}'_{th1}$, while $R' > 0$ for all $\hat{\alpha}'$. The streamline topology for negative $\Delta'$, corresponding to $0 < \hat{\alpha}' < \hat{\alpha}'_{th1}$, is non-spiralling, with the surface streamlines organized by saddles and stable nodes in the equatorial\,($x_1-x_2$) plane, and unstable nodes at the two poles\,(along the $x_3$-axis). The stable and unstable manifolds of the saddles, that connect them to the nodes, divide the drop surface into eight sectors, the three vertices of each sector being a saddle, a stable and an unstable node; unlike the extensional flows examined by \citet{Deepak18a}, however, there is no eight-fold symmetry. From the non-spiralling streamline topologies depicted by the first two unit spheres in Fig.\ref{fig:discandstream}b, one sees that, as $\hat{\alpha}'$ increases towards $\alpha'_{th1}$, the stable node and saddle approach each other, leading to two of the sectors\,(on the unit hemisphere) shrinking in relation to the other two. As $\hat{\alpha}'$ crosses $\hat{\alpha}'_{th1}$, the aforesaid fixed points disappear in a saddle-node bifurcation, with the unstable nodes simultaneously transitioning to unstable foci. The third unit sphere in Fig.\ref{fig:discandstream}b exhibits the degenerate topology at $\hat{\alpha}' = \hat{\alpha}'_{th1}$, where there are four fixed points that include a pair of unstable star nodes, and a pair of degenerate saddle-nodes. The saddle-node bifurcation leads to a spiralling-streamline topology for $\hat{\alpha}' > \alpha'_{th1}$, with all streamlines starting at either of the unstable foci, and approaching the equatorial limit cycle for long times. The spiralling becomes tighter with increasing $\hat{\alpha}'$, with the individual turns of a spiral approaching circles for $\hat{\alpha}' \rightarrow \infty$, corresponding to solid-body rotation. 

For the limiting case of $\epsilon = -2$, $\hat{\alpha}'_{th1} = 0$ and is a point of tangency. The interval of non-spiralling topologies has shrunk to zero, and one starts from the degenerate topology for an axisymmetric extensional flow at $\hat{\alpha}' = 0$, with meridional surface streamlines connecting a pair of unstable star nodes to an equatorial fixed-point ring. For any non-zero $\hat{\alpha}'$, the star nodes become unstable foci, and the fixed ring gives way to an equatorial limit cycle, leading to a spiralling-streamline topology. The $\Delta'-\hat{\alpha}'$ and $R'-\hat{\alpha}'$ plots for this case, with embedded unit-sphere streamline topologies are shown in Fig.\ref{fig:discandstream}e. For the other limiting case, $\epsilon = -1$, the sequence of streamline topologies\,(not shown) remains the same as for $\epsilon\in(-2,-1)$ above, with the only difference being that $R' = 0$ at $\hat{\alpha}' = 0$. The auxiliary flow for $\hat{\alpha}' = 0$ is therefore a planar extension, although the surface-streamline topology remains the same as that for the general case described in the earlier paragraph. For $-1 < \epsilon < -1/2$,  $R'$ has a zero-crossing at a positive $\hat{\alpha}'\,(=\hat{\alpha}'_{th3})$. Although not a planar extension as for $\epsilon = -1$, the auxiliary flow at this $\hat{\alpha}'$ is still a planar hyperbolic flow\,(an eccentric version of the usual planar hyperbolic flows; see \citet{Sabarish22}). Nevertheless, the sequence of surface-streamline topologies remains the same as the ones shown in Figs.\ref{fig:discandstream}a and b. As will be seen in Section \ref{sec:incled_stream_org} below, a qualitative alteration of the surface streamlines arises only when the planar flow is an elliptic or parabolic one. Also note that $\Delta'$ is a monotonically increasing function of $\hat{\alpha}'$, as in Fig.\ref{fig:discandstream}a, only for $-1 < \epsilon < -(\sqrt{3}-1)$. It exhibits a non-monotonic variation in the interval $[0,\hat{\alpha}'_{th1}]$ for $-(\sqrt{3}-1) < \epsilon < -1/2$, starting from a local maximum at $\hat{\alpha}' = 0$, and going through a minimum at a finite $\hat{\alpha}'$. However, since $\Delta'_{max}$ and $\Delta'_{min}$ are both negative, there is no qualitative change in the sequence of streamline topologies.

For case (ii), corresponding to $-1/2 \leq \epsilon < 0$, $\Delta'$ has a simple zero crossing at $\hat{\alpha}' = \hat{\alpha'}_{th1}$ and a doubly degenerate zero at $\hat{\alpha}' = \hat{\alpha}'_{th2}\,(< \hat{\alpha'}_{th1})$; the $R'$ zero-crossing at $\hat{\alpha}' = \hat{\alpha}'_{th3}$ again plays no role. The zero at $\hat{\alpha}' = \hat{\alpha}'_{th2}$ arises from the aforementioned local maximum $\Delta'_{max}$ increasing to zero, as a result of which the $\Delta'$-curve first becomes tangent to the $\hat{\alpha}'$-axis, at $\hat{\alpha}'_{th2} = 0$, for $\epsilon = -1/2$; the point of tangency moves to positive $\hat{\alpha}'$ for $\epsilon > -1/2$. Fig.\ref{fig:discandstream}c shows that, for $\epsilon > -1/2$, $\hat{\alpha}'_{th2}$ divides the interval $[0,\hat{\alpha}'_{th1}]$ into two sub-intervals, with each of these corresponding to a negative $\Delta'$, and thence, to non-spiralling streamline topologies. The point $\hat{\alpha}' = \hat{\alpha}'_{th2}$ itself corresponds to a degenerate meridional topology, albeit a non-canonical one, as seen in Fig.\ref{fig:discandstream}d\,(second unit sphere). Unlike the case $\epsilon = -2,\hat{\alpha}' = 0$ in Fig.\ref{fig:discandstream}e, the great circle of fixed points\,(shown in blue) is now inclined to the diametrical segment that connects the pair of stable nodes. In effect, this fixed-point ring mediates a transition between distinct non-spiralling streamline topologies, with the transition involving an exchange of fixed-point identities\,(between the unstable node and saddle). The subsequent transition from the non-spiralling topology in the second sub-interval $(\hat{\alpha}'_{th2},\hat{\alpha}'_{th1})$, to a spiralling one for $\hat{\alpha}' > \alpha'_{th1}$, is analogous to that already described for $\epsilon\in[-2,-1/2)$ above. For the special case of $\epsilon = -1/2$, when $\hat{\alpha}'_{th2} = 0$, the first sub-interval of non-spiralling streamlines vanishes. One now starts with an axisymmetric meridional topology at $\hat{\alpha}' = 0$, one that is identical to that for $\epsilon = -2, \hat{\alpha}' = 0$\,(Fig.\ref{fig:discandstream}e). In contrast to the latter case, however, one first transitions to a non-spiralling topology with increasing $\hat{\alpha}'$, prior to a spiralling one\,(for $\hat{\alpha}' > \hat{\alpha}'_{th1}$).

Finally, for case (iii) corresponding to $\epsilon = 0$, $R' = 0$ for all $\hat{\alpha}'$, and the auxiliary flow belongs to the one-parameter family of planar linear flows. One starts with planar extension at $\hat{\alpha}' = 0$, culminating in solid-body rotation for $\hat{\alpha}' \rightarrow \infty$. Now, $\Delta' = 4[Q(1+\lambda)^2]^3$, and as shown in Fig.\ref{fig:discandstream}f, is negative for $\hat{\alpha}' < 2$. This corresponds to a non-spiralling streamline topology induced by a planar hyperbolic auxiliary linear flow with $Q < 0$. $\Delta' = Q = 0$ at $\hat{\alpha}' = 2$, a triply degenerate point; the auxiliary flow at this $\hat{\alpha}'$ is simple shear flow, and induces a meridional surface-streamline topology with a fixed-point ring in the $x_1-x_2$ plane. For $\alpha' > 2$, $Q > 0$ and the auxiliary flow is an elliptic linear flow, leading to closed\,(rather than spiralling) surface streamlines; as mentioned in section \ref{1}, the closed streamlines are Jeffery orbits. The $Nu$-calculation for the open-streamline topology, corresponding to $\hat{\alpha}'\in[0,2]$, was carried out by \cite{Deepak18a}. 

\subsubsection{The $C - \tau$ coordinate system} \label{sec:C_tau_aligned}
When expressed in spherical polar coordinates, (\ref{surffield}) and (\ref{alignedvort_Gamma}) lead to the following ODEs governing the surface streamlines:
\begin{align}
    &\frac{d \theta}{dt} = -\frac{(3 \epsilon + (2+\epsilon) \cos 2\phi)\sin 2\theta}{4(1+\lambda)}, \label{ode1}\\
    &\frac{d \phi}{dt} = \frac{(1+\lambda)\hat{\alpha} + (2+\epsilon)\sin 2\phi}{2(1+\lambda)}. \label{ode2}
\end{align}
While the above system is readily solved owing to \eqref{ode2} being decoupled from \eqref{ode1}, we nevertheless adopt the auxiliary-flow-based approach for solving \eqref{ode1}-\eqref{ode2}, since this also works for the fully coupled system obtained for the inclined-vorticity family. The auxiliary velocity gradient tensor is now given by:
\begin{align}
    \hat{\bm{\Gamma}} = \begin{bmatrix}
        -\frac{(1+\epsilon)}{1 + \lambda} &-\hat{\alpha}/2 &0 \\
        \hat{\alpha}/2 &\frac{1}{1+\lambda} &0\\
        0 &0 & \frac{\epsilon}{1 + \lambda}
    \end{bmatrix}, \label{aux1}
\end{align}
and one can solve \eqref{auxfield} exactly to get:
\begin{align}
    &x_1 = \frac{e^{-\frac{(\epsilon + A)}{2(1+\lambda)}t} \left[ \left( (2 + \epsilon + A) - e^{\frac{A}{1 + \lambda}t}(2 + \epsilon - A)\right)x_1^0  - (e^{\frac{A}{1 + \lambda}t} - 1)(1+\lambda)\hat{\alpha} \,x_2^0\right]}{2 A}, \label{x1sol} \\
    &x_2 = \frac{e^{-\frac{(\epsilon + A)}{2(1+\lambda)}t} \left[ \left(e^{\frac{A}{1 + \lambda}t} (2 + \epsilon + A) - (2 + \epsilon - A)\right) x_2^0 + (e^{\frac{A}{1 + \lambda}t} - 1 )(1+\lambda)\hat{\alpha}\,x_1^0\right]}{2 A}, \label{x2sol} \\
    &x_3 = e^{\frac{ \epsilon}{1+\lambda}t} x_3^0, \label{x3sol}
\end{align}
 with $(x_1^0,x_2^0,x_3^0)$ being the location at $t = 0$. Here, $A = \sqrt{(2 + \epsilon)^2 - (1+\lambda)^2 \hat{\alpha}^2} = \sqrt{(\hat{\alpha}'_{th1})^2- \hat{\alpha}'^2}$, and the discussion in section \ref{topology_alignedvort} shows that non-spiralling and spiralling surface-streamline topologies correspond to $A$ being real and imaginary, respectively. Dividing each of the above expressions by $r = \sqrt{x_1^2 + x_2^2 + x_3^2}$, and using $x_1/r = \sin\theta \cos\phi$, $x_2/r=\sin\theta \sin\phi$, $x_3/r = \cos\theta$, one obtains after some manipulation: 
 \begin{align}
&\tan \phi = \frac{x_2^0}{x_1^0}\frac{(A-(2+\epsilon)-(1+\lambda)\hat{\alpha}\frac{x_1^0}{x_2^0}) + e^{\tau}(A+(2+\epsilon)+(1+\lambda)\hat{\alpha}\frac{x_1^0}{x_2^0})}{(A+(2+\epsilon)+(1+\lambda)\hat{\alpha}\frac{x_2^0}{x_1^0}) + e^{\tau}(A-(2+\epsilon)-(1+\lambda)\hat{\alpha}\frac{x_2^0}{x_1^0})}, \label{auxsol1new_int} 
\end{align}
\begin{equation}
\begin{split}
\tan \theta = &\frac{C}{2A} (1+\tan^2 \phi)^{1/2} e^{-\frac{(A + 3\epsilon)}{2A}\tau}x \\
&[(A+(2+\epsilon)+(1+\lambda)\hat{\alpha}\frac{x_2^0}{x_1^0}) + e^{\tau}(A-(2+\epsilon)-(1+\lambda)\hat{\alpha}\frac{x_2^0}{x_1^0})], 
\end{split} \label{auxsol2new_int}
\end{equation}
for the non-spiralling case. $C = x_1^0/x_3^0$ in (\ref{auxsol2new_int}) serves as the surface streamline label, and $\tau = At/(1 + \lambda)$ in (\ref{auxsol1new_int}) and (\ref{auxsol2new_int}) serves as the modified time variable along a given streamline. While not obvious from the above expressions, the actual value of $x_2^0/x_1^0$ only serves to set the origin for time along a given streamline, this being consistent with the autonomous nature of the governing ODEs. Thus, one may cover the entire unit hemisphere by choosing $\tan \phi_0 = x_2^0/x_1^0 =\pm 1$ - the two choices correspond to pairs of diagonally opposite sectors on the unit hemisphere, and lead to the following simpler expressions: 
\begin{align}
&\tan \phi = \frac{(A-B) \pm e^{\tau}(A+B)}{(A+B) \pm e^{\tau}(A-B)}, \label{auxsol1new1} \\
    &\tan \theta = \frac{C}{2A} (1+\tan^2 \phi)^{1/2} e^{-\frac{(A + 3\epsilon)}{2A}\tau}[(A+B) \pm e^{\tau}(A-B)], \label{auxsol2new1}
\end{align}
where $B = (2+\epsilon) + (1+\lambda) \hat{\alpha}$, with both $\tau$ and $C$ having different definitions for the two choices of $\phi_0$ above. It is shown in Appendix \ref{auxiliaryflow:map} that all $x_2^0/x_1^0\in (-\Lambda - (\Lambda^2-1)^{\frac{1}{2}},-\Lambda + (\Lambda^2-1)^{\frac{1}{2}})$ lead to (\ref{auxsol1new1}) and (\ref{auxsol2new1}) with the positive sign in front of the exponential, while $x_2^0/x_1^0\in (-\infty,-\Lambda - (\Lambda^2-1)^{\frac{1}{2}}),(-\Lambda + (\Lambda^2-1)^{\frac{1}{2}},\infty)$ lead to the same expressions but with the negative sign; here, $\Lambda = \frac{2+\epsilon}{\hat{\alpha}(1+\lambda)}$. Within each of these intervals, the only difference the choice of $x_2^0/x_1^0$ makes is that the $\tau$ in the exponential in (\ref{auxsol1new1}) and (\ref{auxsol2new1}) is defined with respect to a different origin, one that is a function of $x_2^0/x_1^0$. 
  
For the spiralling case, one defines $A = \iota A' = \iota \sqrt{\hat{\alpha'}^2 - (\hat{\alpha}'_{th1})^2}$, in which case (\ref{x1sol}) and (\ref{x2sol}) lead to:
\begin{align}
    &\tan \phi = \frac{x_2^0}{x_1^0}\frac{(\mathrm{i}A'-(2+\epsilon)-(1+\lambda)\hat{\alpha}\frac{x_1^0}{x_2^0}) + e^{\mathrm{i}\tau}(\mathrm{i}A'+(2+\epsilon)+(1+\lambda)\hat{\alpha}\frac{x_1^0}{x_2^0})}{(\mathrm{i}A'+(2+\epsilon)+(1+\lambda)\hat{\alpha}\frac{x_2^0}{x_1^0}) + e^{\mathrm{i}\tau}(\mathrm{i}A'-(2+\epsilon)-(1+\lambda)\hat{\alpha}\frac{x_2^0}{x_1^0})}, \label{auxsol1new_spi} 
\end{align}
\begin{equation}
\begin{split}
\tan \theta = &\frac{C}{2\mathrm{i}A'} (1+\tan^2 \phi)^{1/2} \\
&e^{-\frac{(\mathrm{i}A + 3\epsilon)}{2\mathrm{i}A}\mathrm{i}\tau}[(\mathrm{i}A'+(2+\epsilon)+(1+\lambda)\hat{\alpha}\frac{x_2^0}{x_1^0}) + e^{\mathrm{i}\tau}(\mathrm{i}A'-(2+\epsilon)-(1+\lambda)\hat{\alpha}\frac{x_2^0}{x_1^0})],    
\end{split}\label{auxsol2new_spi}
\end{equation}
with $\tau = A't/(1+\lambda)$. It is shown in Appendix \ref{auxiliaryflow:map} that these may be rewritten in the more transparent form:
\begin{align}
    &\tan \phi = \frac{A' \tan(\frac{\tau -\tau_0}{2}) -(2+\epsilon)}{\hat{\alpha}(1+\lambda)}, \label{Spiralling_tau} \\
    &\tan \theta = C (1+\lambda)\hat{\alpha} \left( \frac{(1+\tan^2 \phi)}{1 + (\frac{(2+\epsilon) + (1+\lambda)\hat{\alpha} \tan \phi}{A'})^2}\right)^{1/2} \exp(\frac{3|\epsilon|\tau}{2 A'}), \label{Spiralling_C}
\end{align}
with $\tau_0$ being a function of $x_2^0/x_1^0$.

The relations (\ref{auxsol1new1})-(\ref{auxsol2new1}) and (\ref{Spiralling_tau})-(\ref{Spiralling_C}) define the $C$ and $\tau$ coordinates for the non-spiralling and spiralling cases, respectively. We now explain how the unit hemisphere is mapped out in terms of $C$ and $\tau$; mapping of the other hemisphere follows from inversion symmetry of the surface-streamline pattern, which in turn arises from that of the ambient linear flow. We begin with the non-spiralling scenario where, as evident from Figs.\ref{fig:discandstream}c and d, there are multiple cases to be considered depending on $\hat{\alpha}'$. The significance of $\hat{\alpha}'_{th1}$ for the $(C,\tau)$ mapping is evident from its appearance in $A$ above, while that of $\hat{\alpha}'_{th2}$ follows from vanishing of the factor $(A + 3\epsilon)$, in the argument of the exponential in (\ref{auxsol2new1}), at $\hat{\alpha}' = \hat{\alpha}'_{th2}$.

\begin{figure}   \centering
    \includegraphics[scale = 0.65]{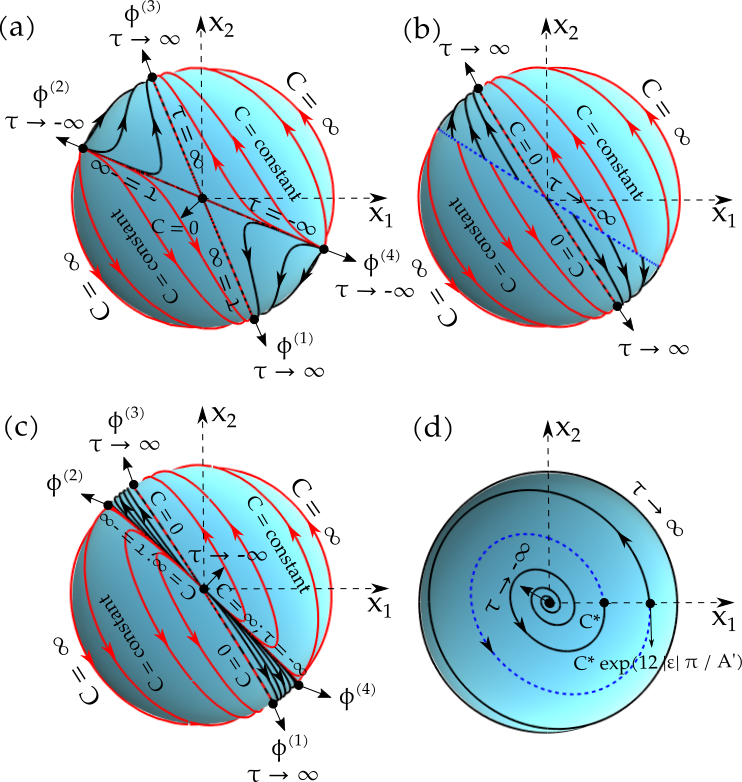}
    \caption[\textwidth]{\justifying{The $C - \tau$ mapping of surface streamlines, for the aligned-vorticity family, for $\hat{\alpha}' < (2 + \epsilon)$\,(non-spiraling regime) shown in (a) for $\hat{\alpha}' < \hat{\alpha}'_{th2}$, in (b) for $\hat{\alpha}' = \hat{\alpha}'_{th2}$, and in (c) for $\hat{\alpha}' > \hat{\alpha}'_{th2}$. Note that $\phi^{(2)}$ and $\phi^{(4)}$ are unstable nodes in (a), and saddle points in (c); the blue dashed line in (b) denotes the unstable fixed-point ring. Subfigure (d) corresponds to $\hat{\alpha}' >(2 + \epsilon)$\,(spiraling regime). The one turn of the spiralling streamline, highlighted by the dashed blue curve, defines the $C$-limits used  in the $Nu$ calculation in section \ref{Nu:inclinedvort}.}}\label{fig:Ctau_aligned}
\end{figure}

For $\hat{\alpha}' < \hat{\alpha}'_{th2} <\hat{\alpha}'_{th1}$, $A, A+3\epsilon > 0$. From (\ref{auxsol1new1}), one then finds $\tan \phi \rightarrow (\frac{A \pm B}{A\mp B})$ for $\tau \rightarrow \pm \infty$ regardless of $C$, and from (\ref{auxsol2new1}), $\theta \rightarrow \frac{\pi}{2}$ for $\tau \rightarrow \pm \infty$; (\ref{auxsol2new1}) also implies $C = \infty$ maps to $\theta = \frac{\pi}{2}$ for any finite $\tau$. This implies there are four fixed points on the great circle\,($\theta = \frac{\pi}{2}$ or $C = \infty$) in which the unit sphere intersects the $x_1-x_2$ plane, and that are defined by $\phi^{(1)} = \tan^{-1}(\frac{A + B}{A - B})$, $\phi^{(3)}= \pi+ \tan^{-1}(\frac{A + B}{A - B})$ and $\phi^{(2)} = \tan^{-1} (\frac{A - B}{A + B})$, $\phi^{(4)} = \pi + \tan^{-1} (\frac{A - B}{A + B})$. $(\phi^{(1)},\frac{\pi}{2})$ and $(\phi^{(3)},\frac{\pi}{2})$ are stable nodes, while  $(\phi^{(2)},\frac{\pi}{2})$ and $(\phi^{(4)},\frac{\pi}{2})$ are unstable nodes. Almost all surface streamlines, including the ones in the $x_1-x_2$ plane, are constant-$C$ trajectories that originate at an unstable node for $\tau \rightarrow  -\infty$, and end at either of the stable nodes in the limit $\tau \rightarrow \infty$. The exceptions are the stable and unstable manifolds of the saddle point at the pole\,($\theta = 0$ which corresponds to $C = 0$ for any finite $\tau$), which connect it to the nodes in the $x_1-x_2$ plane. These are meridional arcs defined by $\phi = \phi^{(1)},\phi^{(3)}$\,(unstable manifold) and $\phi = \phi^{(2)},\phi^{(4)}$\,(stable manifold), with $\theta\in (0,\frac{\pi}{2})$; in $(C,\tau)$ coordinates, these curves correspond to $C = 0,\tau = \pm \infty$. In effect, the stable and unstable manifolds divide the unit hemisphere into four sectors\,(octants), corresponding to the angular intervals $\phi^{(1)} < \phi < \phi^{(2)}$ (Region 1), $\phi^{(2)} < \phi < \phi^{(3)}$ (Region 2), $\phi^{(3)} < \phi < \phi^{(4)}$ (Region 3) and $\phi^{(4)} < \phi < \phi^{(1)}$ (Region 4). 
Fig.\ref{fig:Ctau_aligned}a summarizes the $C-\tau$ mapping of the unit hemisphere for $\hat{\alpha}' < \hat{\alpha}'_{th2}$.

For $\hat{\alpha}' = \hat{\alpha}'_{th2}$, $A > 0, A + 3\epsilon = 0$. The expression for $\tan\phi$ remains unchanged, while that for $\tan \theta$ simplifies on substitution of (\ref{auxsol1new1}) into (\ref{auxsol2new1}), giving
\begin{equation}
\tan \theta = \frac{C}{2^{\frac{1}{2}}A}[(A^2 + B^2)(1+e^{2\tau}) + 2e^\tau (A^2-B^2)]^{\frac{1}{2}}.
\end{equation}
This implies $\tan \theta \rightarrow \frac{C}{2^{\frac{1}{2}}A}(A^2+B^2)^{\frac{1}{2}}$ for $\tau \rightarrow -\infty$. Thus, although $\phi \rightarrow \phi^{(2)},\phi^{(4)}$ for $\tau \rightarrow -\infty$, $\theta$ in this limit can still take all values in the interval $[0,\frac{\pi}{2}]$ for $C\in[0,\infty)$, which traces out the inclined fixed-point ring\,(in blue) on the second unit sphere in Fig.\ref{fig:discandstream}d. All surface streamlines are again constant-$C$ trajectories, but now originate at different points on this ring, while asymptoting to either $(\phi^{(1)},\frac{\pi}{2})$ or $(\phi^{(3)},\frac{\pi}{2})$ for $\tau \rightarrow \infty$. Fig.\ref{fig:Ctau_aligned}b summarizes the $C-\tau$ mapping of the unit hemisphere for $\hat{\alpha}' = \hat{\alpha}'_{th2}$\,(the fixed point ring now depicted as a dashed blue curve).

For $\hat{\alpha}'_{th2} <\hat{\alpha}' < \hat{\alpha}'_{th1}$, $A > 0, A+3\epsilon < 0$. In this case, the exponential in (\ref{auxsol2new1}) is written in the form $e^{\frac{|A + 3\epsilon|}{2A}\tau}$, and one then finds $\theta \rightarrow 0$ for $\tau \rightarrow -\infty$ for any $C$, implying that the unstable nodes are now located at the poles. $\theta \rightarrow \frac{\pi}{2}$ for $\tau \rightarrow \infty$, and the stable nodes therefore still correspond to $\phi = \phi^{(1)}, \phi^{(3)}$ in the equatorial plane. The saddle points are also now in the equatorial plane. Thus, as mentioned in the context of Figs.\ref{fig:discandstream}c and d, the emergence of the fixed-point ring leads to the unstable nodes and saddles exchanging places. The unstable manifold of the saddle point corresponds to all finite $\tau$, with $C= \infty$\,(this ensures $\theta = \frac{\pi}{2}$, with $\phi$ ranging between any two of the $\phi^{(i)}$'s), while the stable manifold corresponds to $C = \infty$, $\tau = -\infty$, with $\theta$ being arbitrary. Fig.\ref{fig:Ctau_aligned}c summarizes the $C-\tau$ mapping of the unit hemisphere for this case.

The $(C,\tau)$-mapping of the unit hemisphere for the spiralling case, corresponding to $\hat{\alpha}' > \hat{\alpha}'_{th1}$, is shown in Fig.\ref{fig:Ctau_aligned}d. Now, $C$ and $\tau$ cannot both range over infinite intervals since this will lead to a single spiralling streamline having multiple\,(indeed, infinitely many) labels. To be consistent with the non-spiralling case above, a given streamline is still taken to correspond to $\tau \in (-\infty,\infty)$ with $C$ fixed. From (\ref{Spiralling_C}), $\tau = -\infty$ is seen to denote the unstable focus\,($\theta = 0$) from which the spiralling streamline originates, with $\tau = \infty$ denoting the equatorial limit cycle\,($\theta = \frac{\pi}{2}$) to which it asymptotes to, for infinite time. The interval of $C$ must then correspond to the finite range of $\theta$ values traversed by any one of the spiralling streamlines over a single turn\,(that is, $\phi$ ranging over a $2\pi$-interval). Choosing a streamline with label $C^*$\,(say) and increasing $\phi$ from $0$ to $2\pi$, (\ref{Spiralling_C}) shows that $C \in [C^*,C^*\mathrm{e}^{12 \epsilon \pi /A'})$; $C =C^*\mathrm{e}^{12 \epsilon \pi /A'}$ corresponding to the original streamline after a complete $2\pi$-turn. Thus, in the spiralling regime, $\tau \in (-\infty,\infty)$, $C \in [C^*,C\mathrm{e}^{12 \epsilon \pi /A'})$ covers the unit hemisphere for any finite nonzero $C^*$; the arbitrariness in the choice of $C^*$ reflects the aforementioned freedom in choosing the streamline when determining the range of $C$. 

The novel aspect about the $C-\tau$ system above is that it is non-orthogonal. The contravariant unit vectors, that is, the unit vectors along the constant coordinate lines, are given by:
\begin{align}
    &\hat{\bm{C}} = \hat{\bm{\theta}}, \label{C_uvec}\\
    &\hat{\bm{\tau}} = \cos \beta \,\hat{\bm{\theta}} + \sin \beta\, \hat{\bm{\phi}}, \label{tau_uvec}
\end{align}
where $\hat{\bm \theta}$ and $\hat{\bm \phi}$ are the unit vectors of a spherical coordinate system defined on the unit sphere with its polar axis aligned with the $x_3$-direction. The skewness angle $\beta$ in (\ref{tau_uvec}), and other related quantities, are defined by:
\begin{align}
    &\sin \beta = \sin \theta \phi_\tau (\theta_\tau^2 + \sin^2 \theta \phi_\tau^2)^{-1/2}, \label{metricfactor1}\\
&h = g_{CC} = \theta_C, \label{metricfactor2}\\
    &k = g_{\tau \tau} = (\theta_\tau^2 + \sin^2 \theta \phi_\tau^2), \label{metricfactor3}\\
    &g_{C \tau} = \theta_C \theta_\tau, \label{metricfactor4}
\end{align}
where $\theta_C$, $\theta_\tau$ etc. denote partial derivatives with respect to the subscripted variables and $g_{CC}$, $g_{\tau \tau}$ and $g_{C \tau}$ are the components of the metric tensor\,\citep{Aris12}. While $\beta = \frac{\pi}{2}$ for an orthogonal coordinate system, it is a function of $C$ and $\tau$ in the present case. From (\ref{C_uvec}), $\hat{\bm{C}}$ is seen to point along the meridional direction regardless of $\hat{\alpha}$, and thence, regardless of whether the surface streamlines have a spiralling or non-spiralling character; $\hat{\bm{\tau}}$ may be shown to be oriented tangent to a surface streamline, which immediately implies the aforementioned non-orthogonality. Further, $\beta \rightarrow 0$ for $\epsilon \rightarrow -2$, $\hat{\alpha} \rightarrow 0$, so that the coordinate system becomes degenerate for a meridional streamline topology - $Nu$ for an axisymmetric extension can therefore only be evaluated the result of a limiting process.

\subsection{Spherical drop in an ambient axisymmetric extension with inclined vorticity} \label{sec:inclinedfamily}

\subsubsection{Surface streamline topology} \label{sec:incled_stream_org}

\begin{figure}
    \centering
    \includegraphics[scale=0.5]{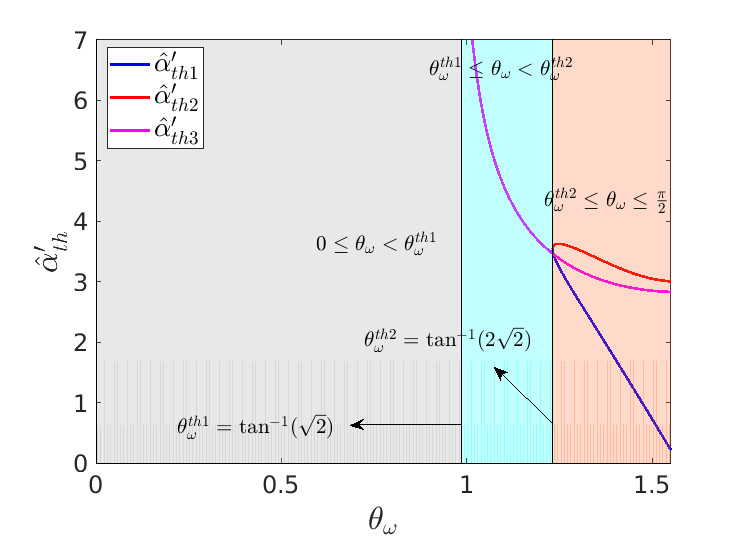}
    \caption[\textwidth]{\justifying{The loci of the roots $\hat{\alpha}'_{th1}$ and $\hat{\alpha}'_{th2}$ of $\Delta' = 0$\,(the curve that demarcates spiralling and non-spiralling streamline topologies), and that of the root $\hat{\alpha}'_{th3}$ of $R' = 0$\,(the planar linear flow locus), plotted as a function of $\theta_\omega$. The plot helps organize the different surface-streamline topologies on the $\hat{\alpha}'-\theta_\omega$  plane.}}
    \label{fig:alphathres_inc}
\end{figure}

\begin{figure}
    \centering
    \includegraphics[scale = 0.33]{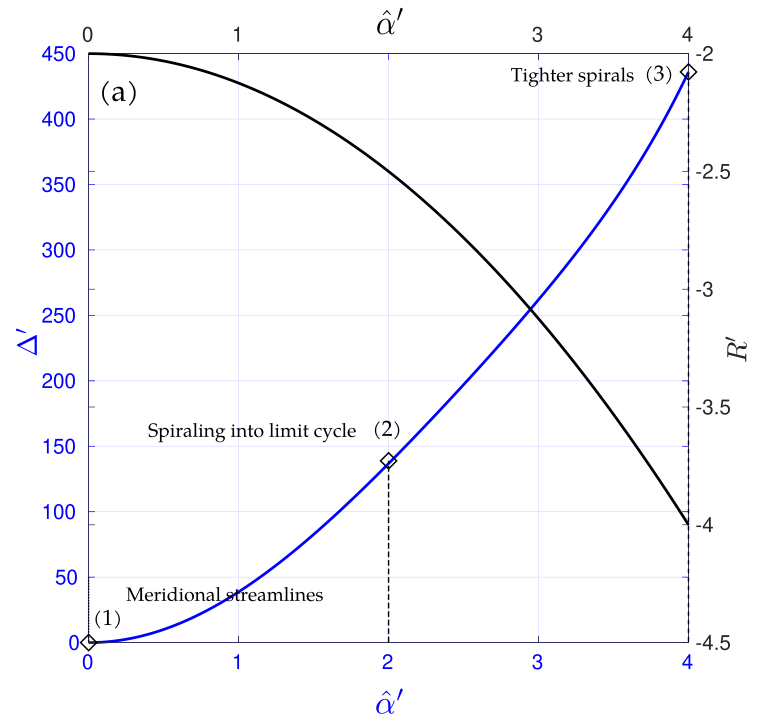}
    \includegraphics[scale = 0.34]{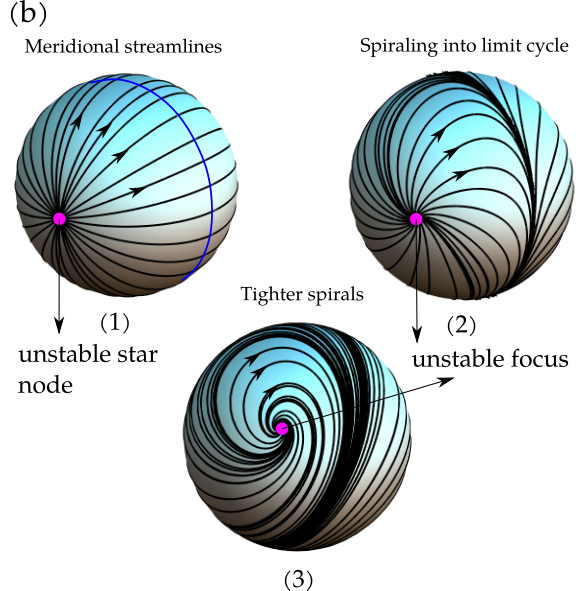}
    \includegraphics[scale = 0.33]{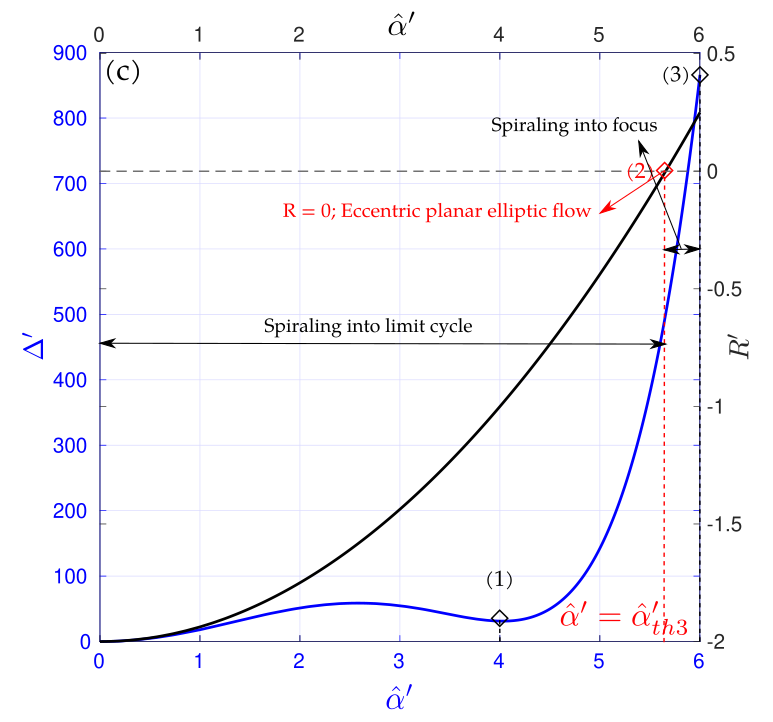}
    \includegraphics[scale = 0.34]{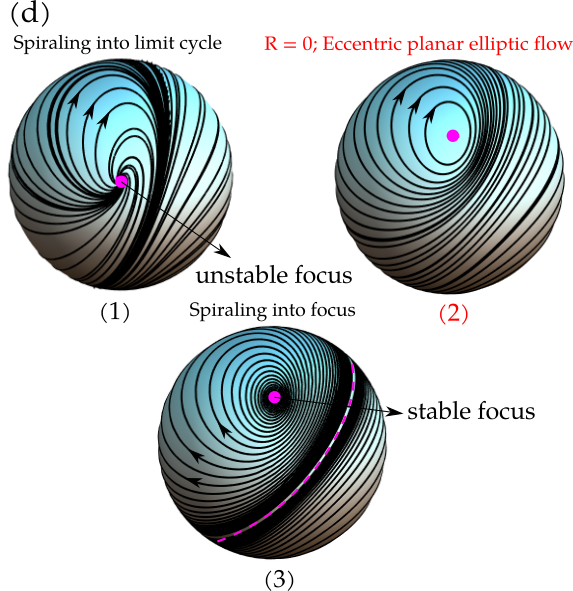}    
    \includegraphics[scale = 0.33]{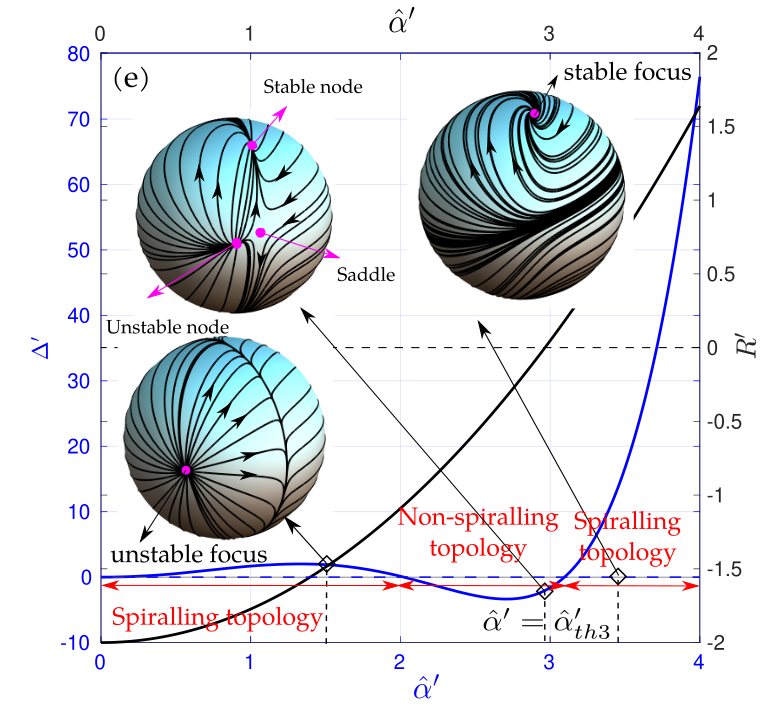}
    \includegraphics[scale = 0.33]{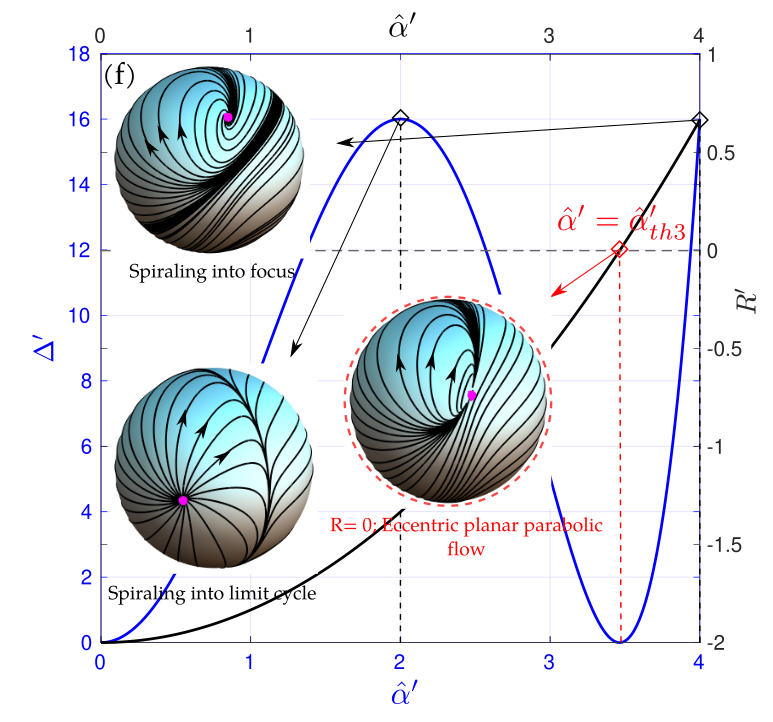}
    \caption[\textwidth]{\justifying{Organization of surface-streamline topologies for the inclined-vorticity family. The figures show plots of $\Delta'$ and $R'$, as a function of $\hat{\alpha}'$, for different $\theta_\omega$: (a) $\theta_\omega = \pi/4$ (case (i)\,$0 \leq \theta_\omega \leq \theta_\omega^{th1}$); (b) shows surface-streamline topologies corresponding to the marked points\,(open diamond symbols) in (a). (c) $\theta_\omega = \pi/3$ (case (ii)\,$\theta_\omega^{th1} < \theta_\omega < \theta_\omega^{th2}$); (d) shows surface-streamline topologies corresponding to the marked points in (c), with a generalized Jeffery-orbit topology mediating the transition between intervals of spiralling-streamline topologies. (e) $\theta_\omega = 4\pi/9$ (case (iii)\,$\theta_\omega^{th2} < \theta_\omega < \pi/2$), with embedded unit-sphere streamline topologies. (f) $\theta_\omega = \theta_\omega^{th2}$ (case (iv)), with embedded unit-sphere streamline topologies; a degenerate parabolic-streamline topology mediates transition between intervals of spiralling streamlines.}}
    \label{fig:discandstream_inc}
\end{figure}

For $\bm \Gamma$ defined by (\ref{inclinedvort_Gamma}), the discriminant and cubic invariant associated with $\hat{\bm{\Gamma}}$ are given by:
\begin{align}
    &\Delta' = \frac{3[(\hat{\alpha}' - 12)^3 + 27(-8 + \hat{\alpha}'^2(1-3\cos^2 \theta_\omega))^2]}{16}, \label{disc1A} \\
    &Q' = \frac{\alpha'^2 - 12}{4}, \label{Q1A} \\
    &R' = \frac{-8 + \hat{\alpha}'^2(1-3\cos^2\theta_\omega)}{4}. \label{R1A}
\end{align}
Proceeding along the same lines as section \ref{topology_alignedvort}, the nontrivial solutions of $\Delta' = 0$ are:
\begin{align}
    &\hat{\alpha}'_{th1,th2} = \frac{3\sqrt{1 + 9\cos^2\theta_\omega(2 - 3 \cos^2 \theta_\omega) \pm \sqrt{(-1 + \cos^2 \theta_\omega)(-1 + 9\cos^2 \theta_\omega)}}}{\sqrt{2}},\label{alpth1A}
\end{align}
while that of $R' = 0$ is given by:
\begin{align}
    &\hat{\alpha}'_{th3} = \frac{4}{\sqrt{-1 -3 \cos(2\theta_\omega)}}. \label{alpth3A}
\end{align}
The $\hat{\alpha}'$-thresholds above are plotted as functions of $\theta_\omega$ in Fig.\ref{fig:alphathres_inc}. Here, $\hat{\alpha}'_{th1}$ and $\hat{\alpha}'_{th2}$ are seen to be real valued only for $\theta_\omega^{th2} \leq \theta_\omega \leq \frac{\pi}{2}$, with $\theta_\omega^{th2}= \tan^{-1} (2\sqrt{2})$;
$\hat{\alpha}'_{th1}=\hat{\alpha}'_{th2} = 2\sqrt{3}$ at $\theta_\omega = \theta_\omega^{th2}$. The planar linear flow locus corresponding to $\hat{\alpha}' = \hat{\alpha}'_{th3}$ is real valued for $\theta_\omega^{th1} \leq \theta_\omega \leq \frac{\pi}{2}$ with $\theta_\omega^{th1} = \tan^{-1} \sqrt{2}$; $\theta_\omega^{th1} < \theta_\omega^{th2}$. This locus starts off at a value between $\hat{\alpha}'_{th1}$ and $\hat{\alpha}'_{th2}$ at $\theta_\omega = \frac{\pi}{2}$, increases with decreasing $\theta_\omega$, intersecting the $\Delta'$-locus at $\theta_\omega = \theta_\omega^{th2}$, before diverging to infinity for $\theta_\omega \rightarrow \theta_\omega^{th1+}$. In light of the above, it is convenient to analyze the surface-streamline topologies, associated with the inclined-vorticity family, in the following distinct $\theta_\omega$-intervals: (i) $0 \leq \theta_\omega \leq \theta_\omega^{th1}$\,(gray region in Fig.\ref{fig:alphathres_inc}), (ii) $\theta_\omega^{th1} < \theta_\omega < \theta_\omega^{th2}$\,(cyan region) and (iii) $\theta_\omega^{th2} < \theta_\omega \leq \pi/2$\,(brown region); $\theta_\omega = \theta_\omega^{th2}$ constitutes case (iv). Figs.\ref{fig:discandstream_inc}a-f show the $\Delta'-\hat{\alpha}'$, $R'-\hat{\alpha}'$ plots and unit-sphere streamline topologies for the aforementioned cases. Note that, in all cases, $\Delta' = 0$\,(point of tangency) and $R' = -2$ at $\hat{\alpha}'=0$, with surface streamlines conforming to the degenerate meridional topology associated with an axisymmetric extensional flow; see left unit sphere in Fig.\ref{fig:discandstream}e. Only the streamline pattern for $\hat{\alpha}' \neq 0$ depends therefore on $\theta_\omega$. 

As shown in Fig.\ref{fig:discandstream_inc}a, $\Delta'$ increases monotonically with increasing $\hat{\alpha}'$ in case (i). The surface-streamlines for any $\hat{\alpha}' \neq 0$ have therefore a spiralling character, being organized by a pair of unstable foci and a limit cycle\,(great circle) in a plane that is inclined to, and bisects, the diameter connecting the foci. The spiralling becomes tighter with increasing $\hat{\alpha}'$, with the streamlines approaching circles for $\hat{\alpha}' \rightarrow \infty$; $R'$ remains negative for all $\hat{\alpha}'$. Note that $\theta_\omega = 0$, when the vorticity vector is aligned with the symmetry axis of the extensional component, was examined in Section \ref{topology_alignedvort}, and corresponds to  $\epsilon = -2$ therein. In this limit alone, the foci locations remain invariant with changing $\hat{\alpha}'$, with the plane of the limit cycle always being normal to the diameter connecting the foci. For all other $\theta_\omega$ in $[0,\theta_\omega^{th1})$, the pair of foci migrate from the symmetry axis of the extension to the vorticity direction, as $\hat{\alpha}'$ increases from zero to infinity, with the limit cycle always being an inclined great circle in between - see the three unit-sphere topologies in Fig.\ref{fig:discandstream_inc}b. The other limiting value, $\theta_\omega = \theta_\omega^{th1}$, also exhibits the same sequence of streamline patterns, except that $R' \rightarrow 0$ for $\hat{\alpha}' \rightarrow \infty$. It will be seen in section \ref{Nu:inclinedvort} that this large-$\hat{\alpha}'$ behavior has an important effect on $Nu$.

For case (ii) shown in Fig.\ref{fig:discandstream_inc}c, $\Delta'$, although non-monotonic, remains positive for all $\hat{\alpha}' \neq 0$. However, $R'$ has a zero-crossing at $\hat{\alpha}' = \hat{\alpha}'_{th3}$, implying a planar auxiliary flow that is now an elliptic linear flow since $Q \propto {\Delta'}^{\frac{1}{3}} > 0$ at this $\hat{\alpha}'$. For $\hat{\alpha}' < \hat{\alpha}'_{th3}$, one has a spiralling topology with surface streamlines going from a pair of unstable foci to a limit cycle in between. For $\hat{\alpha}' > \hat{\alpha}'_{th3}$, the spiralling topology is preserved, but the direction of spiralling is reversed, with the pair of foci now being stable. At $\hat{\alpha}' = \hat{\alpha}'_{th3}$, the surface streamlines are closed curves, as indicated by the second unit sphere in Fig.\ref{fig:discandstream_inc}d. While this is consistent with the nature of the auxiliary flow mentioned above, the eccentric arrangement of the surface streamlines may nevertheless be contrasted with the concentric nature of the Jeffery orbits depicted in Fig.\ref{fig:discandstream}f\,(unit sphere at the top). The planar auxiliary flow at $\hat{\alpha}' = \hat{\alpha}'_{th3}$ is thus a generalization of the usual elliptic flows that are the members of the planar linear flow family with $-1 < \alpha < 0$. The closed streamlines on the unit sphere in Fig.\ref{fig:discandstream_inc}d may be regarded as generalized Jeffery orbits, and are projections of the streamlines of an eccentric elliptic flow\,\citep{Sabarish22}; $\hat{\alpha}(1+\lambda) = \hat{\alpha}'_{th3}$, with $\hat{\alpha}'_{th3}$ given by (\ref{alpth3A}), corresponds therefore to the eccentric-elliptic-flow locus on the $\theta_\omega-\hat{\alpha}$ plane, which was earlier stated in section \ref{sec:setup}. Although the occurrence of eccentric planar hyperbolic flows, for the aligned-vorticity family in Section \ref{topology_alignedvort}, turns out to be incidental from the transport perspective, the occurrence of eccentric elliptic flows is not! It will be shown in Section \ref{sec:Nusselt_calc} that $Nu$ must scale as $Pe^{\frac{1}{3}}$ for such flows. 

For case (ii) above, the non-monotonic variation of $\Delta'$ manifests via the emergence of a local maximum and minimum in the $\Delta'$ vs $\hat{\alpha}'$ curve in Fig.\ref{fig:discandstream_inc}c. Since $\Delta'$ decreases in magnitude for $\hat{\alpha}'$ fixed and with increasing $\theta_\omega$, the aforementioned minimum decreases with increasing $\theta_\omega$, equalling zero at $\theta_\omega = \theta_{\omega}^{th2}$, and becoming negative for larger $\theta_\omega$. Hence, for case (iii) corresponding to $\theta_\omega \in (\theta_\omega^{th2},\frac{\pi}{2})$, $\Delta'$ exhibits a pair of zero-crossings at $\hat{\alpha}'_{th1}$ and $\hat{\alpha}'_{th2}$, being negative for $\hat{\alpha}' \in (\hat{\alpha}'_{th1},\hat{\alpha}'_{th2})$, and being positive on either side of this interval - this is shown in Fig.\ref{fig:discandstream_inc}e. From this figure, one notes that $R'$  also has a zero-crossing; although, this now corresponds to a planar hyperbolic flow, and as for the aligned-vorticity family, is of no particular significance. The streamline topology has a spiralling character in the interval $(0,\hat{\alpha}'_{th1})$, and again in the interval $(\hat{\alpha}'_{th1},\infty)$ with the direction of spiralling now reversed. In contrast to case (ii), where the transition between these intervals occurred across a single $\hat{\alpha}'$ corresponding to a closed-streamline topology, the transition between spiralling-streamline topologies in case (iii) is mediated via a finite interval of non-spiralling topologies for $\hat{\alpha}'\in(\hat{\alpha}'_{th1},\hat{\alpha}'_{th2})$. The non-spiralling streamline pattern in this interval is again organised by six fixed points, but is significantly more skewed than for the aligned-vorticity case - see the middle unit sphere in Fig.\ref{fig:discandstream_inc}e - this is owing to the segment connecting the unstable nodes no longer being constrained to be orthogonal to the plane containing the other two fixed point pairs. The limiting case of $\theta_\omega = \pi/2$ was already seen in the context of the aligned-vorticity family, where it corresponded to $\epsilon = -1/2$. 

Finally, case (iv) in Fig.\ref{fig:discandstream_inc}f corresponds to the transition between cases (ii) and (iii). The $\Delta'$-curve is now tangent to the $\hat{\alpha}'$-axis at the same point that $R'$ has a zero crossing. The point corresponds to the intersection of the two loci in Fig.\ref{fig:alphathres_inc}, implying $\hat{\alpha}'_{th1} = \hat{\alpha}'_{th2} = \hat{\alpha}'_{th3} = 2\sqrt{3}$. $Q'=R' = 0$ at this $\hat{\alpha}'$, with the auxiliary flow having parabolic streamlines. The parabolic linear flow sits at the threshold between eccentric hyperbolic and elliptic linear flows, in the same manner as simple shear flow serves as a transition member connecting the canonical hyperbolic to elliptic flows\,\citep{Sabarish22}. As shown by the unit-spheres embedded in Fig.\ref{fig:discandstream_inc}f, the parabolic auxiliary flow at $\hat{\alpha}' = 2\sqrt{3}$ plays the same role as the eccentric elliptic flow in case (ii), mediating the transition between spiraling-streamline topologies of opposite senses; the encircled unit sphere in this figure shows the degenerate surface-streamline topology associated with this flow. As will be seen in section \ref{sec:Nusselt_calc}, this degenerate topology manifests as a singular point on the $Nu$-curve, although $Nu$ continues to scale as $Pe^{\frac{1}{2}}$ for $Pe \gg 1$.

\subsubsection{The $C - \tau$ coordinate system} \label{sec:C_tau_inclined}
For this family, (\ref{surffield}) and (\ref{inclinedvort_Gamma}) lead to the following coupled ODEs
\begin{align}
    &\frac{d \theta}{dt} = \frac{3}{2(1+\lambda)}  \sin 2\theta - \frac{1}{2} (\hat{\alpha} \sin \theta_\omega \sin \phi), \label{ode1A} \\
    &\frac{d \phi}{dt} = \frac{\hat{\alpha}}{2} (\cos \theta_\omega  - \sin \theta_\omega \cos \phi \cot \theta), \label{ode2A}
\end{align}
for the surface streamlines. With $\hat{\bm \Gamma}$ defined by
\begin{align}
    \hat{\bm{\Gamma}} = \begin{bmatrix}
        \frac{1}{1+\lambda} &-\frac{\hat{\alpha} \cos \theta_\omega}{2} &0 \\
        \frac{\hat{\alpha} \cos \theta_\omega}{2} &\frac{1}{1+\lambda} &-\frac{\hat{\alpha} \sin \theta_\omega}{2} \\
        0 &\frac{\hat{\alpha} \sin \theta_\omega}{2} &\frac{-2}{1+\lambda}
    \end{bmatrix}, \label{gamma_aux_inc}
\end{align}
the streamlines of the auxiliary linear flow, governed by \eqref{auxfield}, may be formally written in the form, $\bm{r}(t) =(\bm{S}^{-1}\hat{\bm{D}}\bm{S})\bm{r}_0$, where $\bm{S}$ is the matrix of (normalized)\,eigenvectors ${\bm x}_i$\,($i = 1-3$), and $\hat{\bm{D}}$ is a diagonal matrix with entries $\hat{D}_{ii} = e^{\xi_i t}$, $\xi_i$ being the eigenvalues of $\hat{\bm \Gamma}$; $\bm{r}_0$ in the above expression is an arbitrary `initial' position vector, and $t \in (-\infty,\infty)$. Note that $\bm{S}^{-1}\hat{\bm{D}}\bm{S}$ is the eigen decomposition of $\exp[\hat{\bm{\Gamma}}t]$. There is no loss of generality in assuming $\hat{\bm \Gamma}$ to be diagonalizable thus, since the cases where it is not correspond to exceptional sets in parameter space; $Nu$ for such cases may be evaluated via a limiting process. The surface streamlines governed by \eqref{ode1A} and \eqref{ode2A} are obtained as $\bm{r}(t)/|\bm{r}(t)|$, $t \in (-\infty,\infty)$.  

To examine the nature of the associated $C-\tau$ mapping
, we revisit the expressions for the aligned-vorticity family given by (\ref{auxsol1new_int}-\ref{auxsol2new_int}) and (\ref{Spiralling_tau}-\ref{Spiralling_C}), which are of the form $\phi \sim f(\tau)$, $\theta \sim f(C,\tau)$, with $\tau$ ranging from minus to plus infinity along each surface streamline. Barring exceptional cases, any point on a streamline can be taken as $\tau = 0$, the origin of time; exceptions exist only in the non-spiralling regime, and are the meridional arcs that connect the fixed points on the $x_3$-axis to the ones in the $x_1-x_2$ plane. If the azimuthal angle corresponding to $\tau = 0$ on a given streamline be $\phi_0$, then the streamline label may be taken as $C \propto \tan \theta_0$, $\theta_0$ being the polar angle of the point $\tau = 0$; the $\tau$-definitions in (\ref{auxsol1new1}) correspond to $\phi_0 = \frac{\pi}{4}$\,(plus sign) and $\frac{3\pi}{4}$\,(minus sign). The choice of polar axis, required to define $\theta_0$ and $\phi_0$, is obvious for this family - it is the $x_3$-axis. With this choice, the fixed point on the $x_3$-axis\,($\theta_0 = 0$) maps to $C = 0$ - this may be either the saddle or unstable node in the non-spiralling regime\,(depending on whether $\hat{\alpha}'$ is less or greater than $\hat{\alpha}'_{th2}$), and is the unstable focus in the spiralling regime. 
The streamlines and fixed points in the equatorial plane\,($\theta_0 = \frac{\pi}{2}$) map to $C = \infty$. 
 
To define $C$ and $\tau$ for the inclined-vorticity family, one first chooses the polar axis of the coordinate system to either pass through the pair of foci in the spiralling regime, or the pair of saddle points in the non-spiralling one - see unit sphere topologies in Fig.\ref{fig:discandstream_inc}e. In contrast to the aligned-vorticity family, the locations of these fixed-point pairs are not fixed. Considering the spiralling regime in Fig.\ref{fig:discandstream_inc}a, for instance, the segment connecting the foci 
coincides with the symmetry axis of the extension\,(the $x_3$-axis) for $\hat{\alpha}'= 0$, and rotates towards the vorticity axis\,(that makes an angle $\theta_\omega$ with the $x_3$-axis) with increasing $\hat{\alpha}'$, coinciding with it for $\hat{\alpha}' \rightarrow \infty$. Further, for any finite $\hat{\alpha}'$, the plane containing the limit cycle\,(spiralling regime), or the remaining four fixed points\,(non-spiralling regime), is not orthogonal to the segment connecting the foci or saddles. As a result, although the above choice of polar axis maps the focus or saddle to $C = \theta_0' = 0$, the great circle in the aforementioned plane will not correspond to $\theta_0' = \pi/2$ for a generic choice of $\phi_0'$. Instead, it would be of the form $\theta'(\phi')$, with $\theta'$ equalling $\frac{\pi}{2}$ only at a pair of $\phi'$'s separated by $\pi$, and neither of which need equal $\phi_0'$. This would mean that the great circle maps to a finite $C$\,(the primes here denote angles defined with respect to the rotated polar axis, as opposed to its original orientation, along the $x_3$-axis, that led to (\ref{ode1A}) and \ref{ode2A})). In order to ensure that the aforementioned great circle always maps to $C = \infty$, similar to the one in the $x_1-x_2$ plane in the aligned-vorticity case, one first finds the polar angle\,($\theta^{'m}_0 = \theta^{'m}(\phi'_0)$) corresponding to it, along the arc $\phi' = \phi'_0$, and then defines $C$ via $\tan \theta'_0 = \frac{C\tan \theta^{'m}_0}{(C + \tan \theta^{'m}_0)}$. For the non-spiralling regime, four different arcs have to be chosen, one corresponding to each octant on the unit hemisphere, because the symmetry leading to the identity of diagonally opposite octants\,(as for the aligned-vorticity case) is now absent. This ensures that the saddle maps to $C = 0$, and that the great circle containing the stable and unstable node pairs maps to $C = \infty$, independent of $\phi_0$. The above procedure is readily extended to the spiralling regime, with only one reference arc now being needed. The result is that the focus is mapped to $C = 0$, and the limit cycle to $C = \infty$.  

All of the above is achieved by the following relation:
\begin{align}
    \bm{r}' = (\bm{S}'^{-1}\bm{\hat{D}}\bm{S}')\begin{bmatrix}
        r_{1,0}' \\
        r_{2,0}' \\
        \frac{\sqrt{r_{1,0}'^2 + r_{2,0}'^2}(C + \tan \theta^{'m}_0)}{C\tan \theta^{'m}_0}
    \end{bmatrix}, \label{auxsolnewA}
\end{align}
where $\bm{r}'_0$ is written in an expanded column-vector form; $\phi_0' = \tan^{-1} \left( \frac{r_{2,0}'}{r_{1,0}'} \right)$. As before, the primed quantities in (\ref{auxsolnewA}) pertain to the rotated coordinate system with its polar axis aligned with the foci or saddle-pair. Note that $\hat{\bm D}$ is invariant to the rotation; the only difference relative to the expression for $\bm{r}(t)$, that appears below (\ref{gamma_aux_inc}), being that $t$ in $\bm{\hat{D}}$ is now written as $\tau$. Knowledge of the invariant arcs connecting the saddle to the stable and unstable nodes, allows one to choose four values for the ratio $r_{2,0}'/r_{1,0}'$, one corresponding to each octant, in the non-spiralling regime. In the spiralling regime, we directly choose $\phi_0'$ such that $\theta_0'^m = \frac{\pi}{2}$, in which case the simpler relation $\tan \theta_0' = C$ ensures that the limit cycle maps to $C = \infty$. and $\bm{r}'_0 = [r_{1,0}', r_{2,0}', \sqrt{r_{1,0}'^2 + r_{2,0}'^2}/C]^T$ in \eqref{auxsolnewA}.
The eigenvalues $\xi_i$ in $\hat{\bm D}$, which are solutions of $\xi^3 + Q'\xi + R' = 0$, are obtained using Cardano's method, being given by $\xi = u + v$, where
\begin{align}
    &u = \left[\frac{-\left( \frac{-8 + \alpha'^2(1-3\cos^2 \theta_\omega)}{4} \right) + \frac{1}{3\sqrt{3}}\sqrt{\frac{3((\alpha'^2 - 12)^3 + 27(-8 + \alpha'^2(1 - 3 \cos^2 \theta_\omega))^2)}{16}}}{2}\right]^{\frac{1}{3}}\begin{bmatrix}
        1 \\
        \omega \\
        \omega^2
    \end{bmatrix}, \label{cardano1} \\
    &v = \left[\frac{-\left( \frac{-8 + \alpha'^2(1-3\cos^2 \theta_\omega)}{4} \right) - \frac{1}{3\sqrt{3}}\sqrt{\frac{3((\alpha'^2 - 12)^3 + 27(-8 + \alpha'^2(1 - 3 \cos^2 \theta_\omega))^2)}{16}}}{2}\right]^{\frac{1}{3}}\begin{bmatrix}
        1 \\
        \omega \\
        \omega^2
    \end{bmatrix}, \label{cardano2} 
\end{align}
with $\omega = \frac{-1 +\sqrt{3}\mathrm{i}}{2}$. The subset of three relevant solutions, from the nine possibilities in (\ref{cardano1}) and (\ref{cardano2}), is found on a case-by-case basis depending on the parameter\,($\theta_\omega$ and $\hat{\alpha}'$) values.

Finally, one may write:
\begin{align}
    &\tan \phi' = \frac{x_2'}{x_1'} = \mathcal{F}(\tau), \label{auxsolnew1A}\\
    &\tan \theta' = \frac{(x_1'^2 + x_2'^2)^{1/2}}{x_3'} = \mathcal{G}(C,\tau), \label{auxsolnew2A}
\end{align}
with ${\bm r}' = [x_1',x_2',x_3']$ given by \eqref{auxsolnewA}, which formally defines the relations for $C$ and $\tau$ for the inclined-vorticity family.

\section{Calculation of the scalar transport rate\,($Nu$)} \label{sec:Nusselt_calc}

Having defined the $C- \tau$ coordinate system for the pair of two-parameter linear flow families in Sections \ref{sec:C_tau_aligned} and \ref{sec:C_tau_inclined}, we use these definitions to solve \eqref{HEQ1} for large $Pe$, so as to obtain the structure of the scalar field within the boundary layer, and thereafter, calculate $Nu$ using \eqref{Nudef}. In what follows, we first provide a general outline of the boundary layer analysis in section \ref{Nu:calcgen}, that leads to a formal expression for $Nu$ in terms of quantities pertaining to the $C-\tau$ coordinate system. Specific results for the aligned and inclined-vorticity families are presented thereafter in sections \ref{Nu:alignedvort} and \ref{Nu:inclinedvort}, respectively.

\subsection{Boundary layer analysis in $C-\tau$ coordinates} \label{Nu:calcgen}
 The surface velocity field in the $(C, \tau)$ coordinate system may be written in the form:
\begin{align}
     {\bm u} \cdot ({\bm I} - {\bm n}{\bm n})\mid_{r=1} = u_C \hat{\bm{C}} + u_\tau \hat{\bm{\tau}},
  \label{u_surf_Ctau}
\end{align}
where $\bm{u}$ is given by \eqref{Sol1}, and
\begin{align}
    &u_C = u_\theta - \frac{u_\phi \theta_\tau}{\phi_\tau \sin \theta}, \label{uCdef} \\
    &u_\tau = \frac{k u_\phi}{\phi_\tau \sin \theta}. \label{utaudef}
\end{align}
The polar and azimuthal velocity components for the linear flows of interest having been defined in (\ref{ode1}-\ref{ode2}) and (\ref{ode1A}-\ref{ode2A}). The convection-diffusion equation in this coordinate system is given by:
\begin{align}
    u_r \frac{\partial \Theta}{\partial r} + \frac{u_C}{h} \frac{\partial \Theta}{\partial C} + \frac{u_\tau}{k} \frac{\partial \Theta}{\partial \tau} = \frac{1}{Pe} \nabla^2 \Theta, \label{HEQCtau}
\end{align}
where $u_r$ is the radial velocity\,(and equals $\bm{u} \cdot \bm{n}$), with expressions for the metric factors having been given in \eqref{metricfactor1}-\eqref{metricfactor4}. 

From (\ref{uCdef}), one can show that $u_C = 0$, as is expected from $C$ acting as a label for the surface streamlines; note that this is consistent with $\hat{\bm \tau}$ in (\ref{u_surf_Ctau}) being tangent to the surface streamlines. The reduction of the surface velocity to a one-component field is crucial to enabling a similarity solution for the boundary layer scalar field, in the same manner as for the simpler flows analyzed in literature\,(refer to examples quoted in section \ref{1}). The difference is that this reduction is readily achievable in an appropriately oriented spherical coordinate system for the aforesaid symmetric flows, while the $C - \tau$ coordinate system achieves the same reduction despite the nontrivial surface streamline topology. With the above simplification, and the recognition that radial diffusion is dominant within the thin boundary layer, one may write (\ref{HEQCtau}) in the form:
\begin{align}
    u_r \frac{\partial \Theta}{\partial r} + \frac{u_\tau}{k} \frac{\partial \Theta}{\partial \tau} = \frac{1}{Pe\, r^2} \frac{\partial}{\partial r}\left( r^2 \frac{\partial \Theta}{\partial r} \right). \label{HEQCtaunew}
\end{align}
Since it is the tangential velocity on, and the radial velocity near, the drop surface that convect the scalar, one defines $y = r-1$, with $u_r$ and $u_\tau$ given by:
\begin{align}
    &u_r = h_r(C,\tau;\{P\},\lambda) y + O(y^2), \label{hr}\\
    &\frac{u_\tau}{k} = h_\tau(\{P\},\lambda) + O(y),\label{htau}
\end{align}
at leading order within the boundary layer. Here, $y \sim O(Pe^{-\frac{1}{2}})$ is of order the drop boundary layer thickness, and $\{P\}$ denotes the set of flow-type parameters - $\{ P \} \equiv \{\epsilon,\hat{\alpha}\}$ and $\{ \hat{\alpha},\theta_\omega \}$ for the aligned and inclined-vorticity families. $h_\tau$ in (\ref{htau}) turns out to be a constant, which renders the boundary layer analysis analogous to the simpler problem of scalar transport from a rotating sphere in an ambient vortical linear flow\citep{Batchelor79}. In spherical coordinates with the polar axis aligned with the ambient vorticity, $k = \sin\theta$, and $u_\phi/\sin\theta$ for this latter problem is constant at leading order on account of the uniform particle rotation\,\citep{SubKoch06a,SubKoch06b}; the constancy being consistent with the near-surface streamlines being tightly wound spirals about the ambient vorticity direction. In the present case, $u_\tau/k$ plays the same role as $u_\phi/\sin\theta$, despite the absence of any rotational symmetry. 

Next, defining the scaled boundary layer coordinate, $Y = Pe^{\frac{1}{2}}y$, \eqref{HEQCtaunew} takes the form:
\begin{align}
    h_r Y \frac{\partial \Theta}{\partial Y} + h_\tau \frac{\partial \Theta}{\partial \tau} =  \frac{\partial^2 \Theta}{\partial Y^2},
\end{align}
 with the analysis proceeding along standard lines hereafter\,\citep{Leal07}. Thus, one defines a similarity variable $\eta = Y/g(C,\tau)$, with $g(C,\tau)$ describing the angular dependence of the boundary layer thickness. Enforcing the similarity ansatz $\Theta \equiv \Theta(\eta)$ leads to:
\begin{align}
    \frac{d^2 \Theta}{d \eta^2} + 2 \eta \frac{d \Theta}{d \eta} = 0, \label{tempode}
\end{align}
governing the scalar field, with:
\begin{align}
    h_\tau g \frac{d g}{d \tau} - h_r g^2 = 2, \label{BLeqn}
\end{align}
governing the boundary layer thickness, and with the boundary conditions for the scalar field given by:
\begin{align}
    &\Theta = 1 \text{ at } \eta = 0, \label{tempBC1}\\
    &\Theta = 0 \text{ at } \eta \rightarrow \infty. \label{tempBC2}
\end{align}
Solving \eqref{tempode} using \eqref{tempBC1}-\eqref{tempBC2}, one obtains:
\begin{align}
    \Theta(\eta) = 1 - \frac{2}{\sqrt{\pi}} \int_0^\eta e^{-s^2}ds. \label{tempsol}
\end{align}
Next, using the substitution $f = g^2/2$ in \eqref{BLeqn} gives:
\begin{align}
    h_\tau\frac{d f}{d \tau} - 2 h_r f = 2.
\end{align}
In terms of the integrating factor,
\begin{align}
    Q = \exp\left(\int_{-\infty}^{\tau} \frac{-2 h_r}{h_\tau} d \tau' \right) \label{IntFact}
\end{align}
the formal solution for $g$ may be written as:
\begin{align}
    g = \frac{2}{h_\tau^{1/2}} \left( Q^{-1} \int_{-\infty}^\tau Q(\tau_1) d\tau_1 \right)^{1/2}, \label{gsol}
\end{align}
where the choice of the lower limit for the $\tau$-integral ensures that the boundary layer thickness is finite at the inlet stagnation points. The finiteness arises because the local flow at these locations is a linear 
extension convecting the scalar towards the drop surface, balancing radially outward diffusion in a boundary layer of a finite thickness. 
Once $g$ and $\Theta$ are determined, one can calculate the Nusselt number as:
\begin{align}
    Nu = - \frac{1}{4 \pi} \int \left(\frac{\partial \Theta}{\partial y} \right)_{y = 0} d\Omega = -\frac{Pe^{\frac{1}{2}}}{4 \pi} \int \left(\frac{\partial \Theta}{\partial Y} \right)_{Y = 0} d\Omega,
\end{align}
which can be rewritten in terms of $\eta$, in the $C-\tau$ coordinate system, as:
\begin{align}
    Nu =& -\frac{Pe^{\frac{1}{2}}}{4 \pi} \int \frac{1}{g(C,\tau)}\left(\frac{d \Theta}{d \eta} \right)_{\eta = 0} d\Omega, \\
    =& \frac{Pe^{\frac{1}{2}}}{2\pi^{\frac{3}{2}}} \int \frac{hk \sin \beta}{g(C,\tau)}dC d\tau, \label{Nuold}
\end{align}
on using \eqref{tempsol}. 

\cite{Deepak18a} obtained a closed form expression for $Q$, as defined by (\ref{IntFact}), for the simpler linear flows that they examined viz.\ the canonical planar linear flows and extensional flows. We now show that the constancy of $h_\tau$ in (\ref{htau}) can be used to obtain $Q$ in closed form, in terms of $C$ and $\tau$, for an arbitrary ambient linear flow. To do so, we begin with the continuity equation in the $(y, C,\tau)$ coordinate system, given by:
\begin{align}
    \frac{\partial (h k \sin \beta \; u_r)}{\partial y} + \frac{\partial (k \sin \beta \; u_C)}{\partial C} + \frac{\partial (h \sin \beta \; u_\tau)}{\partial \tau} = 0, \label{conti}
\end{align}
where we have used that the metric factor for the radial coordinate is unity, with $h$, $k$ and $\beta$ being defined in \eqref{metricfactor1}-\eqref{metricfactor4}. Substituting for $u_r$ and $u_\tau$ from \eqref{hr}-\eqref{htau}, and using $u_C = 0$, one obtains:
\begin{align}
    \frac{\partial (hk \sin \beta\, h_r y)}{\partial y} + \frac{\partial (h k \sin \beta\, h_\tau)}{\partial \tau} = 0. \label{eq:CtauCont}
\end{align}
The constancy of $h_\tau$, and the fact that $h_r$ does not depend on $y$, implies (\ref{eq:CtauCont}) can be rewritten as:
\begin{align}
    (hk \sin \beta) h_r =& -h_\tau \frac{\partial (h k\sin \beta)}{\partial \tau}, \\
    \Rightarrow-\frac{h_r}{h_\tau} =& \frac{\partial}{\partial \tau}\ln(h k\sin \beta).
\end{align}
Using the above in \eqref{IntFact}, one finds:
\begin{align}
    Q = \left[ \frac{h k\sin \beta}{(h k\sin \beta)_{-\infty}}\right]^2, \label{IntFact1}
\end{align}
where $(.)_{-\infty}$ denotes evaluation of the bracketed quantity in the limit $\tau \rightarrow -\infty$, corresponding to the boundary layer inlet(s). 

Now, using (\ref{IntFact1}) and (\ref{gsol}) in (\ref{Nuold}), one obtains:
\begin{align}
    Nu = \frac{Pe^{1/2} h_\tau^{1/2}}{2 \pi^{3/2}} \int_{C} \int_{\tau} \frac{Q^{1/2} (h k \sin \beta)_{-\infty} \; dC d\tau}{2 (Q^{-1} \int_{-\infty}^{\tau}Q(\tau_1) d\tau_1)^{1/2}}. \label{Nuinter}
\end{align}
This can be reduced further, by noting that the $\tau$-integrand reduces to an exact differential, as follows:
\begin{align}
    Nu &= \frac{Pe^{1/2} h_\tau^{1/2}}{2 \pi^{3/2}} \int dC \int_{-\infty}^\infty d\tau \frac{Q (\int_{-\infty}^{\tau}Q(\tau_1) d\tau_1)^{-1/2} (h k \sin \beta)_{-\infty}}{2}, \\
    &= \frac{Pe^{1/2} h_\tau^{1/2}}{2 \pi^{3/2}} \int dC  (h k \sin \beta)_{-\infty} \int_{-\infty}^{\infty} \frac{d}{d \tau}\left( \int_{-\infty}^{\tau} Q(\tau_1) d\tau_1 \right)^{1/2} \!\!d\tau, \\
    &= \frac{Pe^{1/2} h_\tau^{1/2}}{2 \pi^{3/2}} \int dC (h k \sin \beta)_{-\infty}  \left( \int_{-\infty}^{\infty} Q(\tau) d\tau \right)^{1/2}, \\
    & = \frac{Pe^{1/2} h_\tau^{1/2}}{2 \pi^{3/2}} \int dC \left( \int_{-\infty}^{\infty} (hk\sin \beta)^2 d\tau \right)^{1/2}. \label{Nufin}
\end{align}
The upper limit\,($\tau \rightarrow \infty$) of the $\tau$-integral in (\ref{Nufin}) corresponds to the wake location(s) where $g$ diverges. Although, the divergence is integrable and the $Nu$-integral remains finite, implying that the wake contribution contributes at a higher order in $Pe$. The evaluation of $Q$ in closed form above has allowed one to reduce the $Nu$ calculation from a three to a two-dimensional integral. \eqref{Nuinter}. For the canonical planar linear flows, the $\tau$-integral in (\ref{Nufin}) may also be evaluated in closed form, leaving a one-dimensional integral over $C$. In this regard, it is worth noting that the limits for the $C$-integral have not been specified in (\ref{Nufin}), since these depend on the streamline topology identified in sections \ref{sec:alignedfamily} and \ref{sec:inclinedfamily}. 

\subsection{Drop in 3D extensional flows with aligned vorticity} \label{Nu:alignedvort}

We first perform the $Nu$-calculation in the non-spiraling regime\,[$(2+\epsilon) \geq (1+\lambda)\hat{\alpha}$], where open surface streamlines on each unit hemisphere are organized into four octants. The identity of the contributions from diagonally opposite octants implies that one only need determine the contributions from a pair of adjacent octants\,($I$ and $II$, say), with $Nu$ being given by:
\begin{align}
    Nu &= 4 \left[ \frac{Pe^{1/2} h_\tau^{1/2}}{2 \pi^{3/2}} \int_0^\infty \left( \int_{-\infty}^{\infty} [(h k \sin \beta)^2_I + (h k \sin \beta)^2_{II}] \; d\tau \right)^{1/2} dC \right], \label{Nufinopen_int}
\end{align}
where the interval for the $C$-integral is taken to be $[0,\infty)$, on account of the non-spiralling regime. In (\ref{Nufinopen_int}),
\begin{align}
     &(hk \sin \beta)^2_I = \frac{C^2 (2+\epsilon + (1+\lambda)\hat{\alpha})(2+\epsilon - (1+\lambda)\hat{\alpha})^2 e^{\frac{3 \epsilon \tau}{A}}}{[e^{\frac{3 \epsilon \tau}{A}} (2+\epsilon - (1+\lambda)\hat{\alpha}) + 2C^2((2 + \epsilon)\cosh \tau - (1+\lambda)\hat{\alpha})]^3}, \label{eq:hk1} \\
    &(hk \sin \beta)^2_{II} = \frac{C^2 (2+\epsilon + (1+\lambda)\hat{\alpha})^2 (2+\epsilon - (1+\lambda)\hat{\alpha}) e^{\frac{3 \epsilon \tau}{A}}}{[e^{\frac{3 \epsilon \tau}{A}} (2+\epsilon + (1+\lambda)\hat{\alpha}) + 2C^2((2 + \epsilon)\cosh \tau + (1+\lambda)\hat{\alpha})]^3}, \label{eq:hk2}
\end{align}
where we have used the $C-\tau$ definitions in \eqref{auxsol1new1}-\eqref{auxsol2new1} - (\ref{eq:hk1}) and (\ref{eq:hk2}) result from using the plus and minus sign in these definitions. (\ref{Nufinopen_int}) may therefore be written in the form:
\begin{equation}
Nu = 4(Nu_I + Nu_{II}), \label{Nufinopen}
\end{equation}
where $Nu_I$ and $Nu_{II}$ correspond, respectively, to transport rate contributions from octants defined by $\phi^{(1)} \leq \phi \leq \phi^{(2)}$ and $\phi^{(2)} \leq \phi \leq \phi^{(3)}$ (Regions 1 and 2 in Fig.\ref{fig:Ctau_aligned}a; see discussion in Section. 3.1.2). On substituting $h_\tau = A/(1+\lambda)$, these are given by:
\begin{equation}
\begin{split}
    &\frac{Nu_I}{\hat{Pe}^{1/2}}  = \frac{(2+\epsilon + \hat{\alpha}')^{3/4}(2+\epsilon - \hat{\alpha}')^{5/4}}{2 \pi^{3/2}} \times \\
    &\left[\int_{0}^{\infty} C  \left( \int_{-\infty}^{\infty} \frac{e^{\frac{3 \epsilon \tau}{A}}}{[e^{\frac{3 \epsilon \tau}{A}} (2+\epsilon - \hat{\alpha}') + 2C^2((2 + \epsilon)\cosh \tau - \hat{\alpha}')]^3} \; d\tau \right)^{1/2} dC\right],
\end{split} \label{Nufinopen1}
\end{equation}
\begin{equation}
\begin{split}
    &\frac{Nu_{II}}{\hat{Pe}^{1/2}} =  \frac{(2+\epsilon - \hat{\alpha}')^{3/4}(2+\epsilon + \hat{\alpha}')^{5/4}}{2 \pi^{3/2}} \times \\
    &\left[\int_{0}^{\infty} C  \left( \int_{-\infty}^{\infty} \frac{e^{\frac{3 \epsilon \tau}{A}}}{[e^{\frac{3 \epsilon \tau}{A}} ((2+\epsilon) + \hat{\alpha}') + 2C^2((2 + \epsilon)\cosh \tau + \hat{\alpha}')]^3} \; d\tau \right)^{1/2} dC\right]. \end{split} \label{Nufinopen2}
\end{equation}
The $\lambda$-dependence in the above expressions has entirely been incorporated in a re-scaled Peclet number, $\hat{Pe} = Pe/(1+\lambda)$, and $\hat{\alpha}'$. For a given streamline topology, as defined by $\hat{\alpha}'$, $\hat{Pe}$ accounts for the overall slowdown of the surface flow due to the increasing viscosity ratio. The above reflects the exact scaling relation $Nu(Pe;\epsilon,\hat{\alpha},\lambda) \equiv Nu(\hat{Pe};\epsilon,\hat{\alpha}')$, one that remains true for a general linear flow. In the latter case, one may write $Nu(Pe;\{P\},\lambda) \equiv Nu(\hat{Pe};\{ P'\})$, where $\{ P'\} \equiv (\epsilon,\hat{\alpha}',\theta_\omega,\phi_\omega)$.

For the spiraling regime, the $C,\tau$ definitions in \eqref{Spiralling_tau}-\eqref{Spiralling_C} lead to:
\begin{align}
    &(h k \sin \beta)^2 = \frac{C^2 A'^2 ((2+\epsilon) - \hat{\alpha}') e^{\frac{3 \epsilon\tau}{A'}}}{[e^{\frac{3 \epsilon \tau}{A'}} ((2+\epsilon)-\hat{\alpha}' + 2C^2((2 + \epsilon)\cos \tau - \hat{\alpha}')]^3}, \label{IntFact3}  
\end{align}
where $A'$ is real, and was defined in the line preceding \eqref{auxsol1new_spi}. Substituting the above in \eqref{Nufin} gives:
\begin{equation}
\begin{split}
    &\frac{Nu}{\hat{Pe}^{1/2}} =  \frac{A'^{3/2}\sqrt{(2+\epsilon) - \hat{\alpha}'}}{\pi^{3/2}} \times \\
    &\left[ \int_{C^*}^{C^* e^{\frac{|6\epsilon \pi|}{A'}}} C  \left( \int_{-\infty}^{\infty} \frac{e^{\frac{3 \epsilon\tau}{A'}}}{[e^{\frac{3 \epsilon \tau}{A'}} ((2+\epsilon)-\hat{\alpha}') + 2C^2((2 + \epsilon)\cos \tau - \hat{\alpha}')]^3}  \; d\tau \right)^{1/2} dC \right]
    \end{split}\label{Nufinspiral}
\end{equation}
on using $h_\tau = A'/(1+\lambda)$, and the appropriate limits for the $C$-integral; as remarked earlier, $Nu$ is independent of the choice of $C^*$. \eqref{Nufinspiral} contains only one integral that denotes the contribution of the full unit hemisphere. This difference in the number of distinct contributions, between the spiralling and non-spiralling regimes, is also evident from the additional factor of 2 in the denominators of the pre-factors in \eqref{Nufinopen1}) and \eqref{Nufinopen2}, compared to \eqref{Nufinspiral}. 

\begin{figure}
    \centering
    \includegraphics[scale = 0.5]{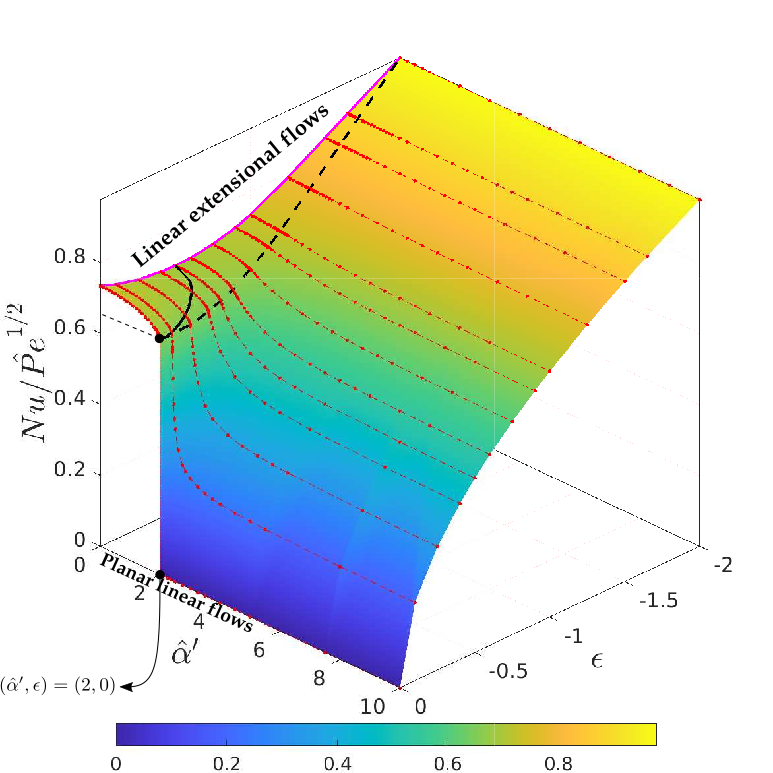} \\
    \includegraphics[scale = 0.275]{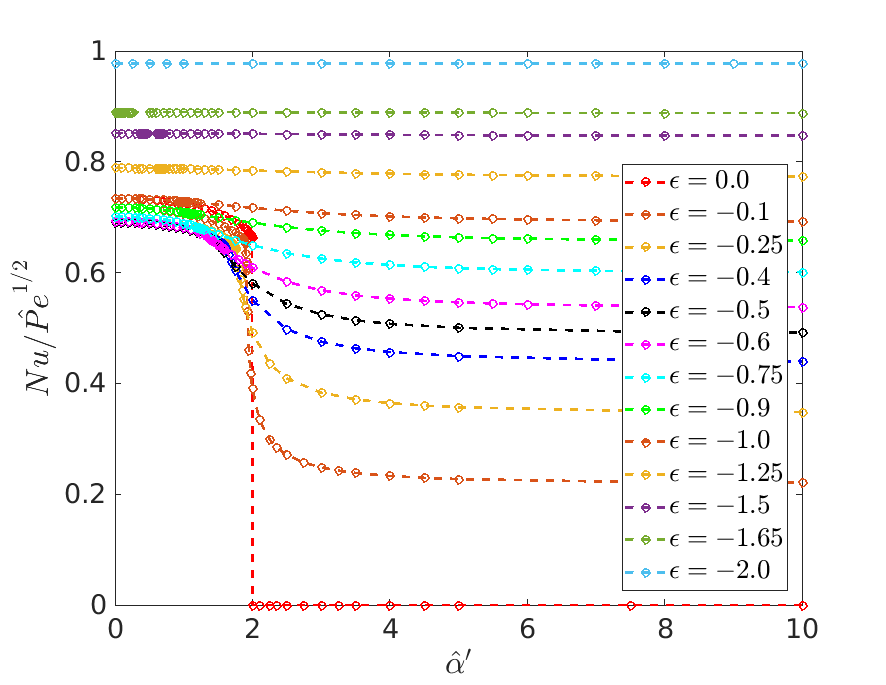}
    \includegraphics[scale = 0.275]{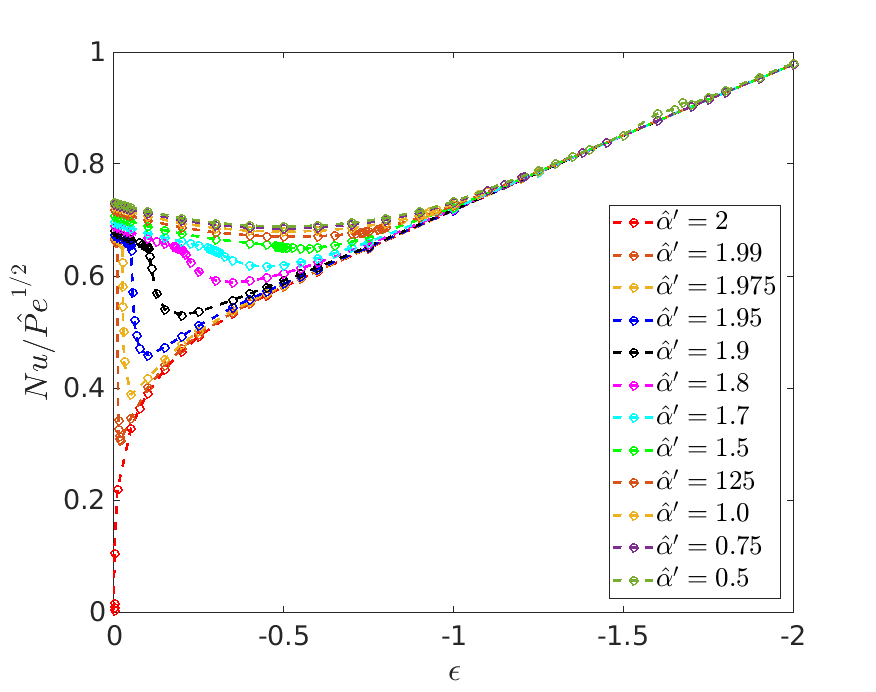}
    \caption[\textwidth]{\justifying{(a) The $Nu/\hat{Pe}^{1/2}$-surface, for a drop in a 3D extensional flow with aligned vorticity, as a function of $\epsilon$ and $\hat{\alpha}' = \hat{\alpha}(1+\lambda)$. The dashed black curve, $\hat{\alpha}' = \hat{\alpha}'_{th1}$, corresponds to the boundary between spiralling and non-spiralling surface-streamline topologies, while the solid black curve, $\hat{\alpha}' = \hat{\alpha}'_{th2}$, corresponds to a degenerate topology with a fixed-point ring; the bounding magenta curve for $\hat{\alpha}' = 0$ corresponds to extensional flows, and goes through a minimum at $\epsilon = -1/2$\,(planar extension). (b) $Nu/\hat{Pe}^{1/2}$ as a function of $\hat{\alpha}'$ for different fixed values of $\epsilon$, showing the approach towards a jump discontinuity\,(at $\hat{\alpha}' = 2$) for $\epsilon \rightarrow 0$. (c) $Nu/\hat{Pe}^{1/2}$ as a function of $\epsilon$, for different fixed $\hat{\alpha}'$ in the interval $[0,2)$. }}
    \label{fig:Nu_aligned}
\end{figure}

The integrals in \eqref{Nufinopen1},\eqref{Nufinopen2}, and \eqref{Nufinspiral} are readily evaluated numerically, allowing one to construct the $Nu/\hat{Pe}^{\frac{1}{2}}$-surface as a function of $\epsilon$ and $\hat{\alpha}'$, as shown in Fig.\ref{fig:Nu_aligned}. The simpler expression at $\hat{\alpha}' =\hat{\alpha}'_{th1}$, corresponding to the boundary between spiralling and non-spiralling topologies, may be obtained from either \eqref{Nufinopen} or \eqref{Nufinspiral}, and is given by:
\begin{align}
    \frac{Nu}{\hat{Pe}^{1/2}} = \frac{4 (2+\epsilon)}{\pi^{3/2}} \int_0^{\infty} C \left( \int_{-\infty}^{\infty}\frac{e^{3\epsilon \tau} d\tau}{[e^{3\epsilon\tau} + 2C^2(1 + \tau^2 (2+\epsilon)^2)]^3} \right)^{1/2} dC, \label{Nufinthres}
\end{align}
with $\epsilon \in [-2,0]$. For any nonzero $\epsilon$, the value obtained from (\ref{Nufinthres}), shown as the dashed black curve in Fig.\ref{fig:Nu_aligned}, matches with that obtained using either \eqref{Nufinopen} or \eqref{Nufinspiral}, for $\hat{\alpha}'$ approaching $\hat{\alpha}'_{th1}$\,(from either side). Thus, for $\epsilon \neq 0$, the $Nu/\hat{Pe}^{\frac{1}{2}}$ varies continuously with varying $\hat{\alpha}'$, despite the qualitative alteration of the surface-streamline topology across $\hat{\alpha}' =\hat{\alpha}'_{th1}$. For $\epsilon = 0$, (\ref{Nufinthres}) reduces to:
\begin{align}
    \frac{Nu}{\hat{Pe}^{1/2}} = \frac{2^{3/4}3^{1/2}}{\pi} \int_0^{\infty} \frac{C^{1/2}}{(1+2 C^2)^{5/4}} dC = 2\sqrt{\frac{3}{\pi}}\frac{\Gamma(3/4)}{\Gamma(1/4)}, \label{Nu:merid}
\end{align}
where $\Gamma(z)$ denotes the Gamma function. This analytical result matches with the one obtained by \cite{Deepak18a}\,(Eq.3.69 therein) for a meridional surface-streamline topology, once the difference in the defintions of $\bm{\Gamma}$ is accounted for. 
Note that, for $\epsilon = 0$, $Nu/\hat{Pe}^{\frac{1}{2}}$ drops discontinuously from (\ref{Nu:merid}) to $0$ for $\hat{\alpha}' > \hat{\alpha}'_{th1}\,(=2)$ owing to the onset of a closed streamline topology. 

As discussed in section \ref{sec:alignedfamily}, a second degenerate topology occurs at $\hat{\alpha}' = \hat{\alpha}'_{th2}$ - see unit sphere with a fixed-point ring in Fig.\ref{fig:discandstream}d. The exponentials in \eqref{Nufinopen1} and \eqref{Nufinopen2} simplify since $A = 3\epsilon$ at this $\hat{\alpha}'$, which allows for the $\tau$-integrals in (\ref{Nufinopen1}) and (\ref{Nufinopen2}) to be evaluated analytically. While the final $C$-integral must still be numerically calculated, this nevertheless serves to validate the evaluation of the full 2D integral. The validation is important particularly because, as mentioned in section \ref{sec:C_tau_aligned}, the stable and unstable manifolds of the saddle point, for $\hat{\alpha}' < \hat{\alpha}'_{th2}$, correspond to $\tau = -\infty$ and $\infty$. As a result, with approach towards the saddle, any finite $\tau$-interval\,(used for numerical integration) will correspond to a progressively shorter stretch of a surface streamline, raising the potential for numerical inaccuracy. We have compared the semi-analytical and numerical results for $Nu$ at $\hat{\alpha}' = \hat{\alpha}'_{th2}$, and used this to calibrate the $\tau$-limits for the calculation at other parameter values.

The shape of the $Nu/\hat{Pe}^{\frac{1}{2}}$-surface in Fig.\ref{fig:Nu_aligned}a changes in a nontrivial manner with increasing $\hat{\alpha}'$. The constant-$\hat{\alpha}'$ contours in Fig.\ref{fig:Nu_aligned}c show $Nu/\hat{Pe}^{1/2}$ to be a non-monotonic function of $\epsilon$ for $\hat{\alpha}' < 2$. This non-monotonic dependence is ‘seeded’ by the bounding curve  at $\hat{\alpha}' = 0$\,(highlighted in magenta), corresponding to linear extensional flows, whose minimum at $\epsilon = -1/2$\,(planar extension) arises from a redundancy of the flow-type parameterization. As explained in section \ref{sec:setup}, to within a change of sign, both $\epsilon \in [-1,-1/2]$ and $[-1/2,0]$ correspond to 3D extensional flows that interpolate between axisymmetric and planar extensions, and for $\hat{\alpha}' = 0$ therefore, the same sequence of $Nu$ values is repeated on either side of $\epsilon = -1/2$. There is no redundancy for nonzero $\hat{\alpha}'$, however, and the $Nu/\hat{Pe}^{\frac{1}{2}}$-minimum moves from $\epsilon = -1/2$ to $\epsilon = 0$ as $\hat{\alpha}'$ increases from $0$ to $2$. The non-monotonic variation gives way to a monotonic increase, with decreasing $\epsilon$, for larger $\hat{\alpha}' > 2$\,(not shown). Fig.\ref{fig:Nu_aligned}b shows the variation of $Nu/\hat{Pe}^{\frac{1}{2}}$ as a function of $\hat{\alpha}'$ with $\epsilon$ fixed. $Nu/\hat{Pe}^{\frac{1}{2}}$ transitions from a (negative)\,jump discontinuity at $\hat{\alpha}' = 2$ for $\epsilon = 0$, to a smooth monotonic decrease with increasing $\hat{\alpha}'$ for $-2 < \epsilon < 0$, to finally being independent of $\hat{\alpha}'$ at $\epsilon = -2$. Note that $\lim\limits_{\hat{\alpha}' \rightarrow 2^-,\epsilon = 0} Nu/\hat{Pe}^{\frac{1}{2}}$ is given by (\ref{Nu:merid}), while $\lim\limits_{\hat{\alpha}' \rightarrow 2^+,\epsilon = 0} Nu/\hat{Pe}^{\frac{1}{2}} = 0$, leading to the aforementioned jump; the $\hat{\alpha}'$-independence for axisymmetric extension is shown below. Fig.\ref{fig:Nu_aligned1} shows $Nu/\hat{Pe}^{\frac{1}{2}}$-surfaces as a function of $\epsilon$ and the unscaled flow-type parameter $\hat{\alpha}$, and helps illustrate the role of changing $\lambda$. Apart from an overall reduction in $Nu$, an increase in $\lambda$ leads to a smaller vorticity threshold corresponding to the jump discontinuity along the $\hat{\alpha}$-axis.
\begin{figure}
    \centering
    \includegraphics[scale = 0.22]{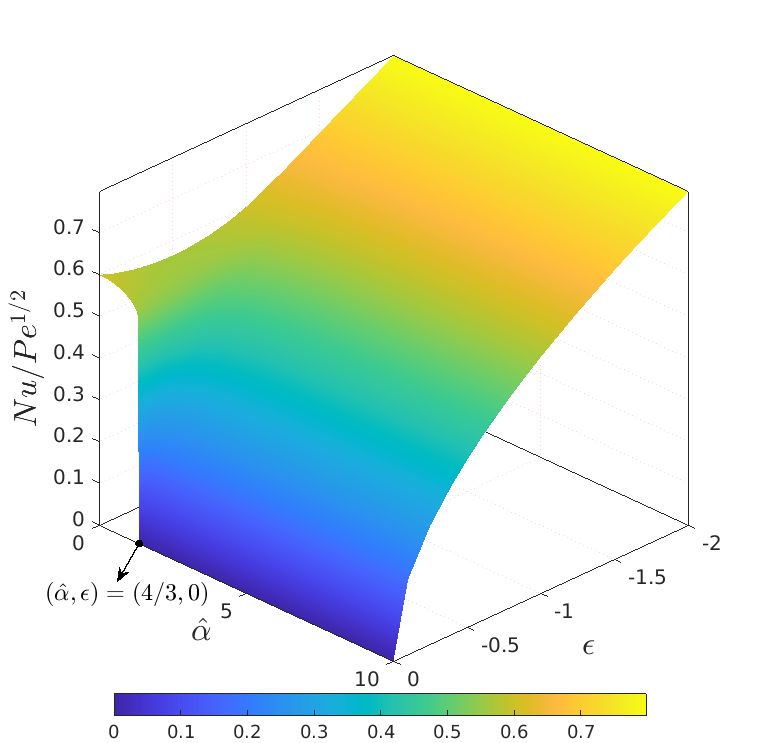}
    \includegraphics[scale = 0.22]{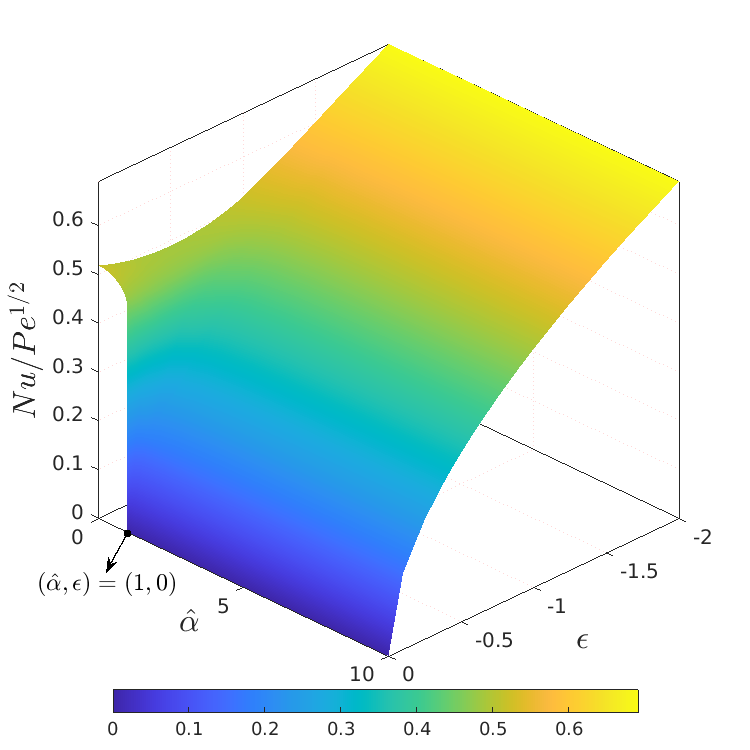}
    \includegraphics[scale = 0.22]{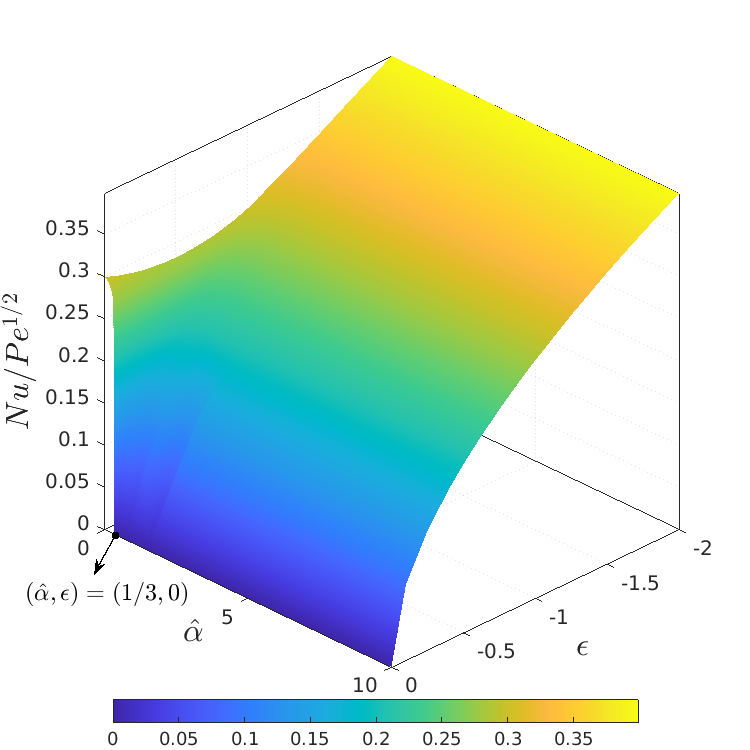}
    \caption[\textwidth]{\justifying{$Nu/\hat{Pe}^{1/2}$-surfaces for a drop, in a 3D extensional flow with aligned vorticity, as a function of $\epsilon$ and $\hat{\alpha}$, and for (a) $\lambda = 0.5$, (b) $\lambda = 1$ and (c) $\lambda =5$.}}
    \label{fig:Nu_aligned1}
\end{figure}
\subsubsection{Special cases} \label{special_cases}
We now connect to results known from the literature, and that were mentioned in section \ref{1}. 
\begin{itemize}
    \item The limiting case $\epsilon = 0$, $\hat{\alpha}' < 2$ corresponds to the subset of canonical planar linear flows with open surface streamlines, and we have verified that the results obtained here are identical to those of \cite{Deepak18a} who calculated $Nu$ for these flows. While the said authors depicted their results as a $Nu/Pe^{\frac{1}{2}}$-surface that is a function of $\alpha$ and $\lambda$\,(Fig.7 therein), with the open-streamline region corresponding to $\lambda < \lambda_c$, use of $\hat{\alpha}'$ allows us to compress this surface into a curve that is one of the boundaries of the $Nu/\hat{Pe}^{1/2}$-surface in Fig.\ref{fig:Nu_aligned}; $Nu$ for the limiting meridional topology, along the curve $\lambda = \lambda_c$, was mentioned above in (\ref{Nu:merid}).

    \item Another limiting case is linear extensional flows, corresponding to $\hat{\alpha}' = 0$, and the $Nu$ for which was also calculated by \cite{Deepak18a}. We have verified that our results over any of the intervals $\epsilon \in [-2,-1]$, $[-1,-1/2]$ or $[-1/2,0]$\,($Pe$ being redefined appropriately, depending on the interval), agree with those of the said authors.
    
    \item The final limiting case is $\epsilon = -2$, an axisymmetric extension with the vorticity vector along the symmetry axis. Since any non-zero $\hat{\alpha}'$ leads to spiralling streamlines in this limit, we start from \eqref{IntFact3}, which takes the form
    \begin{align}
        (h k \sin \beta)^2 = \frac{C^2 e^{12 \tau/\hat{\alpha}'}}{(1 + 2C^2 e^{6 \tau/\hat{\alpha}'})^3},
    \end{align}
 for $\epsilon \rightarrow -2$. Using this and $h_\tau = \hat{\alpha}'/(1+\lambda)$ in \eqref{Nufinspiral}, one obtains:
    \begin{align}
        Nu &= \frac{\hat{Pe}^{1/2}\hat{\alpha}'^{1/2}}{\pi^{3/2}} \int_{C^*}^{C^* e^{|-12\pi/\hat{\alpha}'|}} C \left( \int_{-\infty}^{\infty} \frac{e^{12\tau/\alpha'}}{(1 + 2C^2 e^{6 \tau/\alpha'})^3}\right)^{1/2} dC, \\
        &= \frac{\hat{Pe}^{1/2} \hat{\alpha}'^{1/2}}{\pi^{3/2}} \int_{C^*}^{C^* e^{12\pi/\hat{\alpha}'}} C\left( \frac{\hat{\alpha}'}{48 C^4}\right)^{1/2} dC, \\
        &= \hat{Pe}^{1/2}\sqrt{\frac{3}{\pi}}, \label{Nu_alp0}
    \end{align}
    which is independent of $\hat{\alpha}'$, and thence, the same as the result obtained by \cite{Gupalo72}; see item (ii) in section \ref{1}.\\

\end{itemize}

\subsection{Drop in axisymmetric extension flows with inclined vorticity} \label{Nu:inclinedvort}

The $C$ and $\tau$ coordinates for this family were defined formally, in terms of $\phi_0'$ and $\theta_0'$, by \eqref{auxsolnew1A} and \eqref{auxsolnew2A}.
To determine $Nu$, we first calculate the eigenvalues and eigenvectors of $\hat{\bm \Gamma}$, which allows one to identify the lone real eigenvector associated with the pair of foci in the spiralling regime, or the eigenvector associated with the pair of saddle points in the non-spiralling regime. Choosing a spherical coordinate system with its polar axis aligned with the identified eigenvector, and using the chain rule of differentiation along with the aforementioned coordinate definitions, one may write
\begin{align}
    h k \sin \beta = \frac{\partial \tan \theta'_0}{\partial C} \frac{d \tan \phi'_0}{d \tau} \frac{\tan \theta'_0}{(1+\tan^2 \phi'_0)(1+\tan^2\theta'_0)^{3/2}}, \label{geofact_inc}
\end{align}
for the areal metric that appears in the $Nu$-integral in \eqref{Nufin}. The $\tau$-integral in (\ref{Nufin}) always runs from $-\infty$ to $\infty$. The $C$-integral runs from $0$ to $\infty$ for the non-spiraling case, while for the spiralling case, it runs from $C^*$ to $C^*f(T)$, where $f(T)$ is defined by:
\begin{align}
    f(T) = \frac{\tan \theta'(T,C^*)}{C^* \tan \theta'(0,C^*)},
\end{align}
with $T$ being the $\tau$-interval corresponding to a single turn of a spiralling streamline with label $C^*$, and therefore being satisfying the relation:
\begin{align}
    \phi'(\tau = T) = \phi'_0.
\end{align}
Recall that $f(T) = \mathrm{e}^{|6 \epsilon \pi|/A'}$ for the aligned-vorticity family. The above procedure is carried out for each $(\theta_\omega, \hat{\alpha}')$ pair. Two special cases are useful for validation purposes. The first of these corresponds to $\theta_\omega = 0$, for which $Nu/\hat{Pe}^{1/2}$ equals $\sqrt{3/\pi}$\,\citep{Gupalo72}. The second is $\theta_\omega = \pi/2$, which corresponds to $\epsilon = -1/2$ in the aligned-vorticity family, and for which the semi-analytical result was determined in the manner described in section \ref{sec:alignedfamily}. For both cases, the procedure above yields results in exact agreement with the analytical or semi-analytical predictions.

We now discuss the results of the inclined-vorticity $Nu$ calculation, organizing the results based on our earlier analysis of surface-streamline topologies in section \ref{sec:incled_stream_org}. This requires examination of the intervals (i) $0 \leq \theta_\omega < \theta_\omega^{th1}$, (ii) $\theta_\omega^{th1} < \theta_\omega < \theta_\omega^{th2}$, and (iii) $\theta_\omega^{th2} < \theta_\omega \leq \pi/2$, along with separate consideration of the transition values $\theta_\omega^{th1}$ and $\theta_\omega^{th2}$. Before doing so, we focus on the asymptotic results for $Nu/\hat{Pe}^{1/2}$ in the limit of small and large $\hat{\alpha}'$. The detailed derivations in this regard are relegated to Appendices \ref{SmallNu_asy} and \ref{LargeNu_asy}, respectively, and we only mention the principal conclusions here. $\hat{\alpha}' = 0$ corresponds to axisymmetric extension and $Nu = \sqrt{3/\pi}\hat{Pe}^{1/2}$ in this limit, as already shown in the earlier section, in the context of the aligned-vorticity family\,(see (\ref{Nu_alp0})). The first correction to this result appears at $O(\hat{\alpha}'^2)$, consistent with invariance to a reversal in rotation\,($\hat{\alpha}' \leftrightarrow -\hat{\alpha}'$), and is a function of $\theta_\omega$; the analysis in Appendix \ref{SmallNu_asy} gives $\frac{Nu_0 - Nu}{\hat{Pe}^{1/2}} = \frac{(2-\ln 2)\sin^2 \theta_\omega}{48\sqrt{3 \pi}} \hat{\alpha}'^2$. In the opposite limit, $\hat{\alpha}' \rightarrow \infty$, $Nu$ is determined by $Pe_\omega$, as first shown by \citet{Batchelor79} for a spherical particle in a vortical ambient linear flow. The result of this calculation was stated in section \ref{1}, and shows that the transport rate depends only on the orientation of $\bm \omega$, not its magnitude, in turn implying a dependence on $\theta_\omega$\,(via $E_\omega$) but not $\hat{\alpha}'$. The difference for the drop is that $Nu$ is proportional to $Pe_\omega^{\frac{1}{2}}$ rather than $Pe_\omega^{\frac{1}{3}}$, and the analysis in Appendix \ref{LargeNu_asy} gives $Nu_\infty = \displaystyle\lim_{\hat{\alpha}' \gg 1} Nu = \sqrt{\frac{3|1+3\cos 2\theta_\omega|}{4\pi}}\hat{Pe}_\omega^{\frac{1}{2}}$, with $\hat{Pe}_\omega= Pe_\omega/(1+\lambda)$; note that one recovers the result for axisymmetric extension above for $\theta_\omega = 0$. The aforementioned analytical asymptotes not only serve as a validation for the arbitrary-$\hat{\alpha}'$ $Nu$ calculation, but are also indispensable owing to the nature of spiraling in the said limits. For $\hat{\alpha}' \ll 1$, while the streamlines are tightly wound very near the focus and the equatorial limit cycle, the rate of spiralling is very small in the main portion of the unit hemisphere. The resulting large range of $\theta$ spanned by a single turn of a spiralling streamline implies that the ratio of the integration limits of the $C$-integral\,(which equals $f(T)$), will be exponentially large, making numerical integration difficult. For $\hat{\alpha}' \gg 1$, the tight nature of spiraling implies that the lower\,($\hat{C}$) and upper\,($\hat{C} f(T)$) limits of integration now approach each other (i.e. $f(T) -1 \ll 1$), and the vanishingly small interval of integration again leads to numerical difficulties.

\begin{figure}
    \centering
    \includegraphics[scale = 0.32]{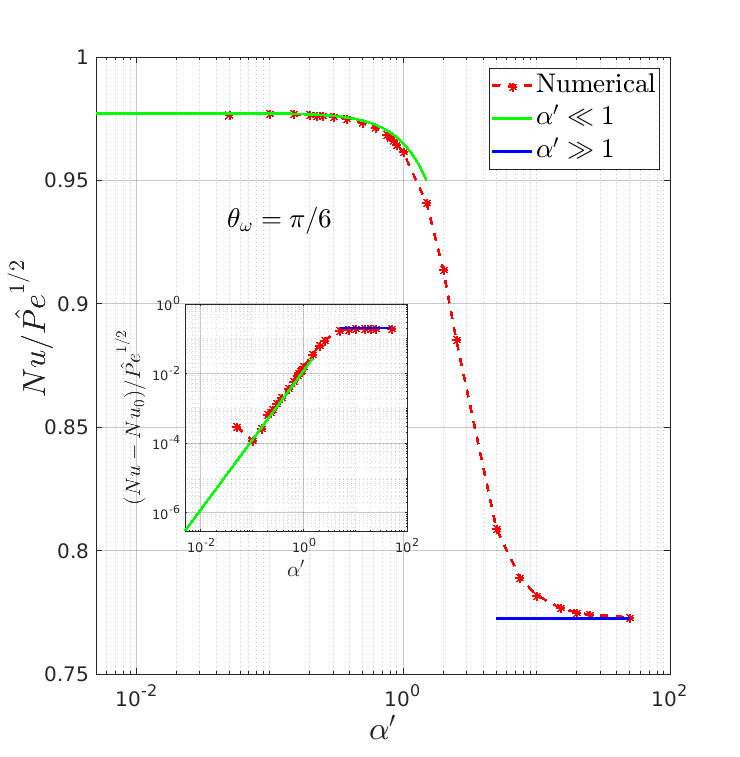}
    \includegraphics[scale = 0.32]{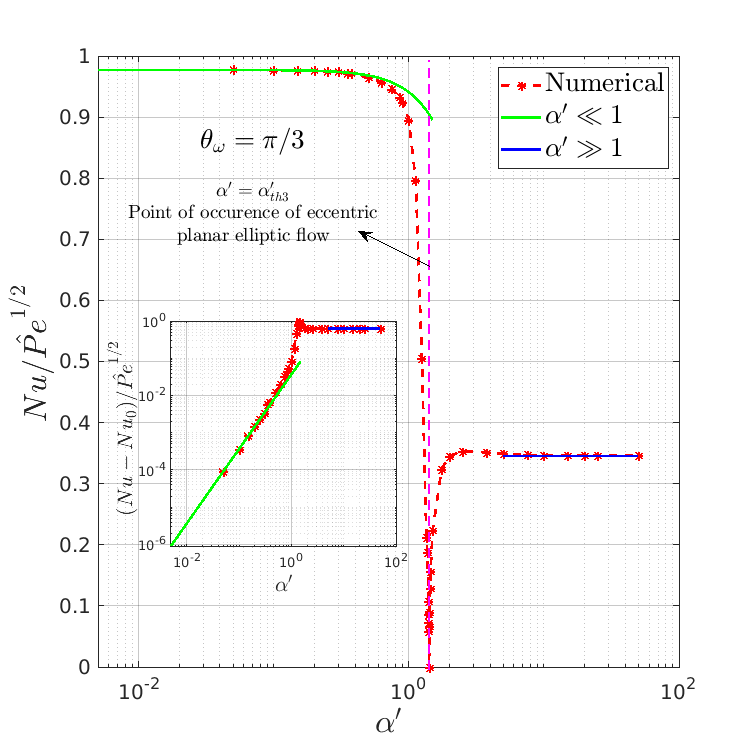}
    \includegraphics[scale = 0.32]{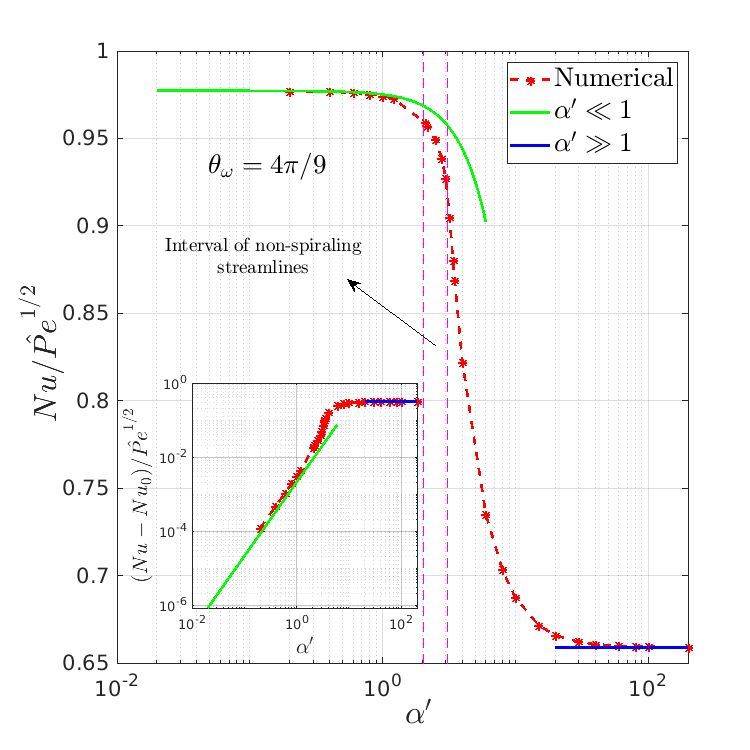}    \includegraphics[scale = 0.32]{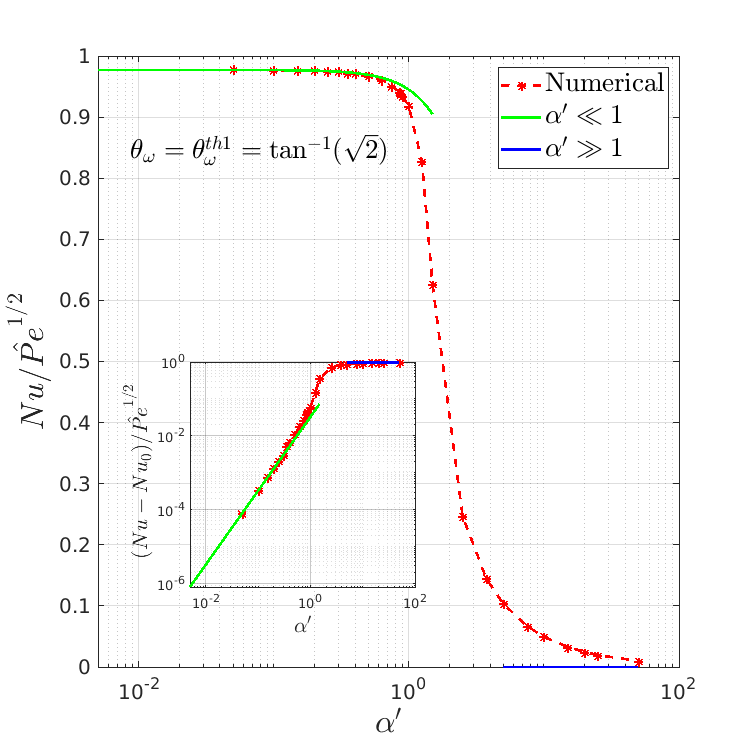}
    \includegraphics[scale = 0.32]{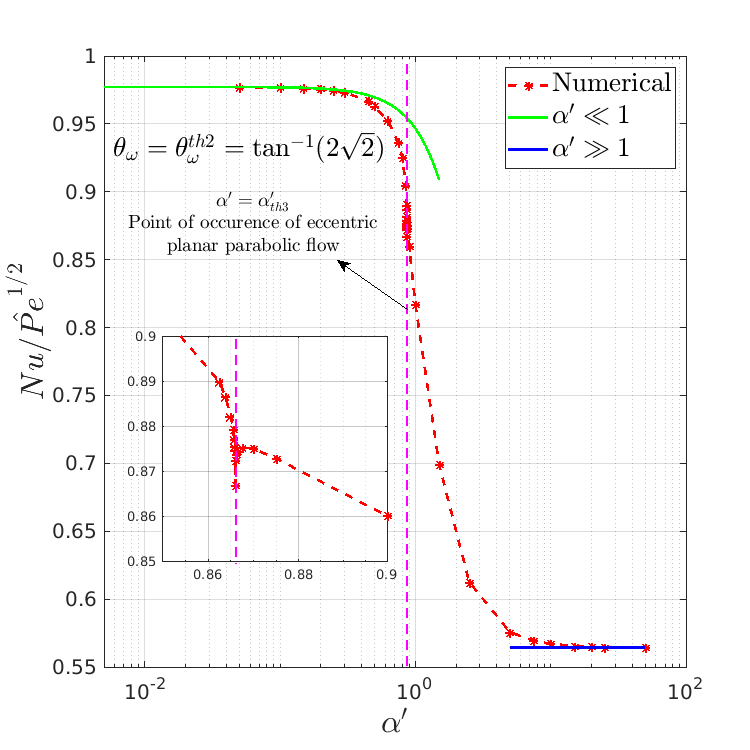}    
    \caption[\textwidth]{\justifying{$Nu/\hat{Pe}^{1/2}$ vs $\hat{\alpha}'$ for (a)\,$\theta_\omega = \pi/6\in [0,\theta_\omega^{th1})$; (b)\,$\theta_\omega = \pi/3\in (\theta_\omega^{th1},\theta_\omega^{th2})$; (c)\,$\theta_\omega = 4\pi/9,\in (\theta_\omega^{th2},\frac{\pi}{2})$; (d)\,$\theta_\omega = \theta_\omega^{th1}$; and (e)\,$\theta_\omega = \theta_\omega^{th2}$. Magenta lines in (b) and (e), corresponding to downward-pointing cusps, mark the occurrence of eccentric elliptic and parabolic surface-streamline topologies, which coincide with a local minimum of $Nu/\hat{Pe}^{1/2}$; the minimum is zero in the former case, and equals $0.8669$ in the latter. The pair of magenta lines in (c) marks the interval, $\hat{\alpha}' \in (2.01,3.09)$, of non-spiralling streamline topologies. The green and blue curves in the main plots are the small and large $\hat{\alpha}'$-asymptotes, respectively; the green line with slope $2$ in the insets is the small-$\hat{\alpha}'$ asymptote on a log scale.}}
    \label{fig:Nu_inclined_ind}
\end{figure}

\begin{enumerate}
    \item For $\theta_\omega = \frac{\pi}{6} \in [0,\theta_\omega^{th1})$, Fig.\ref{fig:Nu_inclined_ind}a shows that $Nu/\hat{Pe}^{1/2}$ monotonically decreases from a zero-$\hat{\alpha}'$ to an infinite-$\hat{\alpha}'$ plateau. Increasing the vorticity-to-extension ratio in this $\theta_\omega$-interval leads to a tighter spiralling of the surface streamlines, as a result of which the magnitude of the ambient vorticity plays a progressively smaller role in transport enhancement. For $\hat{\alpha}' \rightarrow \infty$, it is only the orientation of the vorticity vector that controls the rate of transport, as evident from the expression for $Nu_\infty$ above. The numerical results for $Nu$ match well with the analytical asymptotes in the small and large-$\hat{\alpha}'$ limits. In fact, the small-$\hat{\alpha}'$ asymptote remains close to the numerical curve even for $\hat{\alpha}'$ values larger than unity. This unexpectedly large range of validity is highlighted by the inset figure which plots $(Nu_0 - Nu)/\hat{Pe}^{1/2}$ against $\hat{\alpha}'$ on a logarithmic scale, and also helps confirm the $O(\hat{\alpha}'^{2})$ scaling. We shall see below that the range of validity of the above asymptote decreases for larger $\theta_\omega$, and this is correlated to the emergence of an intervening interval of non-spiralling streamline topologies for $\theta_\omega > \theta_\omega^{th2}$ in Fig.\ref{fig:Nu_inclined_ind}c.
    
    \item For $\theta_\omega \in (\theta_\omega^{th1},\theta_\omega^{th2})$, recall from  Fig.\ref{fig:discandstream_inc}b that surface streamlines have a spiralling character for all $\hat{\alpha}'$, except at $\hat{\alpha}' = \hat{\alpha}'_{th3}$ when they are closed curves\,(generalized Jeffery orbits), with the closed surface-streamline topology expected to lead to a reduced transport rate; the direction of spiralling also reverses across $\hat{\alpha}' = \hat{\alpha}'_{th3}$. Accordingly, for $\theta_\omega = \frac{\pi}{3}$ which lies in the above interval, Fig.\ref{fig:Nu_inclined_ind}b shows that $Nu/\hat{Pe}^{1/2}$ decreases with $\hat{\alpha}'$ starting from the initial plateau, approaching zero for $\hat{\alpha}' \rightarrow \hat{\alpha}'_{th3}$.  
    It increases again thereafter, eventually plateauing at large $\hat{\alpha}'$ after going through a shallow maximum. The plateau value is in agreement with the large-$\hat{\alpha}'$ asymptote above, and remains smaller than the zero-$\hat{\alpha}'$ plateau, thereby confirming an overall reduction\,(with increasing $\hat{\alpha}'$) in the role played by the ambient rate of rotation in transport enhancement. Unlike Fig.\ref{fig:Nu_inclined}a, the $Nu/\hat{Pe}^{1/2}$-variation is now non-monotonic, with the approach to zero at $\hat{\alpha}' = \hat{\alpha}'_{th3}$ manifesting as a downward-pointing cusp on the logarithmic scale; the location of the cusp being marked by a dashed magenta line in Fig.\ref{fig:Nu_inclined_ind}b. As for case (i), the inset highlights the $O(\hat{\alpha}'^2)$ scaling for small $\hat{\alpha}'$.

    \item For $\theta_\omega = \frac{4\pi}{9} \in(\theta_\omega^{th2},\pi/2]$, there is an intervening interval of non-spiraling streamline topologies, seen earlier in Fig.\ref{fig:discandstream_inc}e. Fig.\ref{fig:Nu_inclined_ind}c shows that $Nu/\hat{Pe}^{1/2}$ decreases monotonically with increasing $\hat{\alpha}'$, and that this decrease remains smooth despite the transition between spiralling and non-spiralling surface-streamline topologies; the interval corresponding to the latter is marked by a pair of dashed magenta lines. Thus, similar to the $Nu$-behavior for the aligned-vorticity family\,(for nonzero $\epsilon$) in Fig.\ref{fig:Nu_aligned}, the two zero-crossings in the  $\Delta'-\hat{\alpha}'$ curve in Fig.\ref{fig:discandstream_inc}e do not lead to any signatures as far as the scalar transport rate is concerned. There is also an $\hat{\alpha}'$ value corresponding to a planar hyperbolic flow, corresponding to the zero-crossing of the $R'-\hat{\alpha}'$ curve in Fig.\ref{fig:discandstream_inc}e, but this again has no effect on $Nu$, as is expected from the discussion on the aligned-vorticity scenario. It is worth noting the more rapid departure of the $Nu$-curve from its small-$\hat{\alpha}'$ asymptote for $\hat{\alpha}' \gtrsim 1$, when compared to case (i), owing to impending interval of non-spiralling streamline topologies. 

    \item Finally, we examine the $Nu/\hat{Pe}^{1/2}$-curve at the two transition values. Fig.\ref{fig:Nu_inclined_ind}d shows that, at $\theta_\omega = \theta_\omega^{th1}$, $Nu/\hat{Pe}^{1/2}$ decreases monotonically with $\hat{\alpha}'$, as for cases (i) and (iii), but approaches zero for $\hat{\alpha}' \rightarrow \infty$. The latter is because the factor $(1+3\cos2\theta_\omega)$, in the expression for the large-$\hat{\alpha}'$ asymptote above, equals zero for $\theta_\omega = \theta_\omega^{th1}$. This implies $E_\omega = 0$, and therefore, the absence of boundary-layer-enhanced transport for $\hat{\alpha}' \rightarrow \infty$. The vorticity vector $\bm \omega$ now lies on the surface corresponding to a zero rate-of-stretch, defined by $E_\omega = 0$ and that is a right circular cone for an axisymmetric extension\,\citep{Sabarish22}. As a result, the flow within the boundary layer is a solid-body rotation at leading order; note that this is consistent with $R'$ approaching zero for $\alpha' \rightarrow \infty$\,(see Fig.\ref{fig:discandstream_inc}). 
    
    Fig.\ref{fig:Nu_inclined_ind}e plots $Nu/\hat{Pe}^{1/2}$ for $\theta_\omega = \theta_\omega^{th2}$. The $Nu$-variation with $\hat{\alpha}'$ has a non-monotonic character, with the non-monotonicity arising partly from the cusp at $\hat{\alpha}' = 2\sqrt{3}$\,(see inset). This is similar to Fig.\ref{fig:Nu_inclined_ind}b, although the cusp now terminates at a finite value. For $\hat{\alpha}' > 2\sqrt{3}$, the curve again goes through a local maximum, before asymptoting to a large-$\hat{\alpha}'$ plateau. The point $(\theta_\omega,\hat{\alpha}') \equiv (\theta_\omega^{th2},2\sqrt{3})$ corresponding to the cusp is the degenerate point of intersection of the $\hat{\alpha}'_{th1-2}$ and $\hat{\alpha}'_{th3}$ loci in Fig.\ref{fig:alphathres_inc}. The auxiliary flow at this point is a parabolic linear flow with $R' = \Delta' = 0$; see encircled unit sphere in Fig.\ref{fig:discandstream_inc}e. While all surface streamlines originate from, and end in, either of two fixed points\,(one of which is shown on the said unit sphere), they take an infinite time to do so. They are akin to open surface streamlines in this sense, and one therefore expects $Nu/\hat{Pe}^{1/2}$ to be finite. The finiteness is also consistent with \cite{Deepak18a} finding a finite $Nu$ for a spherical drop in simple shear flow. The latter is the canonical counterpart of a parabolic linear flow\,(again, with $Q' = R' = 0$), and the surface streamlines have a meridional character. The meridians again originate and end in a pair of degenerate fixed points on the flow axis, taking an infinite time to go from one fixed point to the other; the resulting finite $Nu$ was shown, in section \ref{Nu:alignedvort}, to be given by (\ref{Nu:merid}).
\end{enumerate}

\subsubsection{Discussion and Special cases} \label{sec:spcases_inclined}

\begin{figure}
    \centering
    \includegraphics[scale = 0.325]{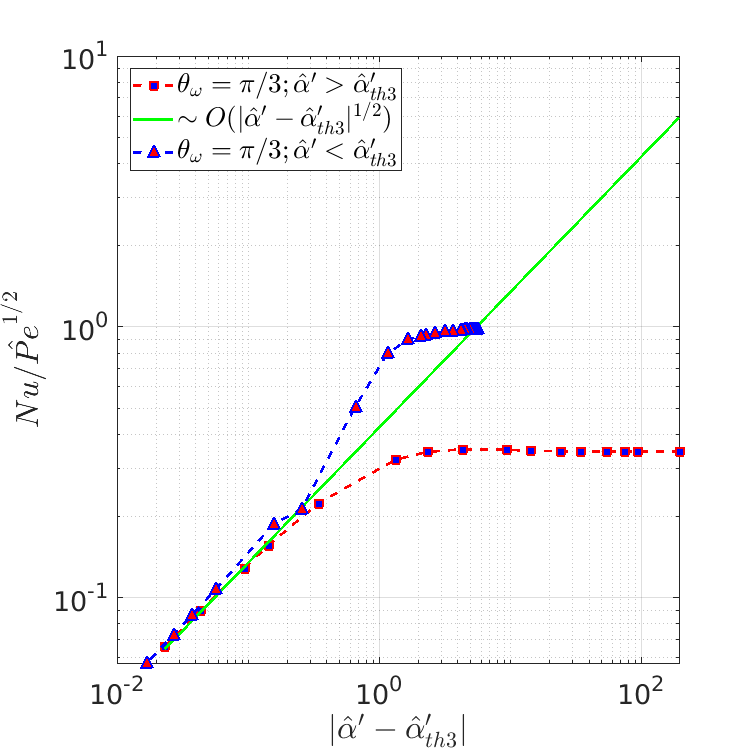}
    \includegraphics[scale = 0.325]{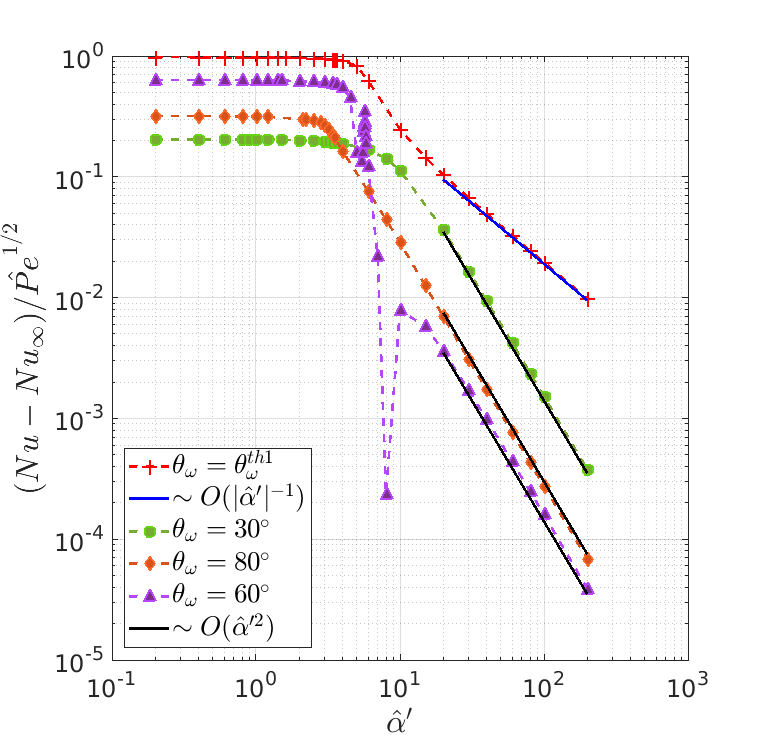} 
    \caption[\textwidth]{\justifying{(a) $Nu/ \hat{Pe}^{1/2}$ vs $|\hat{\alpha}' - \hat{\alpha}'_{th3}|$ for $\theta_\omega = \pi/3$; the red and blue curves show that the $Nu$ values conform to a square-root scaling in the neighborhood of the eccentric elliptic auxiliary flow. (b) $Nu/ \hat{Pe}^{1/2}$ vs $\hat{\alpha}'$ for different $\theta_\omega$ highlights the two different types of large-$\hat{\alpha}'$ behavior: $Nu/ \hat{Pe}^{1/2} \propto |\hat{\alpha}^{'^-1}|$ for $\theta_\omega = \theta_\omega^{th1}$, and $Nu/ \hat{Pe}^{1/2} \propto \hat{\alpha}^{'^-2}$ for $\theta_\omega \neq \theta_\omega^{th1}$.}}
    \label{fig:Nu_inclined_spec}
\end{figure}

We first examine the case where the auxiliary linear flow is an eccentric elliptic flow, corresponding to $\hat{\alpha}' = \hat{\alpha}'_{th3}$ in Fig.\ref{fig:Nu_inclined_ind}b, and for which $Nu/\hat{Pe}^{1/2}$ was seen to approach zero in the limit $Pe \gg 1$, 
As already pointed out, the reduction in transport rate arises because streamlines on the drop surface\,($r = 1$) are closed at the said $\hat{\alpha}'$. Since these surface streamlines may be identified as eccentric versions of the original Jeffery orbits\,\citep{Sabarish21}, one may, in principle, examine the convection-diffusion equation averaged over a Jeffery period. This $\tau$-averaging, by construction, must eliminate the term involving $u_\tau$ at $r = 1$. Further, the boundary layer analysis in section \ref{Nu:calcgen} shows that $u_C$ at $r= 1$ is always zero on account of $C$ being a surface-streamline label\,(an orbit constant in the present case). Thus, the surface slip velocity responsible for the leading order convection within the boundary layer is identically zero at $\hat{\alpha}' = \hat{\alpha}'_{th3}$. One expects the surface slip to be $O(\hat{\alpha}' - \hat{\alpha}'_{th3})$ for linear flows in the neighborhood of the eccentric elliptic flow. A dimensionless measure of the convective terms\,(relative to the diffusive one) for such flows must be based on magnitude of the above slip velocity scale\,($|\hat{\alpha}' - \hat{\alpha}'_{th3}|\dot{\gamma}a$ in dimensional terms), and is given by $\hat{Pe}|\hat{\alpha}' - \hat{\alpha}'_{th3}|$. This in turn implies that $Nu/\hat{Pe}^{1/2} \sim |\hat{\alpha}' - \hat{\alpha}'_{th3}|^{1/2}$ for $\hat{\alpha}' \rightarrow \hat{\alpha}'_{th3}$. Fig.\ref{fig:Nu_inclined_spec}a, which plots $Nu/\hat{Pe}^{1/2}$ against $|\hat{\alpha}' - \hat{\alpha}'_{th3}|$ for $\theta_\omega = \pi/3$\,(the $\theta_\omega$ corresponding to Fig.\ref{fig:Nu_inclined_ind}b), confirms the above scaling for both $\hat{\alpha}'$ greater and less than $\hat{\alpha}'_{th3}$. 

It is important to point out that, unlike the canonical elliptic flows, $Nu$ for a drop in the aforementioned eccentric elliptic (auxiliary)\,flows is not $O(1)$ for $Pe \gg 1$. This is because although the surface streamlines are closed, the near-surface streamlines are not\,\citep{Sabarish22}, and this is in contrast to a drop in canonical elliptic flows. A drop in the latter flows is always enveloped by a closed-streamline region, as a result of which transport becomes diffusion-limited when the scalar boundary layer becomes becomes comparable in thickness to the region of closed streamlines\,\citep{Torza71,Deepak18a,Deepak18b}. That is to say, $Nu \sim O(1)$ for $Pe \rightarrow \infty$ for a drop in a canonical elliptic flow, and this corresponds to the second class of $Nu-Pe$ relationships mentioned in the introduction. Since the near-surface streamlines for a drop in an eccentric elliptic flow have a tightly spiralling character, the scenario is akin to a spherical particle in a vortical ambient flow. Here, on account of sphere rotation, surface streamlines are (trivially)\,circles in the plane perpendicular to $\bm \omega$, but the near-surface streamlines have a tightly spiralling character for any non-zero $E_\omega$. For $Pe \gg 1$, transport within the boundary layer is due to the extension-induced drift across the circular surface streamlines, and therefore, $Nu \propto Pe_\omega^{\frac{1}{3}}$\citep{Batchelor79}. One therefore expects $Nu \propto Pe^{1/3}$ for a drop in eccentric elliptic flows, although a calculation of the pre-factor in this relation is more involved, and would require a generalization of the Jeffery-orbit-based coordinate system used in \citet{Deepak18b}. Importantly, the $1/3$-scaling above implies that the boundary layer transport at $\hat{\alpha}' = \hat{\alpha}_{th3}$ is driven by the near-surface shear associated with the drift across closed surface streamlines. Thus, for large but finite $Pe$, one expects the analysis in section \ref{Nu:calcgen} to break down when the surface slip of $O(|\hat{\alpha}' - \hat{\alpha}'_{th3}|\dot{\gamma}a)$ becomes comparable to the surface-shear-based velocity scale of $O(\dot{\gamma}\delta)$, where the boundary layer thickness, $\delta \sim (D/\dot{\gamma}|\hat{\alpha}' - \hat{\alpha}'_{th3}|)^{\frac{1}{2}}$. 
Equating these two velocity scales gives $|\hat{\alpha}' - \hat{\alpha}'_{th3}| \sim Pe^{-1/3}$, and the scaling relation $Nu/\hat{Pe}^{1/2} \propto |\hat{\alpha}' - \hat{\alpha}'_{th3}|^{\frac{1}{2}}$ obtained in the previous paragraph will therefore only be valid for $|\hat{\alpha}' - \hat{\alpha}'_{th3}| \gg Pe^{-1/3}$.

Next, we examine the manner in which $Nu/\hat{Pe}^{1/2}$ asymptotes to the infinite-$\hat{\alpha}'$ plateau. In Fig.\ref{fig:Nu_inclined_spec}b, $(Nu - Nu_\infty)/\hat{Pe}^{1/2}$ is plotted as a function of $\hat{\alpha}'$ for $\theta_\omega = \theta_\omega^{th1}$, and for all other $\theta_\omega$ examined, which helps highlight two different kinds of limiting behavior. $(Nu-Nu_\infty)/\hat{Pe}^{1/2} \sim O(1/\hat{\alpha}^{'^2})$ for $\theta_\omega \neq \theta_\omega^{th1}$, while $Nu/\hat{Pe}^{1/2} \sim 1/|\hat{\alpha}'|$ for $\theta_\omega = \theta_\omega^{th1}$, when $Nu_\infty = 0$. The former scaling behavior is on account of the perturbation from the infinite-$\hat{\alpha}'$-plateau having a regular character, while being invariant to a reversal in the sense of rotation. For $\theta_\omega = \theta_\omega^{th1}$, the leading order convecting flow within the boundary layer is $O(\hat{\alpha}^{'^2})$, and this along the with the square-root scaling characteristic of the drop boundary layer yields the singular $|\hat{\alpha}'|^{-1}$ scaling, that is again invariant to rotation reversal.

\begin{figure}
    \centering
    \includegraphics[scale = 0.335]{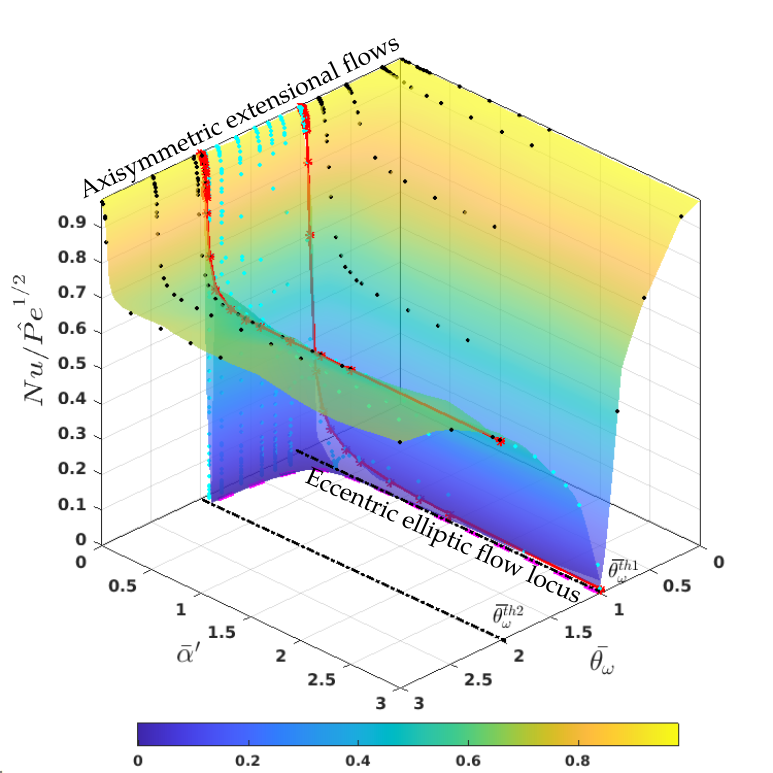} 
    \includegraphics[scale = 0.335]{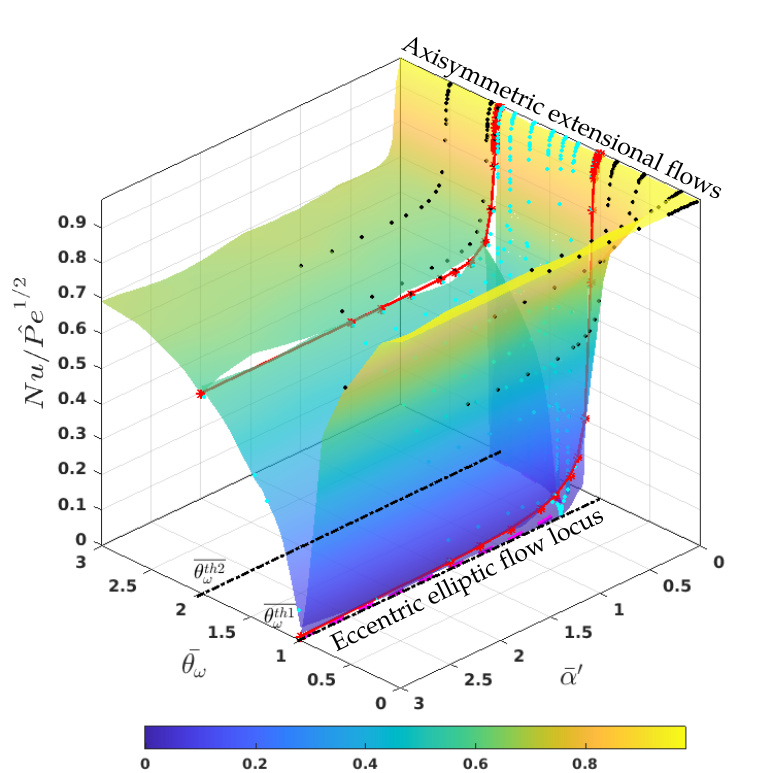}
    \caption[\textwidth]{\justifying{The $Nu/\hat{Pe}^{1/2}$-surface, for a drop in the family of axisymmetric extensions with an inclined vorticity vector, as a function of $\theta_\omega$ and $\hat{\alpha}'$. Towards a clearer depiction, we have used a linear map to compress $\hat{\alpha}' \in [0,100]$ to $[0,3]$, the same length as the $\theta_\omega$-interval; an analogous mapping is used to render the three distinct $\theta_\omega$ sub-intervals equal in length; the flow-type parameters after the mapping are denoted by $\bar{\alpha}'$ and $\bar{\theta}_\omega$. The dashed magenta curve in the $\theta_\omega-\hat{\alpha}'$ plane denotes the eccentric-elliptic-flow locus, $\hat{\alpha}' = \hat{\alpha}'_{th3}$\,[or $16 + \hat{\alpha}^{'^2}(1+3\cos2\theta_\omega) = 0$], along which $Nu/\hat{Pe}^{1/2} = 0$. Black dots correspond to $Nu$-values in the intervals $[0,\theta_\omega^{th1})$ and $(\theta_\omega^{th2},\frac{\pi}{2}]$, and the cyan dots to those in $(\theta_\omega^{th1},\theta_\omega^{th2})$. The pair of red curves are the plots of $Nu/\hat{Pe}^{1/2}$ for $\theta_\omega = \theta_\omega^{th1}$ and $\theta_\omega^{th2}$.}}
    \label{fig:Nu_inclined}
\end{figure}

Having examined the behavior of $Nu/\hat{Pe}^{1/2}$ across a range of $\theta_\omega$, we finally put together all of the curves in Figs.\ref{fig:Nu_inclined_ind}a-e\,(and those at other $\theta_\omega$ not shown)to construct a Nusselt number surface, as was done for the aligned-vorticity case in Fig.\ref{fig:Nu_aligned}. Two different views of this surface are shown in Figs.\ref{fig:Nu_inclined}a and b, and as for the aligned-vorticity case, the $\lambda$-dependence is entirely contained in $\hat{Pe}$ and $\hat{\alpha}'$. The surface starts off from the value, $Nu = \sqrt{\frac{3}{\pi}} \approx 0.9772$, at all points along the $\hat{\alpha}'$ and $\theta_\omega$ axes, this corresponding to axisymmetric extensional flow. The surface then dips to zero along the eccentric-elliptic-flow locus which is shown as a dashed magenta curve in the $\theta_\omega-\hat{\alpha}'$ plane. Recall from the introduction of the inclined-vorticity family in section \ref{sec:setup}, that this locus was defined by $\theta_\omega = \frac{1}{2}\cos^{-1}\left[-\frac{16 + \hat{\alpha}^{'2}}{3 \hat{\alpha}^{'2}}\right]$ - it starts at  $(\theta_\omega,\hat{\alpha}') \equiv (\tan^{-1}(2\sqrt{2}),2\sqrt{3})$, and asymptotes to $\theta_\omega = \tan^{-1}\sqrt{2}$ in the limit $\hat{\alpha}' \rightarrow \infty$. The latter limiting form is consistent with $Nu/\hat{Pe}^{1/2}$ approaching a finite plateau, for $\hat{\alpha}' \rightarrow \infty$, for all $\theta_\omega$\,(Figs.\ref{fig:Nu_inclined_ind}a-c and e) except $\theta_\omega = \theta_\omega^{th1}$ where it approaches zero\,(Fig.\ref{fig:Nu_inclined_ind}d).

\section{Conclusion} \label{sec:conc}

In this effort, we have developed a boundary layer analysis, based on a surface-streamline-aligned non-orthogonal coordinate system\,(the $C$ and $\tau$ coordinates), for determining the transport rate from a spherical drop in an arbitrary ambient linear flow, in the large-$Pe$ limit. The non-dimensional scalar transport rate, $Nu$, scales as $Pe^{\frac{1}{2}}$ in this limit, with the pre-factor in this relationship being a function of the flow-type parameters\,($\{ P \} \equiv [\epsilon,\hat{\alpha},\theta_\omega,\phi_\omega]$) and the viscosity ratio\,($\lambda$). The methodology was illustrated in detail by calculating the $Nu/Pe^{\frac{1}{2}}$-surfaces for a pair of two-parameter linear flow families: (i) Non-axisymmetric extensional flows with vorticity aligned along one of the principal axes\,(the $Nu/Pe^{\frac{1}{2}}$-surface in Fig.\ref{fig:Nu_aligned}), and (ii) Axisymmetric extensional flows with vorticity inclined to the symmetry axis of the extension\,(the $Nu/Pe^{\frac{1}{2}}$-surface in Fig.\ref{fig:Nu_inclined}). Taken together, these families encompass the entire range of surface-streamline topologies pertaining to a spherical drop in an incompressible linear flow. Note that the generic surface-streamline topology in the $(\{P\},\lambda)$-space is either of two types - a non-spiralling streamline pattern organized by diametrically opposite pairs of stable nodes, unstable nodes, and saddle points; and a spiralling pattern organized by a pair of diametrically opposite foci with a limit cycle in between. However, multiple degenerate patterns can occur at the point of transition from a spiralling to a non-spiralling topology. The analysis described here accounts for both of the aforementioned generic topologies with equal felicity, and also for the transitional patterns at the bifurcation points\,(in cases where such patterns yield a $Nu$ of $O(Pe^{\frac{1}{2}})$).

It was mentioned above that the surface-streamline topologies, encountered in our examination of the two-parameter families, are representative of a general linear flow. This is true for the inclined-vorticity family in particular, and may be seen from the fact that the equations governing the surface streamlines for a general linear flow have the same form as those for the above family\,(see (\ref{ode1A}) and (\ref{ode2A}) in section \ref{sec:C_tau_inclined}). One therefore expects the procedure to determine $C$ and $\tau$, outlined in section \ref{sec:C_tau_inclined}, and the associated $Nu$-calculation described in section \ref{Nu:inclinedvort}, to also work for a general linear flow. The ability to, in principle, determine the transport rate in an arbitrary ambient linear flow naturally leads to the question of characterizing scalar transport from a spherical sub-Kolmogorov drop in turbulence. One may envisage storing the Lagrangian sequence of velocity gradient tensors along the trajectory of a neutrally buoyant drop in a turbulent flow, obtained from a direct numerical simulation for instance, and then solving (\ref{HEQ1}), with turbulence appearing as a stochastically varying ambient linear flow in the drop reference frame. This has been done for a sub-Kolmogorov particle in a recent effort\,\citep{Lawson21}, and the results obtained for a spherical particle have been compared to Batchelor's analytical prediction\,\citep{Batchelor80}. However, as pointed out in \citet{Ganesh24}, unlike a rigid particle, the Kolmogorov and boundary layer time scales for a drop are comparable to each other, and this makes a quasi-steady approach, based on calculation of the transport rate in a time-averaged ambient linear flow\,(as in \citet{Batchelor80}), inapplicable. Nevertheless, such a boundary layer calculation, and its comparison to a fully numerical solution, would be a desirable first step towards understanding transport rates from neutrally buoyant drops in turbulence.

One may enquire about the robustness of the $Nu$-surfaces reported here to physical effects not included in our analysis. Examples of such effects are inertia\,(a finite drop Reynolds number $Re$), shape deformation\,(a finite Capillary number $Ca$), and drop or ambient fluid viscoelasticity\,(a finite Deborah number). At any point in the $(\{ P\},\lambda)$ parameter space that does not coincide with a bifurcation locus, one expects the effects of inertia and drop deformation to only lead to perturbative changes to $Nu$ for sufficiently large $Pe$. This is because the generic surface-streamline patterns are structurally stable, and will therefore remain qualitatively unchanged in presence of perturbations such as those due to inertia and deformation. This is not necessarily true for the degenerate streamline patterns along the bifurcation loci. Recall that such patterns include (i) those with meridional streamlines ending in a fixed-point ring\,(the second unit sphere in Fig.\ref{fig:discandstream}d), (ii) those with closed concentric streamlines\,(unit sphere with the Jeffery orbits in Fig.\ref{fig:discandstream}f) or (iii) closed eccentric streamlines\,(unit sphere with the generalized Jeffery orbits in Fig.\ref{fig:discandstream_inc}d), and (iv) the projected streamlines of a parabolic linear flow\,(the encircled unit sphere in Fig.\ref{fig:discandstream_inc}f). Cases (i) and (iv) are robust to sufficiently weak perturbations, but cases (ii) and (iii) are not. Earlier work has shown that even weak inertia\,\citep{Deepak18b} and/or drop deformation\,\citep{Sabarish24,Singeetham24} transform Jeffery orbits into spiralling streamlines, and for sufficiently large $Pe$ such that $Re Pe, Ca Pe \gg 1$, one expects convectively enhanced transport to be driven by spiralling streamlines, with $Nu \propto (Re Pe)^{\frac{1}{2}}, (Ca Pe)^{\frac{1}{2}}$; this is in contrast to the diffusion-limited transport that prevails for $Re = Ca = 0$. The exact result for $Nu$ in the inertial case has been calculated in \citet{Deepak18b}. When surface streamlines are generalized Jeffery orbits, as for members of the inclined-vorticity family along the locus $\theta_\omega = \frac{1}{2}\cos^{-1}\left[-\frac{16 + \hat{\alpha}^{'2}}{3\hat{\alpha}^{'2}}\right]$ with $ \hat{\alpha}' \in [\sqrt{12},\infty)$, the arguments in section \ref{sec:spcases_inclined} showed that $Nu \propto Pe^{\frac{1}{3}}$ for $Pe \gg 1$. Thus, $Nu/Pe^{\frac{1}{2}} \sim O(Pe^{-\frac{1}{6}})$ for such flows, and therefore approaches zero for $Pe \rightarrow \infty$, leading to a singular dip in the $Nu$-surface in Fig.\ref{fig:Nu_inclined}. Similar to the canonical case mentioned above, one expects the generalized Jeffery orbits to also open up, leading to spiralling surface streamlines for finite $Re$ and/or $Ca$, and as a result, $Nu$ is expected to change from a $Pe^{\frac{1}{3}}$ to a $Pe^{\frac{1}{2}}$-scaling. The singular effect of weak inertial effects may be summarized as follows: (a) for small but finite $Re$, the $Nu/Pe^{\frac{1}{2}}$ surface in Fig.\ref{fig:Nu_aligned} will not decrease to zero for $\epsilon \rightarrow 0$ and $\hat{\alpha}' > 2$, but instead saturate at a value of $O(Re^{\frac{1}{2}})$; (b) along analogous lines, the $Nu/Pe^{\frac{1}{2}}$ surface in Fig.\ref{fig:Nu_inclined} will saturate at an $O(Re^{\frac{1}{2}})$ value along the eccentric-elliptic-flow locus; provided $\lambda \sim O(1)$, $Ca$ replaces $Re$ in these estimates when drop deformation is the relevant factor.

Another very important finding, reported possibly for the first time here, is the emergence of a boundary-layer-mediated transport regime for the interior problem. The Langevin simulations in Appendix \ref{App:interior_topo} show that a spherical drop, in an ambient linear flow belonging to either the aligned or inclined-vorticity families, exhibits chaotically wandering interior streamlines; a scenario that is again representative of a general linear flow\,\citep{Stone91}. Further, for cases where the chaos is sufficiently space-filling\,(as judged from Poincare sections in Fig.\ref{fig:inclined_stream}), the scalar concentration profiles exhibit an $O(Pe_i^{-\frac{1}{2}})$ interior boundary layer for sufficiently large $Pe_i$, leading to the aforementioned enhanced transport, with $Nu \propto Pe_i^{1/2}$. The significance of this finding is better appreciated when one notes that virtually all efforts in literature, that examine scalar transport from neutrally buoyant spherical drops in ambient shearing flows, consider highly symmetrized scenarios that invariably lead to closed interior streamlines. As a result, diffusion-limited transport prevails at large $Pe_i$\,\citep{Favelukis13, Favelukis14, Favelukis15, Favelukis16, Favelukis19, Liu18, Liu19}, in turn implying an analogous limitation for the conjugate problem. Thus, for cases examined in the literature, the dominant resistance to transport will eventually shift to the drop phase for sufficiently large $Pe$, independent of the ratio of transport coefficients, with $Nu$ saturating in a $Pe$-independent plateau as a result. Our interior problem simulations imply that the scenario for a general linear flow is different. One expects chaotic interior streamlines to allow $Nu$ to grow as $Pe^{\frac{1}{2}}$ for $Pe \rightarrow \infty$, with the large-$Pe$ transport occurring across boundary layers both outside and immediately within the drop surface. 

On a final note, it is worth contrasting a spherical drop in a linear shearing flow, discussed above, with one in an ambient uniform flow. The interior streamlines in the former case are chaotic even in the Stokes limit, thereby emphasizing the kinematic rather than dynamic origin of the chaos. In contrast, symmetry dictates that surface streamlines in the latter case must always have a meridional character, with interior streamlines being plane closed curves, independent of the Reynolds number, The Hadamard-Rybzinski solution shows that the interior streamlines within a translating spherical drop, in the Stokes limit, are identical to those of the spherical Hill's vortex. An increase in Reynolds number leads, in effect, to a shift\,(away from the midplane) of the central ring within the vortex, with the streamlines still being closed. Thus, the only way in which a nontrivial streamline topology can arise is via a symmetry-breaking instability. Interestingly, this does happen, and the manner in which such an instability arises has been explored in detail in recent efforts\,\citep{Edelmann2017,Gode_2025}. In addition, there are limited results that show that the nontrivial streamline topology after instability onset does eliminate the original diffusion-limited scenario. However, there has been no attempt yet to identify a scaling exponent in the large-$Pe$ regime, and correlate it either with the emergence of an interior boundary layer, or the chaotic nature of streamlines.

\appendix

\section{Equivalence between the  $C-\tau$ definitions from a direct approach, and one based on an auxiliary linear flow framework} \label{auxiliaryflow:map}

The $C-\tau$ definitions, given by \eqref{auxsol1new1} and \eqref{auxsol2new1} for the aligned-vorticity family, were derived based on the auxiliary linear flow framework. Herein, we present an alternate derivation based on directly solving \eqref{ode1} and \eqref{ode2}, which is possible because the equation for $\phi$ is decoupled from that for $\theta$. In adopting this approach, we follow along the lines of \citet{Deepak18a} who first obtained the expressions for $C$ and $\tau$ for the one-parameter family of (canonical)\,planar linear flows.  

The solution of \eqref{ode2} is readily shown to be given by:
\begin{align}
    &\tan \phi = -\frac{A \tanh(\tau/2) +(2+\epsilon)}{\hat{\alpha}(1+\lambda)}, \label{AppA1}
\end{align}
where $\tau = At/(1+\lambda)$; the integration constant involved has been absorbed into $t$, and sets the origin of time. Using (\ref{AppA1}) in (\ref{ode1}), one then obtains:
\begin{align}
    &\tan \theta = C\hat{\alpha} \left[ \frac{(1+\tan^2 \phi)}{1 - (\frac{(2+\epsilon) + (1+\lambda)\hat{\alpha} \tan \phi}{A})^2}\right]^{1/2}\!\!\! \exp(\frac{-3\epsilon\tau}{2A}), \label{AppA2}
\end{align}
where $C$ is a constant of integration\,(after having absorbed a factor of $(1+\lambda)$). Equations (\ref{AppA1}) and (\ref{AppA2}) serve to define $C$ and $\tau$. As mentioned in section \ref{sec:C_tau_aligned}, $A$ is real in the non-spiraling regime, having been defined in the paragraph below (\ref{x3sol}). The relation (\ref{AppA1}) may be inverted to write:
\begin{align}
    &\tau = - 2\tanh^{-1}\left[ \frac{(1+\lambda)\hat{\alpha} \tan \phi + (2+\epsilon)}{A}\right] = \log \left(\frac{\hat{\gamma}_0 - \tan \phi}{\hat{\gamma}_1 + \tan \phi} \right), \label{AppA4}
\end{align}
where $\hat{\gamma}_0 = \frac{A - (2+\epsilon)}{\hat{\alpha}(1+\lambda)}$ and $\hat{\gamma}_1 = \frac{A + (2+\epsilon)}{\hat{\alpha}(1+\lambda)}$. The branch points of the logarithm in (\ref{AppA4}), obtained by setting the numerator and denominator being equated to zero, yield four fixed points in the $x_1-x_2$ plane, given by:
\begin{align}
    \phi^{(1)} =& - \tan^{-1} \hat{\gamma}_1, \pi -\tan^{-1}\hat{\gamma}_1, \\
    \phi^{(2)} =&  \tan^{-1} \hat{\gamma}_0, \pi +\tan^{-1} \hat{\gamma}_0.
\end{align}
It may be shown that $\hat{\gamma}_0 = \frac{A-B}{A+B}$ and $\hat{\gamma}_1 = -\frac{A+B}{A-B}$, so the expressions above are the same as those for the $\phi^{(i)}$\,($i = 1-4$) in Section \ref{sec:C_tau_aligned}. As mentioned therein, provided $A + 3\epsilon \neq 0$, meridional arcs that connect the fixed point at $\theta = 0$, to the four fixed points above, divide the unit hemisphere into four octants. For octants defined by the intervals $\phi^{(1)} < \phi < \phi^{(2)}$\,(Region 1) and $\phi^{(3)} < \phi < \phi^{(4)}$\,(Region 3), the argument of the logarithm in \eqref{AppA4} is positive, while in the remaining two octants, corresponding to Regions 2 and 4, it is negative. In these latter regions, one may rewrite $\tau$ as:
\begin{align}
    \tau =& \log \left( \frac{\tan \phi - \hat{\gamma}_0}{\tan \phi + \hat{\gamma}_1 } e^{\mathrm{i} \pi} \right), \\
    =&  \log \left( \frac{\tan \phi - \hat{\gamma}_0}{\tan \phi + \hat{\gamma}_1} \right) + \mathrm{i} \pi,
\end{align}
where $\frac{\tan \phi - \hat{\gamma}_0}{\tan \phi + \hat{\gamma}_1 }$ is positive. One now defines $\tau =\hat{\tau}+ \mathrm{i}\pi$, with $\hat{\tau}$ real. Along similar lines, the streamline label $C$ is real only in Regions 1 and 3, and one again defines $\hat{C} = iC$ in Regions 2 and 4 with $\hat{C}$ being real. Thus, Regions 1 and 3 on the unit hemisphere are mapped in terms of $(C,\tau)$, with Regions 2 and 4 being mapped in terms of $(\hat{C}, \hat{\tau}) \equiv (\mathrm{i}C,\tau -\mathrm{i}\pi)$; see Fig.\ref{fig:Ctau_aligned}a.

One may now demonstrate the equivalence between the $C-\tau$ definitions in \eqref{AppA1}-\eqref{AppA2}, obtained directly using spherical coordinates, and \eqref{auxsol1new1}-\eqref{auxsol2new1} which were derived based on the auxiliary linear flow framework. In what follows, we show this equivalence for the one of the solutions, the one relating $\tan \phi$ to $\tau$; the equivalence for the other solution (relating $\tan \theta$ to $C,\tau$) can be established along analogous lines. We begin by writing the auxiliary flow solution for $\tan \phi$ from \eqref{x1sol}-\eqref{x2sol} as:
\begin{align}
    \tan \phi = \frac{[A - (2+\epsilon) + e^{\frac{At}{1+\lambda}}(A + (2+\epsilon))]\frac{x_2^0}{x_1^0} + (e^{\frac{At}{1+\lambda}} - 1)(1+\lambda)\hat{\alpha}}{[A + (2+\epsilon) + e^{\frac{At}{1+\lambda}}(A - (2+\epsilon))] - (e^{\frac{At}{1+\lambda}} - 1)(1+\lambda)\hat{\alpha}\frac{x_2^0}{x_1^0}} \label{AppA11}
\end{align}
Next, based on \eqref{AppA1}, we use $x_2^0/x_1^0 = \tan \phi_0 = -\frac{A \tanh (-\tau_0/2) + (2+\epsilon)}{\hat{\alpha}(1+\lambda)} = -\frac{A(e^{-\tau_0} - 1)+ (2+\epsilon)(e^{-\tau_0} + 1)}{(1+\lambda)\hat{\alpha}(e^{-\tau_0} + 1)}$) in \eqref{AppA11}. After simplifications,one obtains:
\begin{align}
\tan \phi = &-\frac{A(e^{\tau-\tau_0} - 1) + (2+\epsilon) (e^{\tau-\tau_0}+1)}{(1+\lambda)\hat{\alpha}(1+e^{\tau-\tau_0})} \\
& = -\frac{A\tanh(\frac{\tau-\tau_0}{2}) + (2+\epsilon)}{\hat{\alpha}(1+\lambda)},
\end{align}
which is the same expression as \eqref{AppA1}, but for the explicit presence of the `initial' time $t_0$\,(defined by $\tau_0 = -At_0/(1+\lambda)$).

To establish the arbitrariness in the choice of $x_2^0/x_1^0$ as far as the surface streamlines are concerned, one starts from \eqref{AppA11}. Taking the limit $\tau\to-\infty$, this becomes
\begin{align*}
    \tan{\phi} &= \frac{\left[A - (2+\epsilon)\right]\frac{x_2^0}{x_1^0}- (1+\lambda)\hat{\alpha}}{\left[A + (2+\epsilon) \right]+ (1+\lambda)\hat{\alpha}\frac{x_2^0}{x_1^0}}, \\
    &= \frac{\left[(\Lambda^2-1)^{\frac{1}{2}} - \Lambda\right]\frac{x_2^0}{x_1^0} - 1}{\left[(\Lambda^2-1)^{\frac{1}{2}} + \Lambda \right]+ \frac{x_2^0}{x_1^0}}, \\
    &= \left[(\Lambda^2-1)^{\frac{1}{2}} - \Lambda\right] \frac{\frac{x_2^0}{x_1^0} - \frac{1}{\left[(\Lambda^2-1)^{\frac{1}{2}} - \Lambda\right]}}{\left[(\Lambda^2-1)^{\frac{1}{2}} + \Lambda \right]+ \frac{x_2^0}{x_1^0}},
\end{align*}
where $\Lambda = \frac{2+\epsilon}{\hat{\alpha}(1+\lambda)}$, as in the main manuscript. On using $\left[(\Lambda^2-1)^{\frac{1}{2}} - \Lambda\right] \left[(\Lambda^2-1)^{\frac{1}{2}} + \Lambda\right] = -1$, the above expression takes the form:
\begin{align*}
    \tan\phi
    &= \left[(\Lambda^2-1)^{\frac{1}{2}} - \Lambda\right] \frac{\frac{x_2^0}{x_1^0} + {\left[(\Lambda^2-1)^{\frac{1}{2}} + \Lambda\right]}}{\left[(\Lambda^2-1)^{\frac{1}{2}} + \Lambda \right]+ \frac{x_2^0}{x_1^0}}, \\
    &= \left[(\Lambda^2-1)^{\frac{1}{2}} - \Lambda\right],
\end{align*}
whose solutions, $\tan^{-1} \left[(\Lambda^2-1)^{\frac{1}{2}} - \Lambda\right]$ and $\pi + \tan^{-1}\left[(\Lambda^2-1)^{\frac{1}{2}} - \Lambda\right]$, correspond to the fixed points $\phi^{(2)}$ and $\phi^{(4)}$. Similarly, for $\tau \to \infty$, one finds $\tan\phi = - \left[(\Lambda^2-1)^{\frac{1}{2}} + \Lambda\right]$,
whose solutions correspond to $\phi^{(1)}$ and $\phi^{(3)}$. The above establishes independence of the fixed-points - the limiting points of all surface streamlines - with respect to $x_2^0/x_1^0$. 

The surface streamlines themselves depend on $x_2^0/x_1^0$, only to the extent of $x_2^0/x_1^0$ determining the particular octant that they belong to\,(regions $1$ and $3$, as opposed to regions $2$ and $4$). To see this, note that inverting (\ref{AppA1}) leads to an expression for $\tau_0$ as a function of $\Lambda$, of the form $\tau_0 = -2 \tanh^{-1}\left[\frac{\Lambda + (x_2^0/x_1^0)}{(\Lambda^2-1)^{\frac{1}{2}}}\right]$. The critical values of $x_2^0/x_1^0$, that mark the transition from one octant to the other, are dictated by the singularities of $\tau_0$. From the above expression, these are $-\Lambda \pm (\Lambda^2-1)^{\frac{1}{2}}$. Further examination shows that, for any $\frac{x^0_2}{x^0_1}$ in the interval $(-\Lambda-(\Lambda^2 - 1)^{1/2}, -\Lambda+(\Lambda^2 -1)^{1/2})$, the auxiliary flow solution yields a streamline in regions 1 and 3, while choosing $\frac{x^0_2}{x^0_1}$ in the intervals $(-\infty, -\Lambda -(\Lambda^2 - 1)^{1/2}), (-\Lambda+(\Lambda^2 -1)^{1/2},\infty)$ leads to streamlines in regions 2 and 4.

\section{Streamline topology and $Nu-Pe$ relationships for the interior problem} \label{App:interior_topo}

Herein, we characterize the flow within a spherical drop immersed in either of the two linear flow families considered here, and then determine the scalar transport rate numerically using Langevin simulations. This is done in the limit that the ambient fluid resistance to transport is negligibly small, the so-called interior problem, so the results serve as a complement to the large-$Pe$ analysis of the exterior problem in the main manuscript. For almost all drop-in-linear-flow cases examined in the literature, from the scalar transport perspective, interior streamlines turn out to be closed curves\citep{Brink50,Chung85,Dewitt93,Leal07,Gupalo72,Gupalo75,Polyanin84,Deepak18a,Singeetham24}. The $Nu-Pe_i$ relationships for these idealized cases therefore conform to the second of two types mentioned in section \ref{1} of the main manuscript - $Nu$ saturates in a diffusion limited plateau for large $Pe_i$, $Pe_i$ here being the interior Péclet number. As a result, even in cases where the exterior streamlines are open, allowing for boundary-layer enhanced transport at large $Pe$, $Nu$ for the conjugate problem will eventually be limited by diffusion-limited transport across closed interior streamlines, as has been explicitly demonstrated for the case of an ambient uniform flow\citep{Rachih_2019}. Closed streamlines imply that the interior scalar field continues to vary over an $O(a)$ length scale even for $Pe_i \gg 1$. Accounting for the $O(a)$ and $O(a Pe^{-\frac{1}{2}})$ scales characterizing the interior and exterior scalar fields, respectively, for $Pe, Pe_i \gg 1$, a comparison of the respective fluxes at the drop surface shows that, for the $Nu \propto Pe^{\frac{1}{2}}$ asymptote obtained via the boundary layer analysis in section \ref{sec:Nusselt_calc} to be valid, the ratio of the drop to ambient fluid diffusivities\,(conductivities) needs to be much larger than $O(Pe^{\frac{1}{2}})$ for $Pe \gg 1$. This is the condition that ensures that scalar variations within the drop are negligible in comparison to those across the exterior boundary layer, allowing for a uniform surface boundary condition in the boundary layer analysis, and is evidently very restrictive. 

Interestingly, the above restriction does not apply for the two-parameter families examined here. Interior streamlines in these cases will be shown to have a chaotically wandering nature below. This, by itself, shouldn't be surprising since interior streamlines for a generic linear flow are expected to be chaotic\citep{Stone91,Ganesh24}. Importantly, the chaos allows for a continued growth of $Nu$ with $Pe$ for the interior problem. The maximally chaotic cases lead to $Nu$ scaling as $Pe^{\frac{1}{2}}$ for $Pe \gg 1$, owing to the emergence of an interior boundary layer with a nearly uniformly mixed bulk. For these cases, one can extend the boundary layer analysis for the exterior problem, to the conjugate problem, by allowing for transport across both interior and exterior boundary layers in the manner discussed in \citet{Deepak18a}. This is a significant departure from the diffusion-limited scenario for the conjugated problem mentioned above, and that prevails for uniform and canonical linear flows.

\subsection{$Nu$ for a drop in $3D$ complex shearing flows} \label{sec:PoincareSection}

Before describing the $Nu-Pe_i$ relationship for a drop in the two-parameter linear flow families, we characterize the interior streamline topology in these cases by means of Poincare sections. The latter were obtained by plotting the points of intersection of streamlines with a chosen plane, the streamlines being obtained from numerically integrating the interior velocity field given by:
\begin{align}
&\hat{\bm{u}} = \bm{\Omega}\cdot \bm{r} + \frac{1}{2(1+\lambda)}(5r^2-3) \bm{E} \cdot \bm{r}  - \frac{1}{(1+\lambda)} (\bm{E}:\bm{rr})\bm{r}, \label{Sol2}
\end{align}
with $\bm{\Omega}$ and $\bm{E}$ given by (\ref{alignedvort_Gamma}) or (\ref{inclinedvort_Gamma}), depending on the linear flow being examined. The integration is done using a fourth order Runge-Kutta scheme, with a fixed time step $dt = 10^{-4}$, for a duration of $10^6$\,(in units of $\dot{\gamma}^{-1}$). 

\begin{figure}
    \centering
    \includegraphics[scale=0.17]{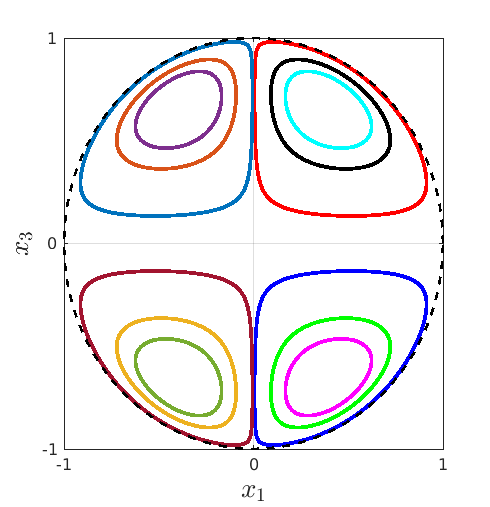}
    \includegraphics[scale=0.17]{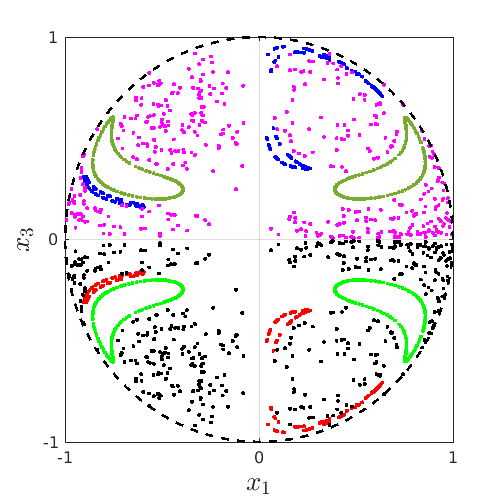}
    \includegraphics[scale=0.17]{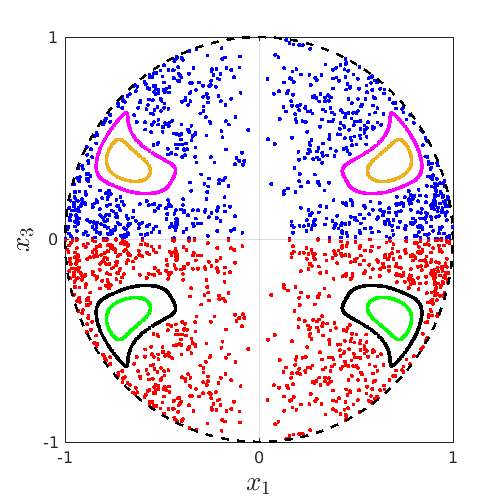}
    \includegraphics[scale=0.17]{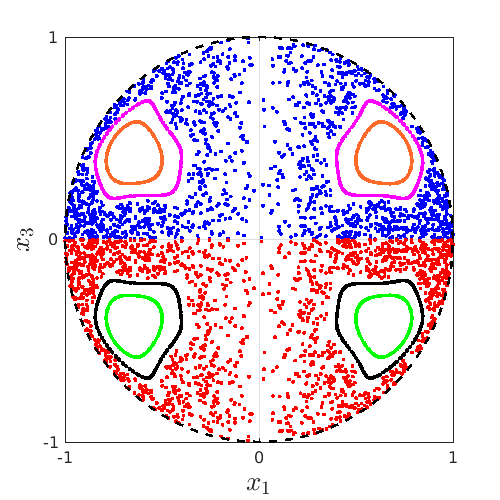}
    \includegraphics[scale=0.17]{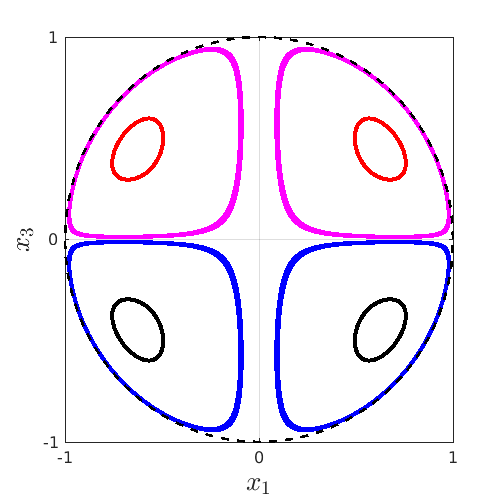}
    \caption[\textwidth]{\justifying{Poincaré sections for a spherical drop in members of the aligned-vorticity family; $\epsilon = -1/4$, $\lambda = 1$. (a) $\hat{\alpha} = 0$, (b) $\hat{\alpha} = 0.5$, (c) $\hat{\alpha} = 1$, (d) $\hat{\alpha} = 2$ and (e) $\hat{\alpha} = 10$. The surface chosen for constructing the sections is the plane $x_2 = 0$, with different colors corresponding to points of intersection of streamlines starting from different initial points. The degree of chaos exhibits a non-monotonic dependence on $\hat{\alpha}$.}}
    \label{fig:aligned_stream}
\end{figure}

Fig.\ref{fig:aligned_stream} plots Poincare sections in the plane $x_2 = 0$, as a function of $\hat{\alpha}$, for members of the aligned-vorticity family; $\epsilon = -1/4, \lambda = 1$. The Poincare section for $\hat{\alpha} = 0$\,(a $3D$ extensional flow) consists entirely of densely filled curves, and corresponds to the regular case where almost all interior streamlines are closed curves. For non-zero $\hat{\alpha}$, a single streamline leads to a scatter of points spread across a finite region, pointing to its chaotic nature; although, the chaos is not space-filling, with chaotic regions always being interspersed with regular islands. The chaotic regions increase in extent to begin with\,(Figs.\ref{fig:aligned_stream}b-c), before shrinking again for sufficiently large $\hat{\alpha}$\,(dominant vorticity) as evident from Figs.\ref{fig:aligned_stream}d and e. The approach to regularity for $\hat{\alpha} \rightarrow \infty$ in Fig.\ref{fig:aligned_stream}e is consistent with the fact that there can be no chaos in solid-body rotation. Note that the plane $x_3 = 0$ is an invariant surface for all values of $\hat{\alpha}$ owing to the alignment of the vorticity vector with one of the principal axes, and therefore, despite the chaos, a fluid particle in the positive-$x_3$ hemisphere never crosses into the negative-$x_3$ one. We have verified the presence of chaotic streamlines for values of $\epsilon$ other than $-1/4$; the exception is $\epsilon = -2$, corresponding to axisymmetric extension with aligned vorticity where the drop is known to foliated by nested invariant tori\,(see below). 
\begin{figure}
    \centering
    \includegraphics[scale=0.22]{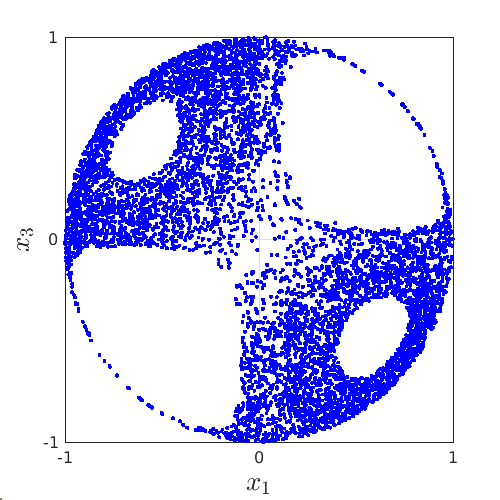}
    \includegraphics[scale=0.22]{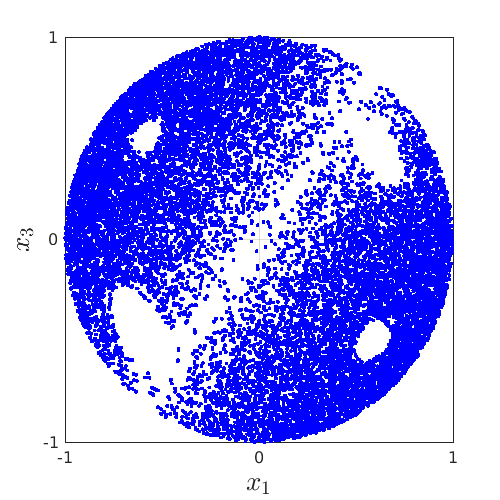}
    \includegraphics[scale=0.22]{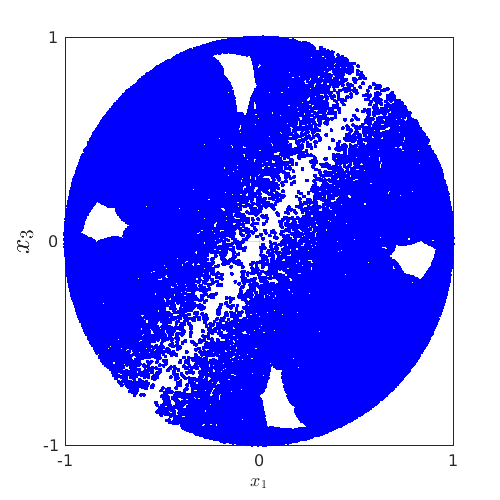}
	  \includegraphics[scale=0.22]{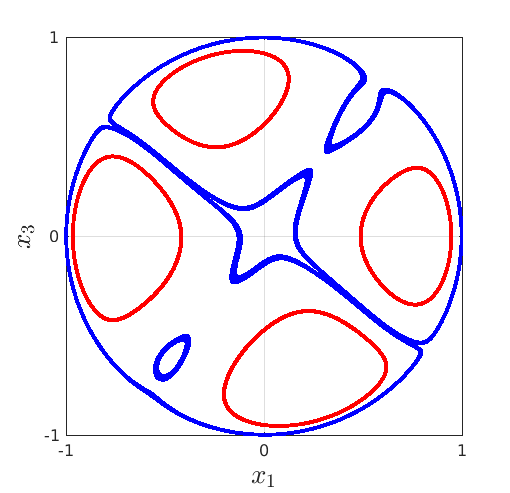}	
    \caption[\textwidth]{\justifying{Poincaré sections for a spherical drop in members of the inclined-vorticity family for $\theta_\omega = 36^\circ$, $\lambda = 1$; (a) $\hat{\alpha} = 0.4$, (b) $\hat{\alpha} = 1$, $\hat{\alpha} = 4$ and $\hat{\alpha} = 16$; the points of intersection in (a-c) correspond to a single chaotic streamline. The section for $\hat{\alpha} = 4$ is almost space-filling.}}
    \label{fig:inclined_stream}
\end{figure}

From the work of \citet{Stone91}, a drop immersed in members of the second linear flow family\,(axisymmetric extensions with inclined vorticity) is known to exhibit chaotic interior streamlines. 
This is confirmed by our calculations, depicted in Figs.\ref{fig:inclined_stream}a-d, which show the sections for $\theta_\omega = 36^\circ$, for varying $\hat{\alpha}$. 
 The Poincare sections in Figs.\ref{fig:inclined_stream}a-c are generated by a single chaotically wandering streamline since, unlike the previous case, there are no invariant planes for $\theta_\omega \neq 0$, and a single streamline can potentially wander over the entire drop interior. The extent of chaos is thus greater, but continues to exhibit a non-monotonic dependence on $\hat{\alpha}$, with chaotic regions in general being interspersed with non-chaotic islands. Although not shown, a similar pattern is observed for a fixed $\hat{\alpha}$, but with varying $\theta_\omega$; the non-monotonic dependence arises here since one limit\,($\theta_\omega = 0$) is a regular scenario, while the other limit\,($\theta_\omega = \frac{\pi}{2}$) corresponds to an aligned vorticity vector, and therefore, to a restricted-chaos scenario\,(as in Fig.\ref{fig:aligned_stream}). Importantly, unlike Fig.\ref{fig:aligned_stream}, there exist parameter values for the inclined-vorticity family where the Poincare sections are almost space-filling. It will be shown in the next subsection that the parameter values corresponding to space-filling Poincare sections lead to $Nu$ growing as $Pe_i^{\frac{1}{2}}$ for sufficiently large $Pe_i$. 
 
The Lagrangian chaos in Figs.\ref{fig:aligned_stream} and \ref{fig:inclined_stream} can be understood from the usual viewpoint of a baseline integrable scenario, with a structurally unstable trajectory configuration, being subject to a non-integrable perturbation\,\citep{wiggins88}. The integrable limit in the present case is an axisymmetric extension with aligned vorticity, which leads to the spherical drop being foliated by invariant tori. The limiting torus consists of the drop surface connected to its center\,(a saddle point) via an equatorial plane of heteroclinic connections. The non-integrable perturbation may be regarded as a superposed planar extension\,(in the plane perpendicular to ${\bm \omega}$) for the aligned-vorticity family, or a deviation of the vorticity vector from alignment for the second family. For small but finite $\hat{\alpha}$, corresponding to small amplitudes  of this perturbation, one can analytically characterize the extent of chaos using the method of averaging\,\citep{Stone91}, and thereby identify the $\theta_\omega$-interval in this limit, corresponding to space-filling Poincare sections, and where $Nu$ scales as $Pe_i^{\frac{1}{2}}$ owing to an interior boundary layer. Herein, our focus is on establishing the fact that the emergence of chaos can indeed lead to such a boundary layer; the detailed simulation program identifying the region in the $\theta_\omega-\hat{\alpha}$ plane, corresponding to $Nu \propto Pe_i^{\frac{1}{2}}$ for $Pe_i \gg 1$, will be reported in a separate study. 

\subsection{Langevin simulations} \label{sec:Langevin}

In this section, we determine the scalar transport rate for the flow fields characterized in section \ref{sec:PoincareSection},  using Langevin simulations. 
Details with regard to these simulations have been laid out in earlier efforts\,\citep{Singeetham24,Ganesh24}, and we will be brief here. The governing equation is, of course, still the convection–diffusion equation for the scalar field $\Theta({\bm x},t)$, (\ref{HEQ1}), with convection now being due to the interior velocity field defined by (\ref{Sol2}), and with $Pe_i$ taking the place of $Pe$. The normalized scalar field satisfies (i) $\Theta = 1$ within the drop at $t = 0$, and (ii) $\Theta = 0$ $\forall\, t \geq 0$ at the drop surface\,($r = 1$), 
this being the absorption boundary condition arising from neglect of the ambient phase transport resistance. Furthermore, one requires that $\Theta$ be finite at all points in the drop interior, including singular points of the coordinate system used; this is evidently satisfied within a numerical framework. The simulations involve integrating the Langevin equations governing the motion of individual tracers:
\begin{align}
    d \bm{r} = Pe_i \hat{\bm{u}}(\bm{r}) dt + \sqrt{2} d{\bm W}(t),
\end{align}
where $d{\bm W}(t)$ is the standard Wiener process, satisfying $<d{\bm W}(t)> = 0$, $<d{\bm W}(t) \cdot d{\bm W}(t)> = dt$. For a sufficiently large initial number of tracers\,($N$, say), the ensemble average of the tracer concentration converges to the solution of (\ref{HEQ1}) with an error of $O(N^{-\frac{1}{2}})$. In all our runs, we start from a uniform distribution of $10^6$ tracers, with the aforementioned absorbing boundary condition implemented at each time step by removing tracers that end up crossing the drop surface. The resulting decrease in the tracer concentration leads to an inherently unsteady scenario for the interior problem, with $Nu$ being determined from the long-time quasisteady state as $Nu = -\frac{2}{3}\frac{1}{\Theta}\frac{d \Theta}{d t} = -\frac{2}{3} \frac{d \log \Theta}{d t}$. In this quasisteady state, the scalar field decreases in time as a simple exponential, and $Nu$ is therefore independent of time\,\citep{Sabarish21,Singeetham24}. Thus, a plot of $Nu$ vs $t$ decreases as $t^{-\frac{1}{2}}$ to begin with, due to the initial transport occurring across a diffusive boundary layer, and eventually fluctuates about a $Pe$-dependent plateau corresponding to the quasisteady state above. All of the $Nu$-values that appear below, in the context of the $Nu–Pe_i$ relationships, are obtained by taking at least decade-long time average in the plateau regime. 
 
\begin{figure}
    \centering
    \includegraphics[scale=0.3]{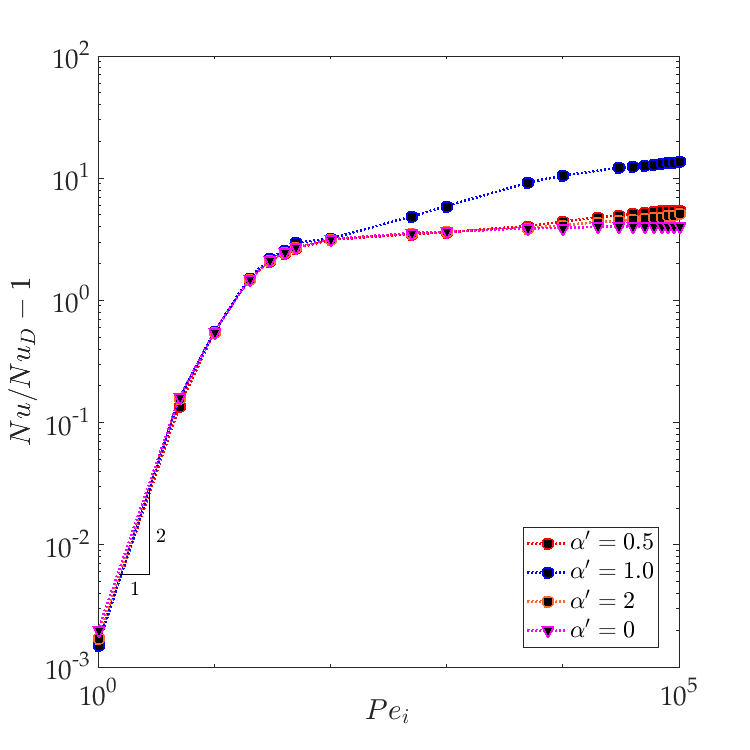}
    \includegraphics[scale=0.3]{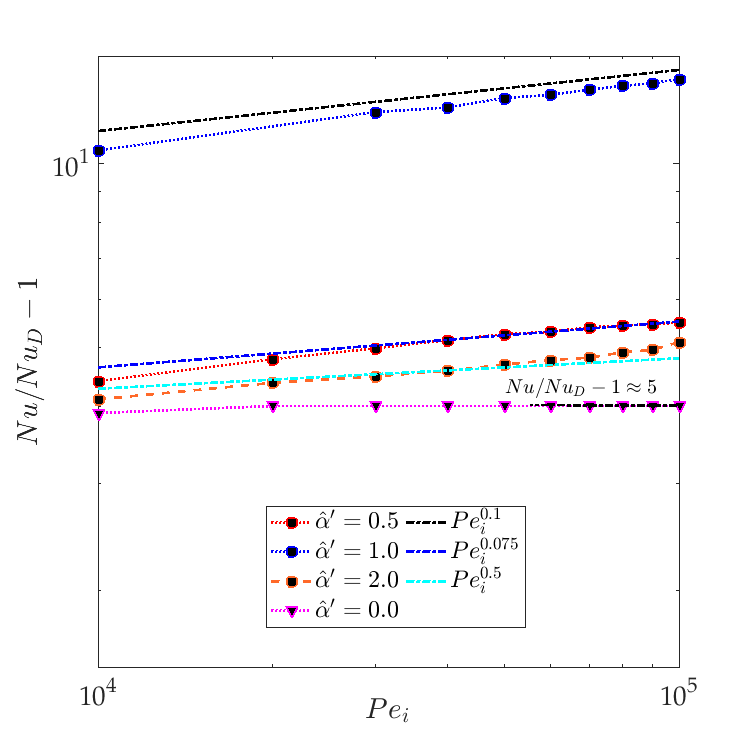}
    \caption[\textwidth]{(a) \justifying{$Nu/Nu_{\text{D}} - 1$ vs $Pe_i$ for different ambient flows belonging to the aligned-vorticity family, with $\epsilon = -0.1/4$, $\lambda = 1$\,(note the initial $Pe_i^2$-scaling). Plot (b) shows the large-$Pe_i$ portion of the $Nu$-curves, and emphasizes the continued growth of $Nu$ in the limit $Pe_i \rightarrow \infty$; the growth exponents are not standard values, and are a function $\hat{\alpha}$.}}
    \label{fig:aligned_Nu_interior}
\end{figure}

Fig.\ref{fig:aligned_Nu_interior} plots $Nu - Nu_\text{D}$ as a function of $Pe_i$ for a spherical drop in 3D extensional flows with aligned vorticity. $Nu_D = 2\pi^2/3$ here is the Nusselt number for pure diffusion\,($Pe = 0$), and the values of the flow-type parameters chosen in Fig.\ref{fig:aligned_Nu_interior} correspond to the Poincare sections in Fig.\ref{fig:aligned_stream}. In all cases, $Nu - Nu_D \sim O(Pe_i^2)$ for $Pe_i \ll 1$ owing to the regular character of the first convective correction, combined with the constraint of flow-reversal invariance; this is in contrast to the exterior problem where the first correction turns out to be $O(|Pe|)$\citep{Ganesh24}. For the integrable case, $\hat{\alpha} = 0$, $Nu$ saturates in a diffusion-limited plateau for $Pe_i \gg 1$ on account of the closed streamlines in the corresponding Poincare section\,(Fig.\ref{fig:aligned_stream}a). In contrast, for $\hat{\alpha} \neq 0$, $Nu$ exhibits a continued increase with $Pe_i$. The magnified view shows that the growth is algebraic at the largest $Pe_i$, with the growth exponent being a function of $\hat{\alpha}$, and its value correlating with the size of the chaotic regions in Fig.\ref{fig:aligned_stream}. Thus, the largest exponent occurs for $\hat{\alpha} = 1$, corresponding to the most chaotic Poincare section in Fig.\ref{fig:aligned_stream}c. All of the exponents differ from, and are smaller than, $1/2$, implying that the large-$Pe_i$ transport scenario is intermediate between the boundary-layer-enhanced and diffusion-limited paradigms. 

\begin{figure}
    \centering
    \includegraphics[scale=0.3]{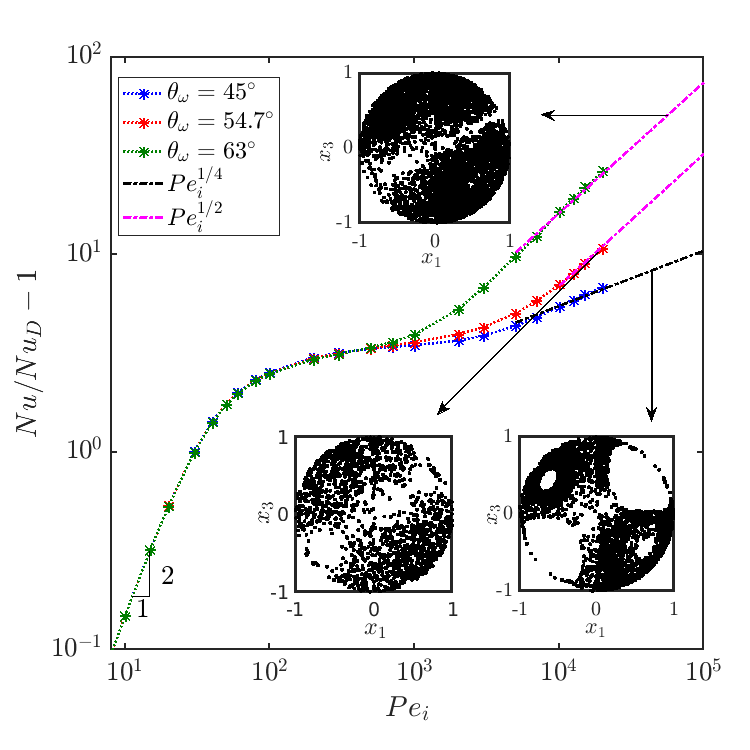}
    \includegraphics[scale=0.3]{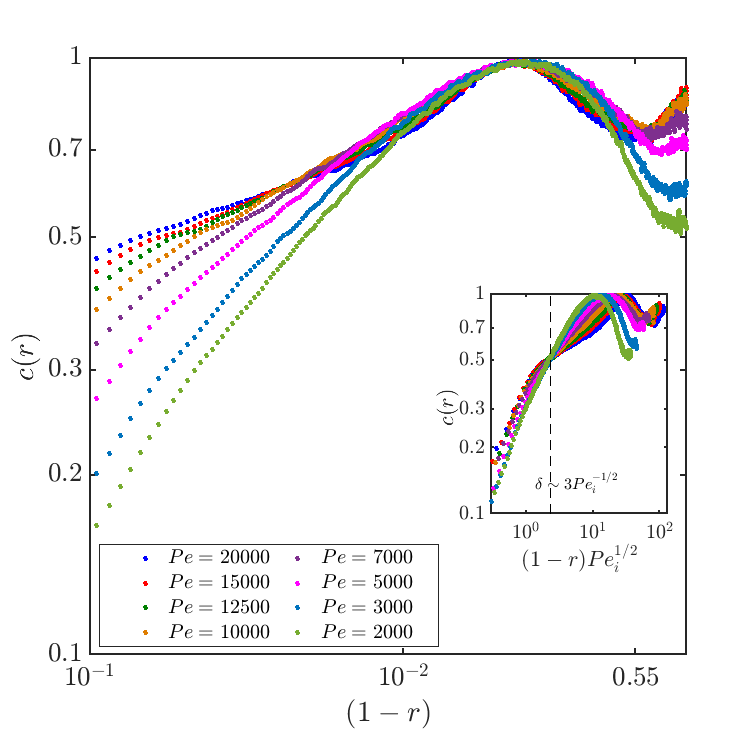}  
    \caption[\textwidth]{\justifying{$Nu/Nu_\text{D} - 1$ vs $Pe_i$ for different ambient flows belonging to the inclined-vorticity family, with $\hat{\alpha} = 0.1$, $\lambda = 1$\,(note the initial $Pe_i^2$-scaling). The insets contrast the space-filling nature of the Poincare sections for $\theta_\omega = 45^\circ$, $54.7^\circ$, and $63^\circ$; $Nu \sim Pe_i^{1/2}$, for $Pe_i \gg 1$, in the latter two cases. (b) Angle-averaged scalar concentration profiles in the quasi steady state, for $\theta_\omega = 63^\circ, \lambda = 1, \hat{\alpha} = 0.1$, as a function of distance from the drop surface; inset shows the same profiles as a function of the scaled boundary layer coordinate $(1 - r)Pe_i^{1/2}$.}}
    \label{fig:inclined_Nu_interior}
\end{figure}

A largely similar trend to Fig.\ref{fig:inclined_Nu_interior} is observed for the plots of $Nu-Nu_\text{D}$ vs $Pe_i$, for the inclined vorticity family. Here again, the large-$Pe_i$ exponents are in general smaller than $1/2$ - this is true, for instance, for the Poincare sections in Fig.\ref{fig:inclined_stream}. However, in contrast to the aligned-vorticity family, there exist parameters when the Poincare sections become completely space-filling\,(given that chaos and regularity are known to co-exist on an infinitely nested hierarchy of scales, the space-filling character here refers to scales of order the drop size), and in which cases $Nu$ exhibits a $Pe_i^{\frac{1}{2}}$-scaling for $Pe_i \gg 1$, in turn implying a boundary-layer-mediated transport similar to the open-streamline exterior problem. An example of this scaling behavior is seen in Fig.\ref{fig:inclined_Nu_interior}a for $\theta_\omega= 63^\circ$, $\hat{\alpha} = 0.1$, where $Nu-Nu_\text{D} \approx Nu \propto Pe_i^{\frac{1}{2}}$ for $Pe_i \gtrsim 10^3$. For $\theta_\omega= 54.7^\circ$, $Nu-Nu_\text{D}$ approaches a $Pe_i^{\frac{1}{2}}$-scaling regime, but for $Pe_i$ closer to $O(10^4)$, this again being consistent with a space-filling Poincare section. In both these cases, one can also establish the emergence of an internal boundary layer by plotting the scalar concentration as a function of $r$, for times long enough to correspond to the aforementioned quasi-steady state. Fig.\ref{fig:inclined_Nu_interior}b plots the long-time angle-averaged concentration profiles for $\theta_\omega = 63^\circ, \hat{\alpha} = 0.1$, as a function of $1-r$, for different $Pe_i$. These plots show the gradual approach of the bulk to uniformity\,(corresponding to a fully mixed drop interior), with increasing $Pe_i$, and the concomitant reduction in the thickness of the near-surface layer where the scalar concentration decreases from the bulk value to zero. The inset shows the collapse of the portions of the concentration profiles near $r = 1$, when plotted as a function of the rescaled radial coordinate $(1 - r)Pe_i^{\frac{1}{2}}$, confirming the existence of an internal boundary layer with a thickness of around $3 Pe_i^{-\frac{1}{2}}$.

A preliminary examination of the Poincaré sections for a general linear flow shows that the interior streamlines continue to have a chaotically wandering character. Fig.\ref{fig:general_stream} below, which depicts these sections for an ambient linear flow with $\epsilon = -0.25$, $\theta_\omega = 36^{\circ}$, $\phi_\omega = 10^{\circ}$, and with varying $\hat{\alpha}$, shows that the chaos is, in fact, slightly more space-filling than that in Fig.\ref{fig:inclined_stream} where the extensional component had an axis of symmetry. The implication is that $Nu$ for the interior problem is expected to scale as $Pe_i^{\frac{1}{2}}$, for sufficiently large $Pe_i$, in most of the 4D parameter space characterizing general linear flows\citep{Ganesh24}.
\begin{figure}
    \centering
    \includegraphics[scale=0.17]{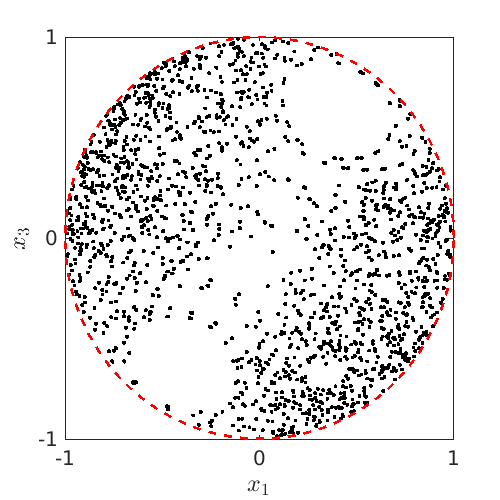}
    \includegraphics[scale=0.17]{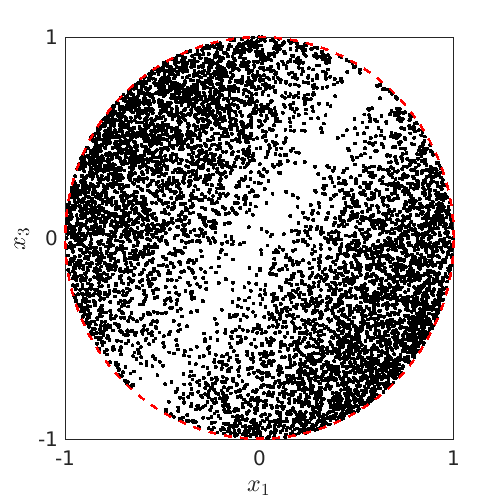}
    \includegraphics[scale=0.17]{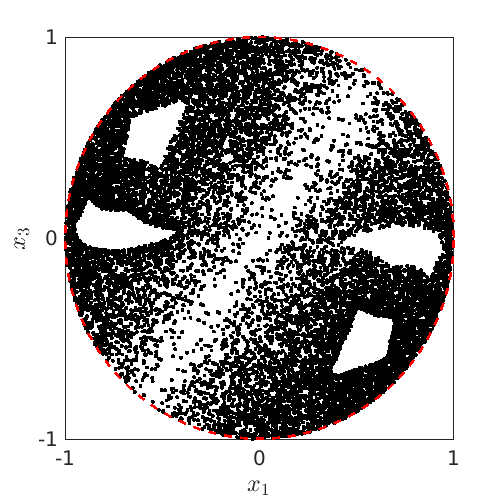}
	  \includegraphics[scale=0.17]{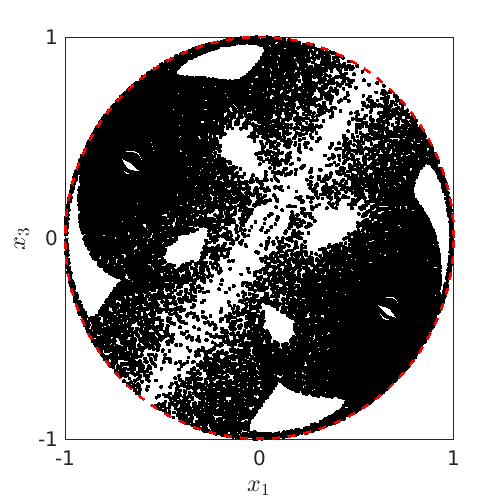}	
      \includegraphics[scale=0.17]{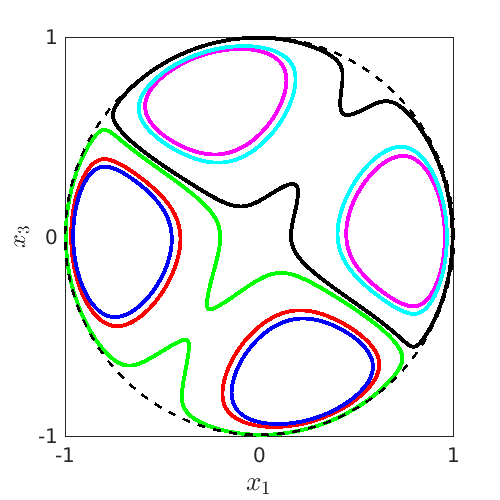}	
    \caption[\textwidth]{\justifying{Poincaré sections for a spherical drop in a general linear flow family member with $\epsilon = -0.25$, $\theta_\omega = 36^{\circ}$, $\phi_\omega = 10^{\circ}$ (a) $\hat{\alpha} = 0.1$, (b) $\hat{\alpha} = 0.5$, $\hat{\alpha} = 1$, $\hat{\alpha} = 2$ and $\hat{\alpha} = 8$; the points of intersection in (a-d) correspond to a single chaotic streamline. The section for $\hat{\alpha} = 0.5$ is almost space-filling.}}
    \label{fig:general_stream}
\end{figure}

\section{The $Nu$ asymptotes for axisymmetric extensional flows\,($\epsilon = -2$) with inclined vorticity}
\subsection{The $Nu$ asymptote for $\hat{\alpha}' \ll 1$} \label{SmallNu_asy}
In the limit of small $\hat{\alpha}'$, we calculate $Nu$ via an expansion in powers of $\hat{\alpha}$, with the leading order term, corresponding to an axisymmetric (biaxial)\,extension ($\hat{\alpha} = 0$), having already been calculated by \cite{Gupalo72}, as mentioned in the main manuscript. To calculate the higher-order terms, one needs to solve an inhomogeneous convection-diffusion equation\,(within the framework of a boundary layer\,(BL) approximation), which requires the Green's function of the pertinent operator. While the latter is not available for an arbitrary linear flow, \cite{Goddard65}, in the context of finding higher-order corrections to $Nu$, for large but finite $Pe$, obtained the Green’s function of the aforementioned operator for a general axisymmetric flow. This Green's function can be used in the present context, the difference being that the forcing functions that lead to the higher order corrections, in our case, arise due to a departure from axisymmetry associated with additional vortical contributions\,(rather than from BL contributions of a smaller order, as in \citet{Goddard65}).

We begin with the convection-diffusion equation, in the BL approximation, written as:
\begin{align}
    P \Theta = 0
\end{align}
where $P$ is the differential operator given by:
\begin{align}
    u_y \frac{\partial }{\partial y} + u_\theta \frac{\partial }{\partial \theta} + \frac{u_\phi}{\sin \theta} \frac{\partial }{\partial \phi} - \frac{1}{Pe}\frac{\partial^2 }{\partial y^2} \label{eqP},
\end{align}
in spherical coordinates, with the polar axis along the axis of symmetry of the ambient extension; $y = r-1$ is the radial distance measured from the drop surface. One now expands the quantities involved as a power series in $\hat{\alpha}$, leading to the following expansions for the differential operator and the BL scalar field:
\begin{align}
    &P = P_0 + \hat{\alpha} P_1 + \hat{\alpha}^2 P_2 + O(\hat{\alpha}^3), \label{exp:P} \\
    &\Theta = \Theta_0 + \hat{\alpha} \Theta_1 + \hat{\alpha}^2 \Theta_2 + O(\hat{\alpha}^3),
\end{align}
which, on substitution, lead to the following hierarchy of equations:
\begin{align}
    O(1): \quad &P_0 \Theta_0 = 0, \label{eq0}\\
    O(\hat{\alpha}): \quad &P_0\Theta_1 + P_1 \Theta_0 = 0, \label{eqalp}\\
    O(\hat{\alpha}^2): \quad &P_0\Theta_2 + P_1 \Theta_1 + P_2 \Theta_0 = 0. \label{eqalpsq}\\
    \cdots
\end{align}
Using (\ref{Sol1}), the velocity field that appears in $P$ is seen to be of the form,
\begin{align}
    u_y = u_y^{(0)} + \hat{\alpha}u_y^{(1)},\label{vel1}\\
    u_\theta = u_\theta^{(0)} + \hat{\alpha}u_\theta^{(1)}, \label{vel2}\\
    u_\phi = u_\phi^{(0)} + \hat{\alpha}u_\phi^{(1)}, \label{vel3}
\end{align}
where the $O(1)$ terms denote the velocity components induced by an ambient axisymmetric extension, and are given by:
\begin{align}
    &u_y^{(0)} = -\frac{1}{\sin \theta}\frac{\partial \psi_s}{\partial \theta}, \\
    &u_\theta^{(0)} = \frac{1}{r\sin \theta}\frac{\partial \psi_s}{\partial r},
\end{align}
in terms of a Stokes streamfunction $\psi_s$, defined as:
\begin{align}
    \psi_s = \left( \frac{(2+5\lambda)}{2(1+\lambda)} - \frac{3 \lambda}{2 r^2 (1+\lambda)} - r^3\right) \cos \theta \sin^2 \theta.
\end{align}
Using the boundary layer approximation, $\psi_s$ simplifies to:
\begin{align}
    \psi_s = \psi_s^{(0)} y + \psi_s^{(1)} y^2 + \cdots
\end{align}
with
\begin{align}
    \psi_s^{(0)} = \frac{3}{1+\lambda} \cos \theta \sin^2 \theta. \label{eqpsis0}
\end{align}
The corresponding radial and angular velocity components in (\ref{eqP}) are given by $u_y = \frac{1}{\sin\theta}\frac{\partial \psi_s^{(0)}}{\partial \theta}y$ and $u_\theta = - \frac{\psi_s^{(0)}}{\sin\theta}$.

The $O(\hat{\alpha})$ terms in (\ref{vel1}-\ref{vel3}) denote the vortical contributions, being given by:
\begin{align}
    &u_y^{(1)} = 0, \label{ur1} \\
    &u_\theta^{(1)} = -\frac{1}{2}\sin \theta_\omega \sin \phi, \label{uth1}\\
    &u_\phi^{(1)} = \frac{1}{2}(\cos \theta_\omega \sin \theta - \sin \theta_\omega \cos \theta \cos \phi), \label{uphi1}
\end{align}
where $\theta_\omega$ is the inclination of the vorticity vector to the axis of symmetry. Note that \eqref{vel1}-\eqref{vel3} are exact, implying that the small-$\hat{\alpha}$ expansion for the velocity field terminates at first order. Therefore, in the expansion of $P$, $P_i = 0\,\forall\,i \geq 2$ and accordingly, the equations in the perturbation hierarchy above reduce to the form $P_0 \Theta_i + P_1 \Theta_{i-1} = 0$ for $i \geq 2$. In particular, the governing equation at $O(\hat{\alpha}^2)$, \eqref{eqalpsq}, reduces to:
\begin{align}
    P_0 \Theta_2 + P_1 \Theta_1 = 0,
\end{align}
the solution of which will yield the $O(\hat{\alpha}'^2)$ correction to the leading order prediction of \citet{Gupalo72}.
 
Using the above definitions, and the boundary layer variable $Y$, defined by $y = Y Pe^{-1/2}$, the governing equation at $O(1)$ becomes,
\begin{align}
    \left(\frac{\partial \psi_s^{(0)}}{\partial \theta} Y \right) \frac{\partial \Theta_0}{\partial Y} - \psi_s^{(0)} \frac{\partial \Theta_0}{\partial \theta} - \sin \theta \frac{\partial^2 \Theta_0}{\partial Y^2} = 0,
\end{align}
with the boundary conditions:
\begin{align}
    &\lim_{Y \rightarrow 0} \Theta_0 = \mathcal{H}(\theta), \\
    &\lim_{Y \rightarrow \infty} \Theta_0 = 0,
\end{align}
where the first condition with the Heaviside function pertains to an isothermal drop surface, while the second one imposes the approach to the ambient value outside the boundary layer. From this
point onward, we closely follow along the lines of \cite{Goddard65} by first using a coordinate
transformation of the form:
\begin{align}
    &z = \psi_s^{(0)} Y, \\
    &t = \int_0^{\theta} \psi_s^{(0)}(\theta') \sin \theta' d \theta'.
\end{align}
Using this transformation, the governing equation can be shown to reduce to:
\begin{align}
    \frac{1}{J} \left[\frac{\partial}{\partial t} - \frac{\partial^2}{\partial z^2}\right] \Theta_0 = 0, \label{eq:transformed}
\end{align}
where $J= [\sin \theta (\psi_s^{(0)})^2]^{-1}$ is the Jacobian of the transformation, with the boundary conditions now given by:
\begin{align}
    &\lim_{z \rightarrow 0} \Theta_0 = \mathcal{H}(t - 0), \\
    &\lim_{z \rightarrow \infty} \Theta_0 = 0.
\end{align}
From (\ref{eq:transformed}), $P_0$ in the new coordinate system is seen to be the diffusion operator, for which the Green's function satisfying the boundary conditions can be obtained using the method of images. An expression for this Green's function was already reported by \cite{Goddard65} for the general case where $\psi_s \propto Y^n$. A drop in an axisymmetric extension corresponds to $n=1$, and the Green’s function for this case is given by:
\begin{align}
    G(z,t;z^*,t^*) =& \frac{(z z^*)^{1/2}}{2(t - t^*)} \exp \left[-(\zeta^2 + \zeta^{*2}) \right] I_{\frac{1}{2}} (2 \zeta \zeta^*), \label{Greens_transf}
\end{align}
with $\zeta = z/(2(t - t^*)^{1/2})$, $\zeta^* = z^*/(2(t - t^*)^{1/2})$ and with $I_{\frac{1}{2}}$ being the modified Bessel function of the first kind, of order $1/2$\,(note that the general expression, Eq.(4.9) in \cite{Goddard65}) has a typographical error, with the denominator missing the quantity $\tau = t - t^* $). 

Using (\ref{Greens_transf}), $\Theta_0$ is given by:
\begin{align}
    \Theta_0(z,t) = \int_{0}^t \left[\frac{\partial G(z, t;z^*,t^*)}{\partial z^*} \right]_{z^* = 0} \mathcal{H}(t^*) dt^*,
\end{align}
The integral above can be evaluated to give:
\begin{align}
    \Theta_0 = \frac{\Gamma(1/2,\zeta^2)}{\Gamma(1/2)}, \label{The0}
\end{align}
where $\Gamma(1/2,\zeta^2)= \int_{\zeta^2}^{\infty} e^{-s} s^{(1/2) - 1} ds$ is the incomplete Gamma function. The resulting Nusselt number, $Nu^{(0)}$, is given by:
\begin{align}
    \frac{Nu^{(0)}}{Pe^{1/2}} = -\frac{1}{2 \pi} \int_0^{t_m} \int_0^{2 \pi} \left(\frac{\partial \Theta_0}{\partial z} \right)_{z = 0} dt \; d \phi = \sqrt{\frac{3}{(1+\lambda) \pi}},
\end{align}
which can be rewritten as:
\begin{align}
    \frac{Nu^{(0)}}{\hat{Pe}^{1/2}} = \sqrt{\frac{3}{\pi}},
\end{align}
which, of course, matches with the result of \cite{Gupalo72}.

We now use (\ref{The0}) to evaluate the forcing function, $-P_1 \Theta_0$, in \eqref{eqalp}. First, substituting the first order velocity components from \eqref{ur1}-\eqref{uphi1} into \eqref{eqP} gives:
\begin{align}
    P_1 = -\frac{\sin\theta_\omega}{2} \sin \phi \frac{\partial }{\partial \theta} + \frac{1}{2}\left(\cos\theta_\omega - \sin\theta_\omega \cot \theta \cos \phi \right)\frac{\partial }{\partial \phi},
\end{align}
which may be rewritten in terms of the new variables $z$ and $t$ as:
\begin{align}
    P_1 = -\frac{\sin\theta_\omega}{2} \sin \phi \left( \left(\frac{\dot{\psi_s^{(0)}} z}{\psi_s^{(0)}} \right)\frac{\partial }{\partial z} + (\psi_s^{(0)} \sin \theta)\frac{\partial }{\partial t} \right) + \frac{1}{2}\left(\cos\theta_\omega - \sin\theta_\omega \cot \theta \cos \phi \right)\frac{\partial }{\partial \phi}. \label{eqP1}
\end{align}
Thus, the governing equation at $O(\hat{\alpha})$ is given by:
\begin{align}
    P_0 \Theta_1 = - P_1 \Theta_0 = q(z,t), \label{goveqalp}
\end{align}
where $P_1$ and $\Theta_0$ are given by \eqref{eqP1} and \eqref{The0}, respectively, with the boundary conditions now being homogeneous, and given by:
\begin{align}
    \lim_{z \rightarrow 0} \Theta_1 = \lim_{z \rightarrow \infty} \Theta_1 = 0.
\end{align}
The formal solution of \eqref{goveqalp} can be written as\,\citep{Goddard65}:
\begin{align}
    \Theta_1 = \int_0^{t} \int_0^{\infty} G(z^*,t^*,z,t) J(t^*) q(t^*,z^*)dz^* dt^*,
\end{align}
which, after several steps of manipulations, yields:
\begin{align}
    \Theta_1(z,t) = \frac{e^{-\zeta^2}\left(2 \sqrt{3 t} - 2 t \sqrt{1+\lambda} -\psi_s^{(0)}\sqrt{1+\lambda} \right)z \eta \sin \phi}{4 \sqrt{\pi (1+\lambda)} t^{3/2} \psi_s^{(0)}}. \label{eqThe1}
\end{align}
The Nusselt number contribution at this order is given by:
\begin{align}
    \frac{Nu^{(1)}}{Pe^{1/2}} = -\frac{1}{2 \pi} \int_0^{t_m} \int_0^{2 \pi} \left(\frac{d \Theta_1}{d z} \right)_{z = 0} dt \; d \phi = 0,
\end{align}
as is expected on account of the invariance to rotation reversal; one expects the $Nu$-expansion to proceed in even powers of $\hat{\alpha}$.

At $O(\hat{\alpha}^2)$, one has:
\begin{align}
    P_0 \Theta_2 = -P_1 \Theta_1 = q_1(z,t),
\end{align}
with $P_1$ and $\Theta_1$ given by \eqref{eqP1} and \eqref{eqThe1}, and the formal solution again given by:
\begin{align}
    \Theta_2(z,t) = \int_0^{t} \int_0^{\infty} G(z^*,t^*,z,t) J(t^*) q_1(t^*,z^*)dz^* dt^*. \label{GF:theta2}
\end{align}
While the integral in (\ref{GF:theta2}) could not be evaluated analytically, 
$Nu^{(2)}$, which only involves $\left(\frac{d \Theta_1}{d z} \right)_{z = 0}$, can nevertheless be evaluated in closed form as:
\begin{align}
    \frac{Nu-Nu^{(0)}}{\hat{Pe}^{1/2}} = \frac{Nu^{(2)}}{\hat{Pe}^{1/2}}=& \frac{(\ln 2-2)\sin^2 \theta_\omega}{48\sqrt{3 \pi}} \hat{\alpha}'^2, \\
    =& f(\theta_\omega) \hat{\alpha}'^2, \label{Nu:smallalpha} 
\end{align}
where the $\lambda$-dependence has been absorbed into $\hat{\alpha}'$ and $\hat{Pe}$. Note that the coefficient $f(\theta_\omega)$ in (\ref{Nu:smallalpha}) is negative. Barring exceptional surface-streamline topologies, this is consistent with an increased rotational component leading to a reduced transport rate; see Figs.\ref{fig:Nu_aligned} and \ref{fig:Nu_inclined_ind}. Fig.\ref{fig:AppCD}(a) plots $|f(\theta_\omega)|$ as a function of $\theta_\omega$. The coefficient equals zero at $\theta_\omega = 0$, consistent with the expected $\hat{\alpha}$-independence of $Nu$ in the aligned-vorticity case, and increases monotonically\,(in magnitude) to a maximum at $\theta_\omega = \pi/2$. The asymptotic prediction given by (\ref{Nu:smallalpha}) has been used to supplement the numerical $Nu$-curves in Fig.\ref{fig:Nu_inclined_ind}.

\subsection{The $Nu$ asymptote for $\hat{\alpha}' \gg 1$} \label{LargeNu_asy}

In the limit $\hat{\alpha}' \rightarrow \infty$, the ambient flow is a solid-body rotation at leading order. The associated closed surface streamlines imply that the smaller $O(1/\hat{\alpha}')$ extensional component drives the convectively enhanced transport, and this latter contribution is calculated here. As mentioned in the main mansucript, the scenario of a dominant rotation occurs for a spherical particle in any ambient vortical flow at large $Pe$. While \citet{Batchelor79} examined this problem in the Stokes limit for a vortical linear flow, \citet{SubKoch06b} analyzed the subset of ambient planar linear flows, with the extensional component orthogonal to the plane arising from fluid inertia, and therefore being $O(Re)$. Following \citet{SubKoch06b}, we adopt a spherical coordinate system with its polar axis along the vorticity vector, so the velocity field given by (\ref{ur1}-\ref{uphi1}) may be rewritten as:
\begin{align}
    u_r =&\, u_r^{(1)}, \label{ur:2} \\
    u_\theta =&\, u_\theta^{(1)}, \label{utheta:2} \\
    u_\phi =&\, \hat{\alpha} u_\phi^{(0)} + u_\phi^{(1)}. \label{uphi:2}
\end{align}
The $O(\hat{\alpha})$ term in (\ref{uphi:2}) is the same as that in the small-$\hat{\alpha}'$ expansion given in the earlier Appendix, except that it is now the leading order contribution. Further, owing to the new choice for the polar axis, this contribution only has an azimuthal component given by $u_\phi^{(0)} = -\frac{1}{2}(1+y)\sin \theta$. The smaller $O(1)$ terms in (\ref{ur:2}-\ref{uphi:2}) are the leading order axisymmetric-extension contributions in (\ref{ur1}-\ref{uphi1}), and are now functions of $\theta_\omega$. Within the boundary layer framework, the convection-diffusion equation takes the form:
\begin{align}
    h_rY\frac{\partial \Theta}{\partial Y} + h_\theta \frac{\partial \Theta}{\partial \theta} + \frac{[\hat{\alpha}(-\sin\theta/2) + h_\phi]}{\sin \theta}\frac{\partial \Theta}{\partial \phi} = \frac{\partial^2 \Theta}{\partial Y^2}, \label{App:BLeqn}
\end{align}
where we have used that $u_r^{(1)} = h_rY$, $u_\theta^{(1)} = h_\theta$, and $u_\phi = \hat{\alpha}\frac{-\sin\theta}{2} + h_\phi$ close to the drop surface. Here, 
\begin{align}
\begin{split} 
&h_r(\theta,\phi;\theta_\omega,\lambda) = 
    \frac{3}{16 \alpha  (\lambda +1) \left(\cos 2 \theta_{\omega
}+3\right)} \left[-6 \cos ^2\phi  \left(\cos 4 \theta _{\omega }+3\right)+4 (3 \cos 2 \phi -5) \cos 2 \theta_{\omega }\right.\\
   &\left.- 3 \cos 2 \theta  \left(20 \cos 2 \theta _{\omega }+3 \cos
   4 \theta _{\omega }-8 \sin 2 \phi \sin 2 \theta _{\omega } \sin \theta _{\omega }-8 \cos 2 \phi \sin ^4\theta _{\omega }+9\right) + \right. \\
   &\left. +96 \sin 2 \theta \sin
   \phi \sin \theta _{\omega } \cos ^3\theta _{\omega } \sqrt{\tan ^2\theta_{\omega }+2} \right.\\
   &\left.-48 \cos \phi \sin \theta _{\omega } \left(\sin \phi \sin 2
   \theta _{\omega }-2 \sin 2 \theta \cos ^2\theta _{\omega } \sqrt{\tan ^2\theta _{\omega }+2}\right)\right],
\end{split}\\
\begin{split}
 &h_\theta(\theta,\phi;\theta_\omega,\lambda) = \frac{3}{16 \alpha  (\lambda +1) \left(\cos \left(2 \theta _{\omega }\right)+3\right) } \times \\
 &\left[-4 \sin ^2(\theta ) \sin (\phi ) \left(6 \sin \left(2 \theta _{\omega }\right)+\sin \left(4 \theta _{\omega }\right)\right) (\tan ^2\left(\theta _{\omega }\right)+2)^{-1/2} \right.\\
   &\left.+8 \cos ^2(\theta ) \left(5 \sin \left(\theta _{\omega }\right)+\sin \left(3 \theta
   _{\omega }\right)\right) \left(\sin (\phi ) \cos \left(\theta _{\omega }\right)+\cos (\phi )\right) (\tan ^2\left(\theta _{\omega }\right)+2)^{-1/2} \right.\\
   &\left.-8 \sin ^2(\theta ) \cos (\phi ) \left(5 \sin \left(\theta _{\omega }\right)+\sin \left(3 \theta _{\omega
   }\right)\right) (\tan ^2\left(\theta _{\omega }\right)+2)^{-1/2} \right.\\
   &\left. +\sin (2 \theta ) \left(20 \cos \left(2 \theta _{\omega }\right)+3 \cos \left(4 \theta _{\omega }\right)-8 \sin (2 \phi ) \sin \left(2 \theta _{\omega
   }\right) \sin \left(\theta _{\omega }\right)-8 \cos (2 \phi ) \sin ^4\left(\theta _{\omega }\right)+9\right)\right],   
\end{split}
\end{align}
with $h_\phi \equiv h_\phi(\theta,\phi;\theta_\omega)$, although its explicit form is not needed here.

Using the expansion $\Theta = \Theta_0 + \hat{\alpha}'^{-1}\Theta_1 + \dots$ in (\ref{App:BLeqn}), one obtains at leading order, 
\begin{equation}
    \frac{\partial \Theta_0}{\partial \phi} = 0,
\end{equation}
implying $\Theta_0 \equiv \Theta_0(r,\theta)$. Therefore, using $\Theta = \Theta_0(r,\theta) + \hat{\alpha}^{-1}\Theta_1(r,\theta,\phi)$, one obtains at first order:
\begin{align}
    h_rY\frac{\partial \Theta_0}{\partial Y} + h_\theta \frac{\partial \Theta_0}{\partial \theta} - \frac{1}{2}\frac{\partial \Theta_1}{\partial \phi} =\frac{\partial^2 \Theta_0}{\partial Y^2}.
\end{align}
Next, averaging over the $\phi$-coordinate via $\frac{1}{2\pi}\textstyle\int_0^{2\pi}(.)d\phi$ eliminates $\Theta_1$, and leads to the following equation for $\Theta_0$:
\begin{align}
    \frac{1}{2\pi}\left( \int_0^{2\pi} h_r d\phi \right)Y\frac{\partial \Theta_0}{\partial Y} + \frac{1}{2\pi} \left( \int_0^{2 \pi} h_\theta d \phi \right) \frac{\partial \Theta_0}{\partial \theta} = \frac{\partial^2 \Theta_0}{\partial Y^2}, \label{Eqn:Theta0}
\end{align}
where
\begin{align}
\bar{h}_r =&\frac{1}{2\pi} \int_0^{2 \pi} h_r d \phi = -\frac{3}{8(\lambda+1)}(1 + 3 \cos 2\theta )(1 + 3 \cos 2 \theta_\omega), \label{hrbar} \\
\bar{h}_\theta =&\frac{1}{2\pi} \int_0^{2 \pi} h_\theta d \phi = \frac{3}{8(\lambda+1)} \sin 2 \theta(1 + 3 \cos 2 \theta_\omega). \label{hthetabar}
\end{align}

\begin{figure}
    \centering
    \includegraphics[scale=0.36]{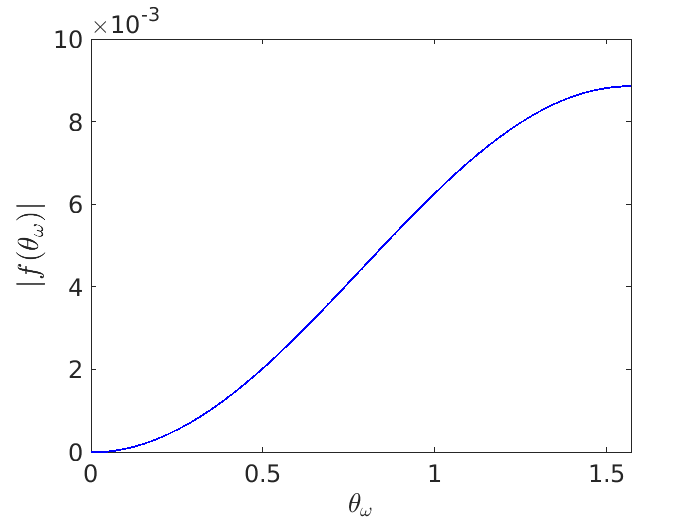}
    \includegraphics[scale=0.36]{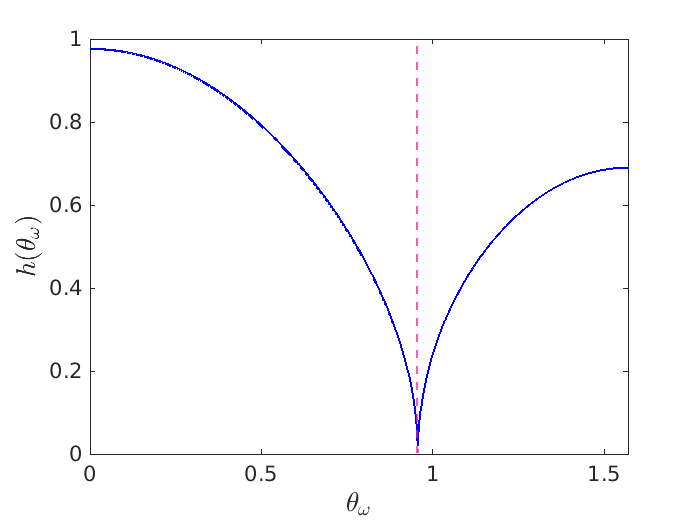}    
    \caption[\textwidth]{\justifying{Plot of (a) the small-$\hat{\alpha}'$ coefficient $f(\theta_\omega)$ and (b) the large-$\hat{\alpha}'$ coefficient $h(\theta_\omega)$, as a function of $\theta_\omega$. These coefficients characterize the zero and infinite-$\hat{\alpha}'$ plateaus in Fig.\ref{fig:Nu_inclined_ind}. With increasing $\theta_\omega$, $f$ increases monotonically from zero to a maximum at $\theta_\omega = \frac{\pi}{2}$; the dashed red line in (b), where $h = 0$ marks a closed surface-streamline topology.}}
    \label{fig:AppCD}
\end{figure}

Equation (\ref{Eqn:Theta0}) shows that, for large $\hat{\alpha}$\,(or $\hat{\alpha}'$, since $\lambda$ is fixed), the leading order scalar field within the boundary layer is governed by a $\phi$-averaged extensional flow. The analysis from this point onward is standard. Assuming $\Theta_0 \equiv \Theta_0(\eta)$ in terms of the similarity variable $\eta = Y/g(\mu)$, with $\mu = \cos \theta$, and with $g(\mu)$ characterizing the polar-angle dependence of the axisymmetrized boundary layer thickness, \eqref{Eqn:Theta0} reduces to,
\begin{align}
    \frac{d^2 \Theta_0}{d \eta^2} + 2\eta \frac{d \Theta_0}{d \eta} = 0, \label{BLeqnN}   
\end{align}
subject to the boundary conditions:
\begin{align}
    &\Theta_0 = 1 \; \text{at } \eta = 0, \\
    &\Theta \rightarrow 0 \; \text{for } \eta \rightarrow \infty.
\end{align}
The boundary layer thickness satisfies
\begin{align}
    \bar{h}_r g^2 + \frac{\bar{h}_\theta}{2}(1-\mu^2)^{1/2}\frac{d g^2}{d \mu} = -2, \label{BLeqnNS}
\end{align}
with $g$ being finite at the inlet for the axisymmetrized near-surface flow in (\ref{Eqn:Theta0}), and defined by (\ref{hrbar}-\ref{hthetabar}). For this case, as mentioned in section \ref{sec:incled_stream_org}, there is a change in the sense of $\phi$-averaged extension\,(from uniaxial to biaxial), and thence, a swapping of the inlet and wake locations across $\theta_\omega = \theta_\omega^{th1} = \tan^{-1}\sqrt{2}$. Thus, the BL thickness $g(\mu)$ must be finite at $\mu = 1$\,(or $\theta = 0$) for $\theta_\omega \leq \theta_\omega^{th1}$ and at $\mu = 0$ (or $\theta = \pi/2$) for $\theta_\omega > \theta_\omega^{th1}$. While the  scalar field is given by:
\begin{align}
    \Theta_0(\eta) = 1 - \frac{2}{\sqrt{\pi}} \int_0^\eta e^{-t^2} dt,
\end{align}
in both regimes, $g(\mu)$, which solves \eqref{BLeqnNS}, has different expressions:
\begin{equation}
    g(\mu) = \begin{cases}
        &\sqrt{\frac{4(1+\lambda)}{3|(1+3 \cos 2\theta_\omega)|}}\left( \frac{1}{\mu}\right) \text{ for } \theta_\omega < \theta_\omega^{th1},  \\
        &\sqrt{\frac{4(1+\lambda)}{3|(1+3 \cos 2\theta_\omega)|}}\left(\frac{\mu^2 - 2}{(\mu^2 - 1)^2}\right)^{1/2} \text{ for } \theta_\omega > \theta_\omega^{th1}. 
    \end{cases}.
\end{equation}
The invariance to ambient flow reversal implies that either of the above expressions may be used to calculate $Nu$ over the entire range of $\theta_\omega$. Using the biaxial one, one obtains:
\begin{align}
    Nu &= Pe^{\frac{1}{2}}\displaystyle\int_{-1}^{1} \frac{d\mu}{g(\mu)}\frac{d\Theta_0}{d\eta}|_{\eta =0}\\
    &= \frac{2\hat{Pe}^{1/2}}{\Gamma(1/2)} \left(\sqrt{\frac{4}{3|1+3 \cos 2\theta_\omega|}} \right)^{-1} \int_0^1 \mu \; d \mu, \\
        &=\sqrt{\frac{3(1+3 \cos 2\theta_\omega)}{4 \pi}} \hat{Pe}^{1/2} \\
    &= h(\theta_\omega) \hat{Pe}^{1/2}, \label{Nu:inftyalpha}
\end{align}
where $h(\theta_\omega)$ is plotted as a function of $\theta_\omega$ in  Fig.\ref{fig:AppCD}(b). It goes to zero at $\theta_\omega^{th1} = \tan^{-1} \sqrt{2}$. As explained in section \ref{sec:incled_stream_org}, this coincides with $\lim_{\hat{\alpha}' \rightarrow \infty} R' = 0$, and corresponds to an eccentric elliptic linear flow topology. $h(\theta_\omega)$ equals $(3/\pi)^{\frac{1}{2}}$ for $\theta_\omega = 0$, which is the same as the result of \cite{Gupalo72} for $\hat{\alpha}' = 0$, this being consistent with the independence of $Nu$ with respect to $\hat{\alpha}'$ for the aligned-vorticity case\,(see bounding curve  of the $Nu$-surface in Fig.\ref{fig:Nu_aligned} for $\epsilon = -2$).

The dependence on $\theta_\omega$ in (\ref{Nu:inftyalpha}) could have been inferred apriori using
the same arguments as \cite{Batchelor79} who showed that convectively enhanced transport is driven by the component of the extension along the vorticity vector, $E_\omega = \bm{E}:\bm{\omega \omega}/|\bm{\omega}|^2$, which comes out to be $(1 + 3 \cos 2 \theta_\omega)/4$. Thus, knowing the aligned-vorticity result, $(3/\pi)^{\frac{1}{2}}\hat{Pe}^{1/2}$, the one for an arbitrary $\theta_\omega$, given by (\ref{Nu:inftyalpha})  may simply be written as  $(3/\pi)^{\frac{1}{2}}\hat{Pe}_\omega^{1/2}$.

\bibliographystyle{jfm}

\end{document}